\documentclass[reprint,notitlepage,amsmath,amsfonts,amssymb,floatfix,nolongbibliography,prb,aps,superscriptaddress]{revtex4-2}
\usepackage{graphicx}
\usepackage{color}
\usepackage{braket}
\usepackage{ulem}

\definecolor{EmoryGold}{RGB}{181, 133, 0}

\newcommand{\JW}{JW}
\newcommand{{\BK}}{BK}
\newcommand{{\parity}}{$\Pi$}
\newcommand{{\rotatedparity}}{$\Pi_R$}
\newcommand{\sierpinski}{Sierpinski}
\begin{document}
\title{Fermionic Mean-Field Theory as a Tool for Studying Spin Hamiltonians}
\author{Thomas M. Henderson}
\email{thomas.henderson@rice.edu}
\affiliation{Department of Chemistry, Rice University, Houston, Texas 77005, USA}
\affiliation{Department of Physics and Astronomy, Rice University, Houston, Texas 77005, USA}

\author{Brent Harrison}
\affiliation{Department of Physics and Astronomy, Dartmouth College, Hanover, New Hampshire 03755, USA}

\author{Ilias Magoulas}
\affiliation{Department of Chemistry and Cherry Emerson Center for Scientific Computation, Emory University, Atlanta, Georgia 30322, USA}

\author{Jason Necaise}
\affiliation{Department of Physics and Astronomy, Dartmouth College, Hanover, New Hampshire 03755, USA}

\author{Andrew M. Projansky}
\affiliation{Department of Physics and Astronomy, Dartmouth College, Hanover, New Hampshire 03755, USA}

\author{Francesco A. Evangelista}
\affiliation{Department of Chemistry and Cherry Emerson Center for Scientific Computation, Emory University, Atlanta, Georgia 30322, USA}

\author{James D. Whitfield}
\affiliation{Department of Physics and Astronomy, Dartmouth College, Hanover, New Hampshire 03755, USA}
\affiliation{AWS Center for Quantum Computing, Pasadena, California 91125, USA}

\author{Gustavo E. Scuseria}
\affiliation{Department of Chemistry, Rice University, Houston, Texas 77005, USA}
\affiliation{Department of Physics and Astronomy, Rice University, Houston, Texas 77005, USA}
\date{\today}

\begin{abstract}
The Jordan--Wigner transformation permits one to convert spin $1/2$ operators into spinless fermion ones, or vice versa.  In some cases, it transforms an interacting spin Hamiltonian into a noninteracting fermionic one which is exactly solved at the mean-field level.  Even when the resulting fermionic Hamiltonian is interacting, its mean-field solution can provide surprisingly accurate energies and correlation functions.  Jordan--Wigner is, however, only one possible means of interconverting spin and fermionic degrees of freedom.  Here, we apply several such techniques to the XXZ and $J_1\text{--}J_2$ Heisenberg models, as well as to the pairing or reduced BCS Hamiltonian, with the aim of discovering which of these mappings is most useful in applying fermionic mean-field theory to the study of spin Hamiltonians.
\end{abstract}

\maketitle

\section{Introduction}
It is textbook material \cite{Batista2001,Nishimori2011} that the Jordan--Wigner (JW) transformation \cite{Jordan1928} can convert certain interacting Hamiltonians of spin 1/2 systems into noninteracting (therefore exactly solvable) Hamiltonians of spinless fermions.  Perhaps less appreciated is that even when the {\JW} transformation produces an interacting fermionic Hamiltonian, it can nevertheless be solved at the mean-field level with moderate computational cost and good accuracy \cite{Goncalves2005,Verkholyak2006,Kitaev2009,Verkholyak2010,Bardyn2012,Zvyagin2013,Greiter2014,Gebhard2022,Henderson2022,Henderson2024a}.

Historically, the {\JW} mapping was the first transformation used to relate spin 1/2 objects to fermions. Recently, the prospect of simulating fermionic systems on quantum devices has reignited interest in such mappings \cite{Bravyi2002,Cao2019,Bauer2020,McArdle2020}, largely due to the parallels between qubits and spin 1/2 systems. Representing many-fermion problems on a quantum computer requires a mapping from fermions to qubits. Although the inverse {\JW} transformation is a natural choice for this purpose, the past several years have seen the development of a variety of %alternative mappings
alternatives \cite{Bravyi2002,Verstraete2005,Seeley2012,Havlicek2017,Bravyi2017,Setia2019,Jiang2020,Picozzi2023,Liu2024,Harrison2023,OBrien2024,Harrison2024b}.

In this work, we focus on a subset of these alternative mappings and apply them in the other direction.  That is, given a transformation converting spinless fermions to spin 1/2 objects, we can invert the mapping to convert spin 1/2 objects to spinless fermions, thereby enabling the solution of interacting spin Hamiltonians using fermionic methods.  Of course, the inversion of these mappings is not an entirely trivial task and the resulting fermionic Hamiltonians can be very exotic, but we aim to survey the landscape and discern which, if any, of the many spin-to-fermion mappings we will consider might be particularly useful.

To help orient the reader, we will begin with a quick survey of the spin-to-fermion transformations that we will consider in this work.  We certainly cannot do full justice to this topic here, but we hope to provide a reasonably concise summary of these ideas.  Likewise, because the Hamiltonians we obtain after the spin-to-fermion transformation are generally quite unusual, we must employ equally unusual mean-field methods to solve them, and we also provide a brief introduction to the general Hartree--Fock--Bogoliubov--Fukutome (HFBF) mean-field theory \cite{Fukutome1977,Fukutome1977b,Moussa2018,Henderson2024b} we use to tackle these problems.

\section{Spin-to-Fermion Transformations}
The Hilbert space for a spin 1/2 object is spanned by two states: $\ket{\uparrow}$ and $\ket{\downarrow}$.  The same is true for the Hilbert space of a spinless fermion, spanned by states $\ket{0}$ and $\ket{1}$ in the occupation number representation.  Since the Hilbert spaces have the same size, we can map states for a single spinless fermion onto states for a single spin 1/2, and vice versa.  The only real difficulty in extending these ideas to map states for $M$ fermions onto states for $M$ spins is that operators acting on different fermions anticommute, while operators acting on different spins commute.  This idea was discussed by Jordan and Wigner in 1928 \cite{Jordan1928}, but as we have said, a plethora of different mappings have since been introduced.  Here, we want to summarize those which we will explore in this work.

\subsection{Notation, and Majorana and Pauli Operators}
It will prove helpful to establish our basic concepts and notation first.  The bare fermionic creation and annihilation operators for level $p$ will be denoted as $c_p^\dagger$ and $c_p$, respectively, and they interconvert the two fermionic states, via
\begin{subequations}
\begin{align}
c_p^\dagger \ket{0_p} &= \ket{1_p},
\\
c_p \ket{1_p} &= \ket{0_p}.
\end{align}
\end{subequations}
These operators obey canonical anticommutation relations:
\begin{subequations}
\begin{align}
\{c_p,c_q\} &= 0,
\\
\{c_p,c_q^\dagger\} &= \delta_{pq},
\end{align}
\end{subequations}
where $\{A,B\} = AB + BA$ is the anticommutator. The number operator $n_p = c_p^\dagger \, c_p$ determines whether level $p$ is occupied or empty.

We can draw a close analogy between these fermionic operators and the spin operators.  The spin raising and lowering operators, $S_p^+$ and $S_p^-$, interconvert the two spin states:
\begin{subequations}
\begin{align}
S_p^+ \ket{\downarrow_p} &= \ket{\uparrow_p},
\\
S_p^- \ket{\uparrow_p} &= \ket{\downarrow_p}.
\end{align}
\end{subequations}
Where the fermionic states $\ket{0_p}$ and $\ket{1_p}$ are eigenstates of the number operator $n_p$ with eigenvalues 0 and 1, respectively, the spin states $\ket{\uparrow_p}$ and $\ket{\downarrow_p}$ are instead eigenstates of the operator $S_p^z$ with respective eigenvalues $+1/2$ and $-1/2$.  However, the spin operators obey $\mathfrak{su}(2)$ commutation rules:
\begin{subequations}
\begin{align}
[S_p^+,S_q^-] &= 2 \, \delta_{pq} \, S_p^z,
\\
[S_p^z,S_q^\pm] &= \pm \delta_{pq} \, S_p^\pm.
\end{align}
\end{subequations}
We also have the $x$ and $y$ spin operators, which we can obtain from
\begin{equation}
S_p^\pm = S_p^x \pm \mathrm{i} \, S_p^y,
\end{equation}
which together with $S_p^z$ satisfy
\begin{equation}
[S_p^i,S_q^j] = \mathrm{i} \, \delta_{pq} \, \epsilon_{ijk} \, S_p^k,
\end{equation}
%IM: In my mind, "run over" implies summation. If the following change is inappropriate, feel free to revert back to the original.
where $i,j,k \in \{x,y,z\}$.  Since we are working purely with spin 1/2, we also have that
\begin{equation}
\left(S_p^i\right)^2 = 1/4.
\end{equation}

While the creation and annihilation operators and the spin operators may be more familiar, they are less convenient for our purposes.  Instead, it is easier to use Majorana operators and Pauli operators.  We define the Majorana operators $\gamma_{1,p}$ and $\gamma_{2,p}$ as
\begin{subequations}
\begin{align}
\gamma_{1,p} &= c_p + c_p^\dagger,
\\
\gamma_{2,p} &= \mathrm{i} \, \left(c_p - c_p^\dagger\right).
\end{align}
\end{subequations}
These operators are Hermitian (e.g., $\gamma_{1,p}^\dagger = \gamma_{1,p}$), with $\gamma_{1,p}^2 = \gamma_{2,p}^2 = 1$, and the different Majorana operators all anticommute with one another.  Notice that
\begin{equation}
\mathrm{i} \, \gamma_{1,p} \, \gamma_{2,p} = 1 - 2 \, n_p.
\end{equation}

Similarly, we can define Pauli operators, $\sigma_p^i$, from
\begin{equation}
S_p^i = \frac{1}{2} \, \sigma_p^i.
\end{equation}
Like the Majoranas, the Pauli operators are Hermitian and are involutions ($O^2 = 1$).  Paulis on different sites commute, while for Paulis on the same site we may use
\begin{equation}
\sigma_p^i \, \sigma_p^j = \delta_{ij} \, 1 + \mathrm{i} \, \epsilon_{ijk} \, \sigma_p^k.
\end{equation}

The fact that both Majorana and Pauli operators are Hermitian involutions simplifies converting between fermions and spins; specifically, by taking advantage of the mathematical properties of Pauli and Majorana operators, we reduce the process of finding the inverse transformation to a matrix inversion problem.  In practice, having used Pauli and Majorana operators to derive spin to fermion mappings, we must still convert the Hamiltonian from spin operators to Pauli operators (by including appropriate factors of two), then map to Majoranas, and finally convert back to the usual creation and annihilation operators which form the basis of our numerical implementation.

\subsection{The Jordan--Wigner Transformation}
Based on the similarities we have noted between the spin operators on the one hand and the fermionic operators on the other, Jordan and Wigner \cite{Jordan1928} suggested a mapping which, when transforming fermions to spins, we may write as
\begin{subequations}
\begin{align}
c_p^\dagger &\underset{\text{JW}}{\mapsto} S_p^+ \, \phi_p^z,
\\
c_p &\underset{\text{JW}}{\mapsto} S_p^- \, \phi_p^z,
\\
\phi_p^z &= \prod_{q<p} (-2 \, S_q^z),
\end{align}
\end{subequations}
which just means that for states we map
\begin{subequations}
\begin{align}
\ket{0_p} &\underset{\text{JW}}{\mapsto} \ket{\downarrow_p},
\\
\ket{1_p} &\underset{\text{JW}}{\mapsto} \ket{\uparrow_p}.
\end{align}
\end{subequations}
The Hermitian operator $\phi_p^z$ is the {\JW} string and is responsible for yielding the correct anticommutation relationships.  Note that in the quantum computing community, it is common to map $\ket{0_p}$ to $\ket{\uparrow_p}$, which means that the creation operator $c_p^\dagger$ maps to $S_p^- \, \hat{\phi}_p^z$, where $\hat{\phi}_p^z$ does not have signs in defining the {\JW} strings.
%swap roles of $S_p^+$ and $S_p^-$, which has the virtue of eliminating the minus signs in defining the {\JW} strings.
%IM: Alternatively, we can say that in quantum computing \ket{0_p} maps to \ket{\uparrow_p} and \ket{1_p} to \ket{\downarrow_p}.

This transformation can be inverted to map spin operators to fermions:
\begin{subequations}
\begin{align}
S_p^+ &\underset{\text{JW}}{\mapsto} c_p^\dagger \, \tilde{\phi}_p,
\\
S_p^- &\underset{\text{JW}}{\mapsto} c_p \, \tilde{\phi}_p,
\\
S_p^z &\underset{\text{JW}}{\mapsto} n_p - \frac{1}{2},
\\
\tilde{\phi}_p &= \prod_{q<p} \mathrm{e}^{\mathrm{i} \, \pi \, n_q} = \prod_{q<p} \left(1 - 2 \, n_q\right).
\end{align}
\end{subequations}

The {\JW} transformation has the virtue of simplicity.  Moreover, because it maps the total $S^z$ operator into the total number operator $N$ (modulo an irrelevant shift) it converts fermionic Hamiltonians with number symmetry into spin Hamiltonians with $S^z$ symmetry and vice versa.

We note that the {\JW} transformation depends on the labeling of the sites or fermions, because in the strings for level $p$ we have products over $q<p$.  This dependence is of no practical significance if the Hamiltonian is solved exactly, but when mapped Hamiltonians are solved approximately, the result may depend on the ordering of the spins or fermions. One can take advantage of this additional degree of freedom and choose the ordering that minimizes a given cost function, such as the number of Pauli operators in the transformed Hamiltonian \cite{Chiew2023}. Alternatively,  there is an extension of the {\JW} transformation \cite{Wang1990} which, at the mean-field level, eliminates this dependency \cite{Henderson2024a}.

Finally, we have given the {\JW} transformation in terms of bare fermion and spin operators.  Because we find it easier to express the parity transformation in terms of Majorana and Pauli operators, it may be helpful to do the same here.  One finds
\begin{subequations}
\label{Eqn:JW}
\begin{align}
\gamma_{1,p} &\underset{\text{JW}}{\mapsto} \sigma_p^x \, \phi_p^z,
\\
\gamma_{2,p} &\underset{\text{JW}}{\mapsto} \sigma_p^y \, \phi_p^z,
\\
\phi_p^z &= \prod_{q<p} \left(-\sigma_q^z\right),
\\
\sigma_p^x &\underset{\text{JW}}{\mapsto} \gamma_{1,p} \, \tilde{\phi}_p,
\\
\sigma_p^y &\underset{\text{JW}}{\mapsto} \gamma_{2,p} \, \tilde{\phi}_p,
\\
\sigma_p^z &\underset{\text{JW}}{\mapsto} -\mathrm{i} \, \gamma_{1,p} \, \gamma_{2,p},
\\
\tilde{\phi}_p &= \prod_{q<p} \left(\mathrm{i} \, \gamma_{1,q} \, \gamma_{2,q}\right).
\end{align}
\end{subequations}

\subsection{The Parity Transformation}
The parity ($\Pi$) transformation \cite{Seeley2012} is in some sense the dual companion of the {\JW} transformation.  In the {\JW} case, we map the occupation of fermionic level $p$ into the direction of spin $p$.  Because fermions anticommute, occupation numbers alone are not enough to distinguish states, and we must also determine sign information, or parity.  In the {\JW} case, the sign information through site $p$ is found by looking at the occupations of all levels $q < p$.  Thus, in {\JW}, \textit{occupation} is encoded locally but \textit{parity} is encoded nonlocally.  In contrast, the {\parity} transformation encodes parity information locally, but then must encode occupancy nonlocally.

Specifically, we have
\begin{subequations}
\begin{align}
\gamma_{1,p} &\underset{\Pi}{\mapsto} \psi_p^x \, \sigma_{p-1}^z,
\\
\gamma_{2,p} &\underset{\Pi}{\mapsto} -\mathrm{i} \, \psi_p^x \, \sigma_p^z,
\\
\psi_p^x &= \prod_{q \ge p} \sigma_q^x.
\end{align}
\end{subequations}

Inverting this transformation yields
\begin{subequations}
\begin{align}
\sigma_p^x &\underset{\Pi}{\mapsto} \mathrm{i} \, \gamma_{2,p} \, \gamma_{1,p+1},
\\
\sigma_p^y &\underset{\Pi}{\mapsto} \mathrm{i} \, \tilde{\phi}_p \, \gamma_{1,p} \, \gamma_{1,p+1},
\\
\sigma_p^z &\underset{\Pi}{\mapsto} \mathrm{i} \, \tilde{\phi}_p \, \gamma_{1,p} \, \gamma_{2,p}.
%IM: Why not simply \tilde{\phi}_{p+1}?
%TMH: No good reason, of course -- I just think it's easier to see the su(2) if the strings are the same.
\end{align}
\end{subequations}
With $M$ sites, counting from 1, we must define $\sigma_0^z$ and $\gamma_{1,M+1}$, which we take to be, respectively,
\begin{subequations}
\begin{align}
\sigma_0^z &\equiv 1,
\\
\gamma_{1,M+1} &\equiv \Pi = \tilde{\phi}_{M+1} = \mathrm{e}^{\mathrm{i} \, \pi \, N}.
\label{Eqn:DefNumberParity}
\end{align}
\end{subequations}
The operator $\Pi$ is the global number parity operator and acts on number eigenstates to return eigenvalues of $\pm 1$, depending on whether the state has an even (eigenvalue $+1$) or an odd (eigenvalue $-1$) number of particles.  The presence of this number parity operator leads us to use the symbol $\Pi$ to refer also to the parity transformation.

Note that the {\JW} transformation singles out the $z$ axis for special treatment, in that $\sigma_p^z$ is the Pauli operator which does not get a string when it is mapped, while the {\parity} transformation instead singles out the $x$ axis.  Of course these choices of axes are merely a matter of convention, and can be rotated arbitrarily.  It will be useful to refer to a ``rotated parity'' transformation which, like the {\JW} mapping, singles out the $z$ axis for special treatment.  This rotated {\parity} transformation, denoted by $\Pi_R$, just has $\{\sigma_p^x,\sigma_p^y,\sigma_p^z\} \to \{\sigma_p^z,\sigma_p^x,\sigma_p^y\}$ so that
\begin{subequations}
\label{Eqn:PiRot}
\begin{align}
\gamma_{1,p} &\underset{\Pi_R}{\mapsto} \psi_p^z \, \sigma_{p-1}^y,
\\
\gamma_{2,p} &\underset{\Pi_R}{\mapsto} -\mathrm{i} \, \psi_p^z \, \sigma_p^y,
\\
\psi_p^z &= \prod_{q \ge p} \sigma_k^z,
\\
\sigma_p^x &\underset{\Pi_R}{\mapsto} \mathrm{i} \, \tilde{\phi}_p \, \gamma_{1,p} \, \gamma_{1,p+1},
\\
\sigma_p^y &\underset{\Pi_R}{\mapsto} \mathrm{i} \, \tilde{\phi}_p \, \gamma_{1,p} \, \gamma_{2,p},
\\
\sigma_p^z &\underset{\Pi_R}{\mapsto} \mathrm{i} \, \gamma_{2,p} \, \gamma_{1,p+1},
\\
\sigma_0^y &\equiv 1.
\end{align}
\end{subequations}

Note that the {\parity} transformation converts a single Pauli into a product of two Majoranas, possibly times a string.  Since the string is itself a product of an even number of Majoranas, it seems like the {\parity} transformation should convert each Pauli to a fermionic operator which conserves number parity (since operators which are products of an even number of Majoranas conserve number parity).  Unfortunately, at the $M^{\text{th}}$ site, we instead have, for the {\parity} transformation, $\sigma_M^x \mapsto \mathrm{i} \, \gamma_{2,M} \, \Pi$, and this operator changes an even number parity state into an odd number parity state, and vice versa.  Something similar happens for $\sigma_M^y$.  This means that spin Hamiltonians which contain these two operators will typically transform into fermionic Hamiltonians which break number parity symmetry.

\subsection{The Bravyi--Kitaev and Sierpinski Transformations}
%\notes{F}{I feel like the text below is repetitive and could be condensed. See the following edited version}

%\edited{
The {\JW} and {\parity} transformations both use the spin state to represent the information of the occupied orbitals in the corresponding fermionic states, but they do so with a sub-optimal cost due to the need for long strings of Pauli operators to handle occupation changes and phase retrieval.
In both cases, evaluating changes in occupation and fermionic phases scales as $M$, for $M$ spin states.
However, we can choose our encoding so that it reduces the lookup and update cost associated with the phase and occupation number of fermionic states. The Bravyi--Kitaev (BK) transformation \cite{Bravyi2002,Seeley2012} achieves this goal with $\mathcal{O}(\log_2(M))$ scaling in the following way.
%}

%\remove{
%The {\JW} and {\parity} transformations both use the spin state to represent the information of the occupied orbitals in the corresponding fermionic states. The {\JW} transformation uses each spin degree of freedom to directly encode the occupation of the corresponding fermionic mode. In this mapping, the change in occupation of the $p$th fermionic mode is carried out via $\sigma_p^{\pm}$ (spin flip) and a string of $\sigma_q^z$ operators acting on all sites before $p$ to extract the fermionic sign (phase). Conversely, in the {\parity} transformation, each spin state encodes the partial parity of the corresponding fermionic mode, i.e., the partial sum modulo 2 of the occupation numbers. Here, the change in occupation of the $p$th fermionic mode is carried out via a string of $\sigma_q^x$ operators with $q \geq p$ to update the parity of the corresponding spin states, and a single $\sigma_p^z$ to obtain the fermionic phase. \add{However, we ...} We can choose our encoding so that it reduces the lookup and update cost associated with \add{the} phase and occupation number of fermionic states. The Bravyi--Kitaev (BK) transformation \cite{Bravyi2002,Seeley2012} achieves this \add{goal} with $\log_2(M)$ scaling,\notes{F}{Shuldn't we use big-O notation here as done later in the text?} where $M$ is the number of spin states, in the following way.
%}

%\add{
The BK transformation improves the encoding efficiency by iteratively bisecting sets of fermionic modes to store partial sums in spin states. This approach enables efficient parity checks for any set of modes while minimizing the number of spins that need to be flipped during raising or lowering operations \cite{bk_prop}.
%}
%\remove{
%We iteratively bisect sets of fermionic modes to store in the spin states as partial sums, to guarantee an efficient route for checking the total parity of any set of modes, while minimizing the number of parity spins which will need to be flipped for any particular raising or lowering operation \cite{bk_prop}.
%}
%\edited{
The transformation is captured by a matrix $\beta_M^{\text{BK}}$, which linearly maps Fock basis states $\ket{f}$ to encoded spin states $\ket{\sigma}^{\text{BK}}$ using addition modulo $2$.
This matrix is illustrated below for $M=8$:%}
\begin{equation} \label{bktransf}
\ket{\sigma}^{\text{BK}} \equiv \beta_{M}^{\text{BK}} \ket{f} :=
\begin{pmatrix} 1 &   &   &   &   &   &   &   \\
                1 & 1 &   &   &   &   &   &   \\      
                  &   & 1 &   &   &   &   &   \\
                1 & 1 & 1 & 1 &   &   &   &   \\
                  &   &   &   & 1 &   &   &   \\
                  &   &   &   & 1 & 1 &   &   \\
                  &   &   &   &   &   & 1 &   \\
                1 & 1 & 1 & 1 & 1 & 1 & 1 & 1 \\
\end{pmatrix} \begin{pmatrix} n_1 \\ n_2 \\ n_3 \\ n_4 \\ n_5 \\ n_6 \\ n_7 \\ n_8 \end{pmatrix},
\end{equation}
resulting in
\begin{equation}
\begin{pmatrix} \sigma_1 \\ \sigma_2 \\ \sigma_3 \\ \sigma_4 \\ \sigma_5 \\ \sigma_6 \\ \sigma_7 \\ \sigma_8 \end{pmatrix} = \begin{pmatrix} n_1 \\ n_1 + n_2 \\ n_3 \\ n_1 + n_2 + n_3 + n_4 \\ n_5 \\ n_5 + n_6 \\ n_7 \\ n_1 + n_2 + n_3 + n_4 + n_5 + n_6 + n_7 + n_8 \end{pmatrix}.
\end{equation}

All the information needed to replicate the group action of the fermionic raising and lowering operators is accessible from these spins.
%\notes{F}{I feel like the direction is slightly shifting here but the reader is not told where we are headed? Are we now trying to reformulate BK in a more general framework that encompasses other methods? I feel like this also around Eq 22 and 23, where we introduce notation that can be potentially confused with the one used for the parity mapping.}
%\remove{For each $i$, there are three sets of spin indices to consider: the Update set $U(i)$, the Parity set $P(i)$, and the Flip set $F(i)$. The first two sets are self-explanatory, being the spin sites needed to flip if the occupation of fermionic mode $i$ is changed and the minimum set needed to query to compute the parity of fermionic modes up to $i-1$, respectively. The Flip set is the smallest set of additional spin states which must be queried to determine whether the $i$th spin state is equal to the $i$th fermionic occupation number or its opposite. }
%\add{
For each mode $i$, there are three sets of spin indices to consider.
The Update set $U(i)$, comprises the spin sites needed to flip if the occupation of fermionic mode $i$ is changed.
The Parity set $P(i)$, the minimum set of spins required to compute the parity of fermionic modes up to $i-1$.
The Flip set $F(i)$, i.e., the smallest set of additional spin states needed to determine whether the $i$th spin state is equal to the $i$th fermionic occupation number or its opposite.
%}
%\notes{F}{Can we show which of these is which for the $M=8$ example? That would be helpful.}
For ease of notation, an operator with the subscript set $\mathcal{S}$ will be taken to mean ``the tensor product of this operator on each element in $\mathcal{S}$'' where appropriate; e.g., $(\sigma^z)_\mathcal{S} = \prod_{s\in \mathcal{S}}(\sigma^z)_s$. 

To know whether to associate $c^{\dagger}_i$ with $S^+_i$ or $S^-_i$, consider the projection operators onto spaces of even or odd parity over a set $\mathcal{S}$:
\begin{equation}
\begin{split}
E_\mathcal{S} = \frac{1}{2} (I + (\sigma^z)_\mathcal{S}) \qquad O_\mathcal{S} = \frac{1}{2} (I - (\sigma^z)_\mathcal{S})
\end{split}
\end{equation}
and incorporate the knowledge from the Flip set to guarantee that the correct association is made with the new {\BK} spin raising and lowering operators:
\begin{equation}
\begin{split}
    \Pi^+_i &= S^+_i \otimes E_{F(i)} - S^-_i \otimes O_{F(i)}\\
            &= \frac{1}{2}\left((\sigma^x)_i \otimes (\sigma^z)_{F(i)} - (\mathrm{i} \, \sigma^y)_i\right)\\
    \Pi^-_i &= S^-_i \otimes E_{F(i)} - S^+_i \otimes O_{F(i)}\\
            &= \frac{1}{2}\left((\sigma^x)_i \otimes (\sigma^z)_{F(i)} + (\mathrm{i} \, \sigma^y)_i\right)
\end{split}
\end{equation}

Finally, with properly constructed {\BK} spin raising and lowering operators, and Update, Parity, and Flip sets, it is possible to define the fermionic creation and annihilation operators, as
\begin{equation}
\begin{split} 
c^{\dagger}_i &\underset{\text{BK}}{\mapsto} (\sigma^z)_{P(i)} \otimes (\sigma^x)_{U(i)} \otimes \Pi^+_i,
\\
c_i &\underset{\text{BK}}{\mapsto} (\sigma^z)_{P(i)} \otimes (\sigma^x)_{U(i)} \otimes \Pi^-_i.
\end{split}
\end{equation}

This formalism is completely generic. The only thing that determines the representation of the encoding is the occupation storage matrix $\beta_M$. For {\JW}, this is the identity matrix, while for the {\parity} transformation, it is the lower-diagonal matrix with all 1s. 

The bisections in the construction of the {\BK} encoding matrix $\beta^{\text{BK}}_M$ ensure $O(\log_2(M))$ scaling for all three Update, Parity, and Flip sets. Meanwhile, the straightforward nature of the {\JW} encoding matrix $\beta^{\text{JW}}_M$ allows for $O(1)$ size Update and Flip sets, but then requires an $O(M)$ size Parity set. Similarly, the {\parity} encoding matrix $\beta^{\Pi }_M$ allows for $O(1)$ size Parity and Flip sets, but then requires an $O(M)$ size Update set.

The ternary tree encoding \cite{Jiang2020} improves this yet further by building anticommuting spin operators to act as Majoranas with only $O(\log_3(M))$ weight, which the authors prove is optimal. However, this construction works directly on building the operators, and thus is not guaranteed to map Slater determinants in the Fock space to elementary spin basis states; for example, the fermionic vacuum state may be mapped to an arbitrary superposition of spin configuration states. In the recently introduced Sierpinski tree (S) encoding \cite{Harrison2024b}, the authors extend the partial sum framework seen in {\BK} and define a $\beta^{S}_M$ which yields an encoding which also has the optimal $O(\log_3(M))$ weight, but with the added advantage that Slater determinants in the Fock space are guaranteed to map to spin basis states.  In this case, the partial sums are allocated according to trisections, reflecting the fact that we get three anticommuting Pauli operators per spin site. This is illustrated for $M = 13$:
\begin{equation}
    \beta^{\text{S}}_M := 
    \begin{pmatrix}
        1 \\
        1 & 1 & 1 \\
          &   & 1 \\
          &   &   & 1 \\
        1 & 1 & 1 & 1 & 1 & 1 & 1 & 1 & 1 \\
          &   &   &   &   & 1 \\
          &   &   &   &   &   & 1 \\
          &   &   &   &   &   & 1 & 1 & 1 \\
          &   &   &   &   &   &   &   & 1 \\ 
          &   &   &   &   &   &   &   &   & 1 \\
          &   &   &   &   &   &   &   &   & 1 & 1 & 1 & \\
          &   &   &   &   &   &   &   &   &   &   & 1 & \\
          &   &   &   &   &   &   &   &   &   &   &   & 1
    \end{pmatrix}
\end{equation}
This allows us to get Update, Parity, and Flip sets all of size $O(\log_3(M))$, an improvement still on {\BK} which is provably optimal when $2M + 1$ is a power of 3.

\section{Fermionic Mean-Field Methods}
Fermionic mean-field theory is a surprisingly complex topic.  Frequently when one hears the term, one envisions Hartree--Fock (HF), in which the bare fermionic operators $c$ and $c^\dagger$ are replaced by a transformed set of operators $a$ and $a^\dagger$, respectively, where
\begin{equation}
a_p^\dagger = \sum_q U_{qp} \, c_q^\dagger
\end{equation}
and where the matrix $\boldsymbol{U}$ is unitary so that the transformation remains canonical.  The mean-field state is then the product of an occupied set of $N$ creation operators acting on the physical vacuum $\ket{-}$:
\begin{equation}
\ket{\Phi_\mathrm{HF}} = \prod_{i=1}^N a_i^\dagger \ket{-},
\end{equation}
and the resulting energy, defined as an expectation value, is minimized with respect to $\boldsymbol{U}$.  It is convenient to represent $\boldsymbol{U}$ as the exponential of an antihermitian one-body matrix $\boldsymbol{\tau}$, and this is closely related to the Thouless representation \cite{Thouless1960} of the HF wave function, in which we write
\begin{subequations}
\begin{align}
\ket{\Phi_\mathrm{HF}} &= \mathrm{e}^{\tau} \ket{\Phi_0},
\\
\tau &= \sum_{ia} \left(\tau_i^a \, c_a^\dagger \, c_i - h.c.\right),
\\
\ket{\Phi_0} &= \prod_i c_i^\dagger \, \ket{-},
\end{align}
\end{subequations}
where $\ket{\Phi_0}$ is some suitably chosen reference; the index $i$ runs over levels occupied in $\ket{\Phi_0}$ and $a$ runs over levels empty in $\ket{\Phi_0}$.

It is frequently more convenient to use a non-unitary representation instead:
\begin{subequations}
\begin{align}
\ket{\Phi_\mathrm{HF}} &= \mathcal{N} \, \mathrm{e}^{T} \ket{\Phi_0},
\\
T &= \sum_{ia} T_i^a \, c_a^\dagger \, c_i,
\end{align}
\end{subequations}
where $\mathcal{N}$ is a normalization constant.  So long as $\braket{\Phi_0|\Phi_\mathrm{HF}} \ne 0$, this non-unitary approach is possible, though in practice we require the overlap $\braket{\Phi_0|\Phi_\mathrm{HF}}$ to be sufficiently large so as to stave off numerical difficulties.

\begin{figure*}[t]
\includegraphics[width=0.45\textwidth]{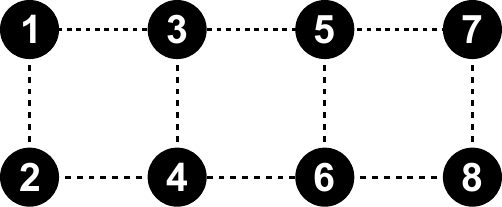}%
\hfill
\includegraphics[width=0.45\textwidth]{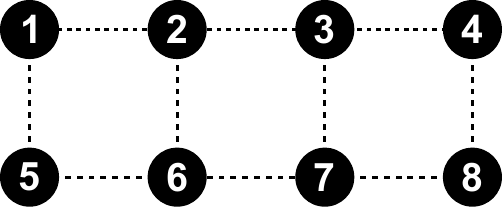}
\caption{Labeling schemes for quasi-1D lattices.  Left: the ``$2 \times n$'' labeling for $n=4$.  Right: the ``$n \times 2$'' labeling for $n=4$.
\label{Fig:Labeling}}
\end{figure*}

When the interaction is purely repulsive, HF is the optimal fermionic mean-field \cite{Bach1994}.  Interactions with attractive components, however, can lead to mean-field solutions with spontaneous number symmetry breaking.  This mean-field theory is Hartree--Fock--Bogoliubov (HFB), and it is characterized by quasiparticle operators
\begin{equation}
\alpha_p^\dagger = \sum_q \left(U_{qp} \, c_q^\dag + V_{qp} \, c_q\right),
\end{equation}
where $\boldsymbol{U}$ and $\boldsymbol{V}$ are components of the unitary Bogoliubov transformation matrix  ${\boldsymbol{W} = \left( \begin{smallmatrix}\boldsymbol{U} & \boldsymbol{V}^\star \\ \boldsymbol{V} & \boldsymbol{U}^\star\end{smallmatrix}\right)}$.  The corresponding mean-field wave function is
\begin{equation}
\ket{\Phi_\mathrm{HFB}} = \prod_{p=1}^M \alpha_p \ket{-}
\end{equation}
and we obtain $\boldsymbol{W}$ by minimizing the expectation value of the Hamiltonian.  We can write the HFB mean-field state in terms of either a unitary Thouless transformation or a non-unitary transformation of, for example, the physical vacuum:
\begin{subequations}
\begin{align}
\ket{\Phi_\mathrm{HFB}} &= \mathrm{e}^{Z} \ket{-} = \mathcal{N} \, \mathrm{e}^{\mathcal{T}} \ket{-},
\\
Z &= \sum_{p<q} \left(Z_{pq} \, c_p^\dagger \, c_q^\dagger - h.c.\right),
\\
\mathcal{T} &= \sum_{p<q} \mathcal{T}_{pq} \, c_p^\dagger \, c_q^\dagger.
\end{align}
\end{subequations}

In addition to its use in number-conserving Hamiltonians, HFB is also the natural mean-field for fermionic Hamiltonians which lack number symmetry but have number parity symmetry.  The number parity operator is a symmetry of, so far as we are aware, all physical fermionic Hamiltonians.  Frequently, however, it is not a symmetry of spin Hamiltonians which have been transformed to fermions.  When the number parity operator is not a symmetry of the fermionic Hamiltonian, we must step beyond more standard mean-field theories \cite{Henderson2024b}.  One approach, pioneered by Colpa \cite{Colpa1979}, is the addition of a single extra fermionic degree of freedom, not used in the Hamiltonian but present in the Hilbert space, which permits us to then use standard HFB.  Alternatively, we can use a mean-field directly constructed for these parity-violating Hamiltonians.  We call this mean-field HFBF \cite{Fukutome1977,Fukutome1977b,Moussa2018,Nishiyama2019,Henderson2024b} and it is defined in analogy to HF and HFB.  We can use a  canonical transformation with a nonlinear component to define quasiparticle creation and annihilation operators, with a unitary transformation matrix.  We can instead choose a Thouless parameterization, as we will prefer to do here, writing
\begin{subequations}
\begin{align}
\ket{\Phi_\mathrm{HFBF}} &= \mathcal{N} \, \left(1 + t\right) \, \mathrm{e}^{\mathcal{T}} \ket{-},
\\
t &= \sum t_p \, c_p^\dagger,
\end{align}
\end{subequations}
or using an equivalent unitary form.

As a practical matter, for both HFB and HFBF, we use as a reference not the physical vacuum $\ket{-}$ but some well-chosen reference $\ket{\Phi_0}$, just as we do for HF.  In this case, we employ a quasiparticle transformation so that $\mathcal{T}$ and $t$ are excitation operators acting on $\ket{\Phi_0}$:
\begin{subequations}
\begin{align}
\mathcal{T} &= \sum_{i<j} \mathcal{T}_{ij} \, c_i \, c_j + \sum_{ia} \mathcal{T}_{ai} \, c_a^\dagger \, c_i + \sum_{a<b} \mathcal{T}_{ab} \, c_a^\dagger \, c_b^\dagger,
\\
t &= \sum_i t_i \, c_i + \sum_a t_a \, c_a^\dagger,
\end{align}
\end{subequations}
where $i$ and $j$ ($a$ and $b$) index levels occupied (empty) in $\ket{\Phi_0}$.  Note that in making this quasiparticle transformation, we use bare fermion annihilation operators for the levels occupied in $\ket{\Phi_0}$ rather than creation operators.

Although the various mean-field methods we need can be implemented in a self-consistent field code with polynomial scaling, we have chosen instead for simplicity to implement all of them in a full configuration interaction code, using the conjugate gradient algorithm to minimize the mean-field energy with respect to the Thouless parameters.  We have chosen a non-unitary Thouless representation, for which the analytic gradient of the wave function and energy is straightforward.  Implementing these various techniques in an exact diagonalization scheme facilitates comparison between the various fermionizations, as we can implement each spin operator in a straightforward way, whereas in a self-consistent field code we would require either high-order density matrices or the use of a nonorthgonal version of Wick's theorem \cite{Balian1969,Chen2023}.  Unfortunately, this choice limits us to about a dozen sites for practical calculations.

\begin{figure}
\includegraphics[width=\columnwidth]{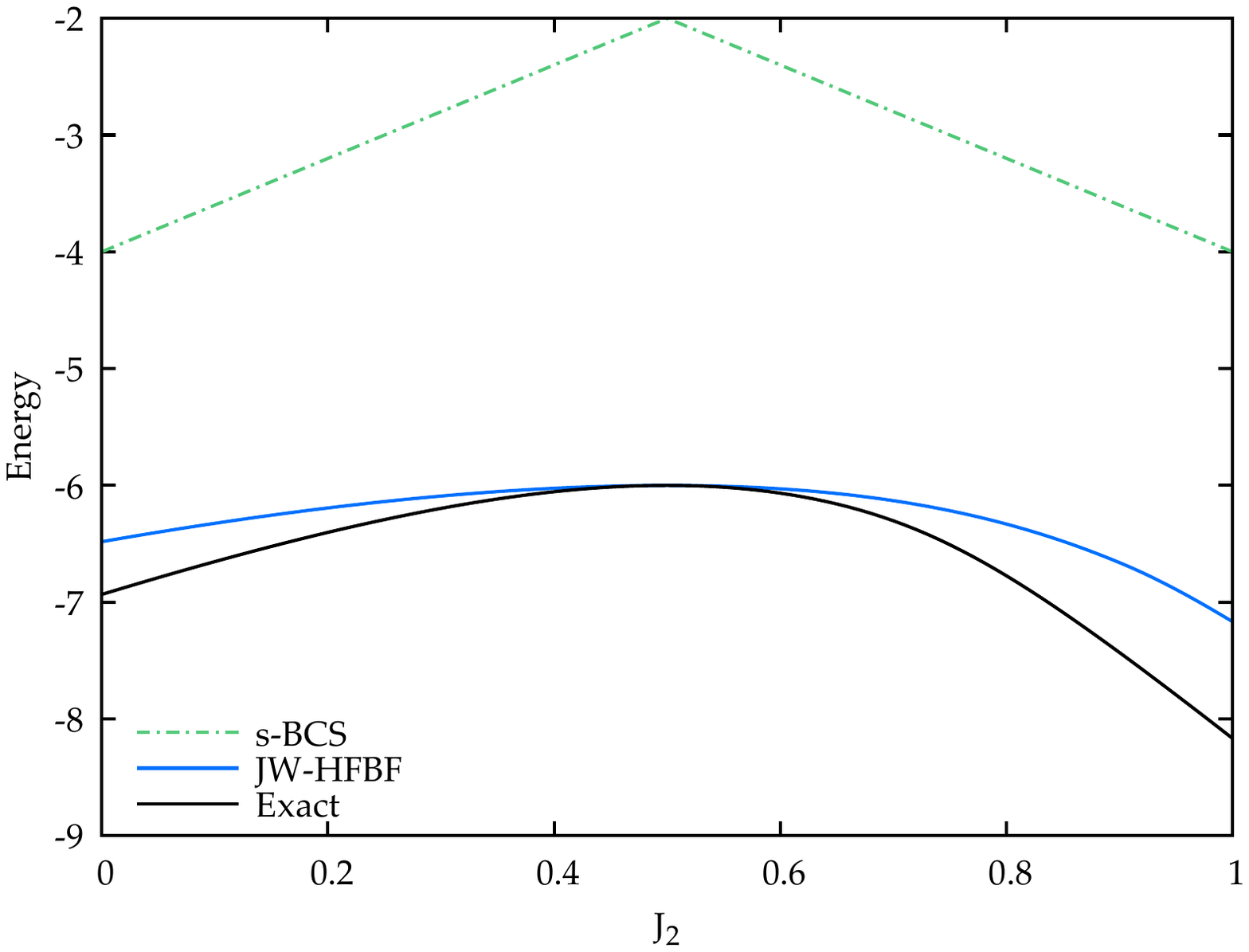}
\caption{Total energies in the $2 \times 4$ $J_1\text{--}J_2$ model with periodic boundary conditions.  We compare the spin mean-field (``s-BCS'') and the fermionic mean-field in the Jordan-Wigner--transformed Hamiltonian (``JW-HFBF'') to the results of exact diagonalization (``Exact'').
\label{Fig:J1J2TotalEnergies}}
\end{figure}

\begin{figure*}[t]
\includegraphics[width=0.45\textwidth]{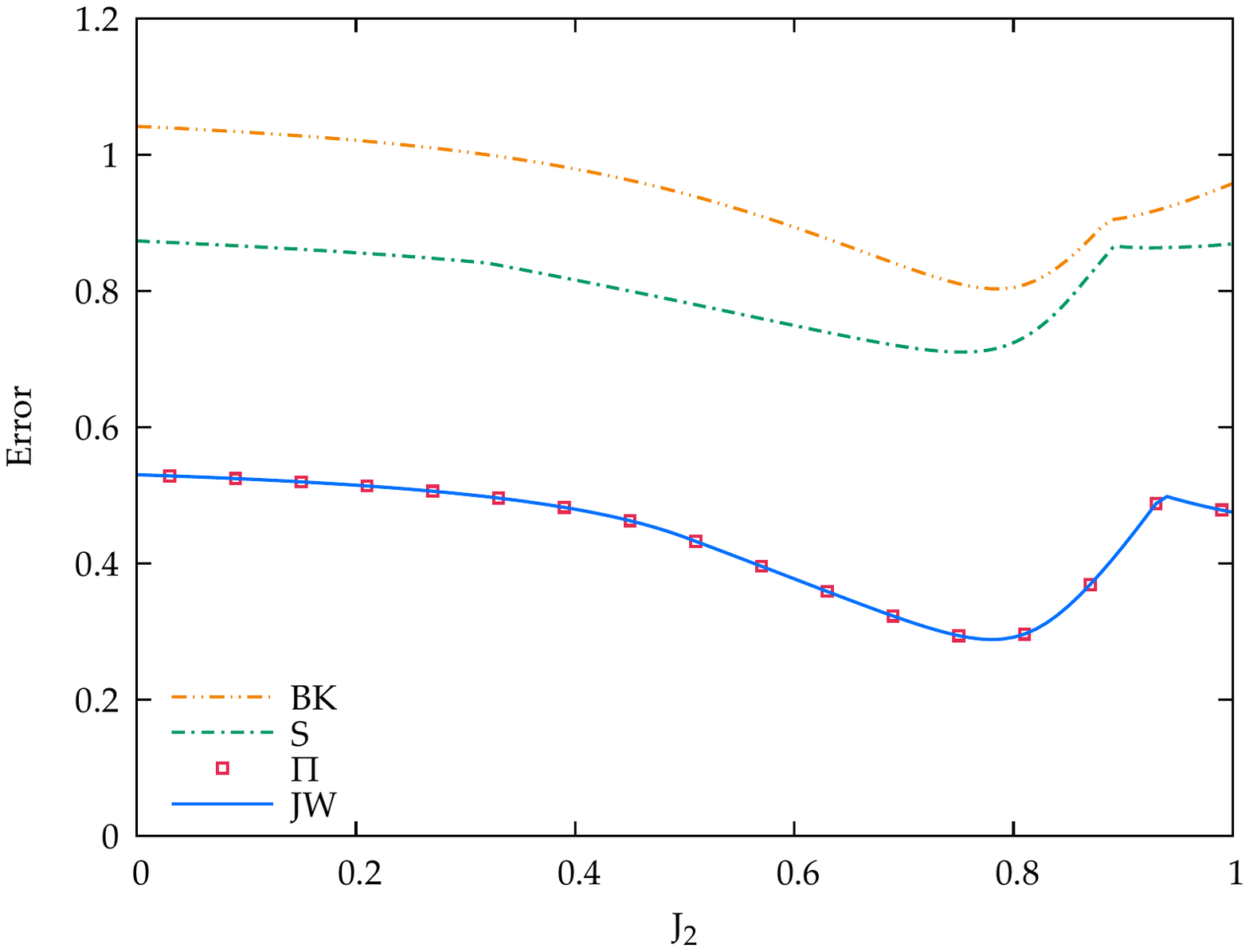}%
\hfill
\includegraphics[width=0.45\textwidth]{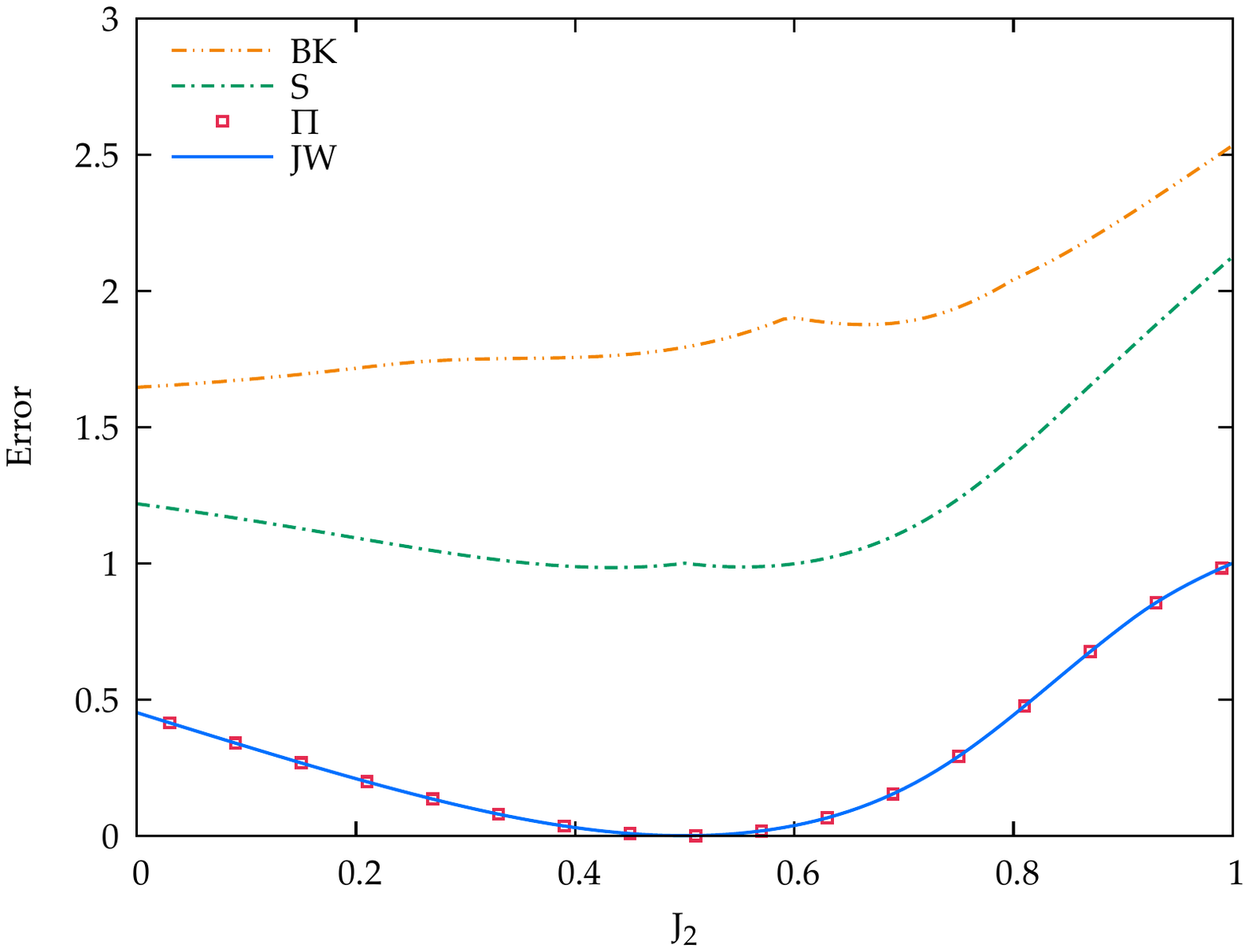}
\\
\includegraphics[width=0.45\textwidth]{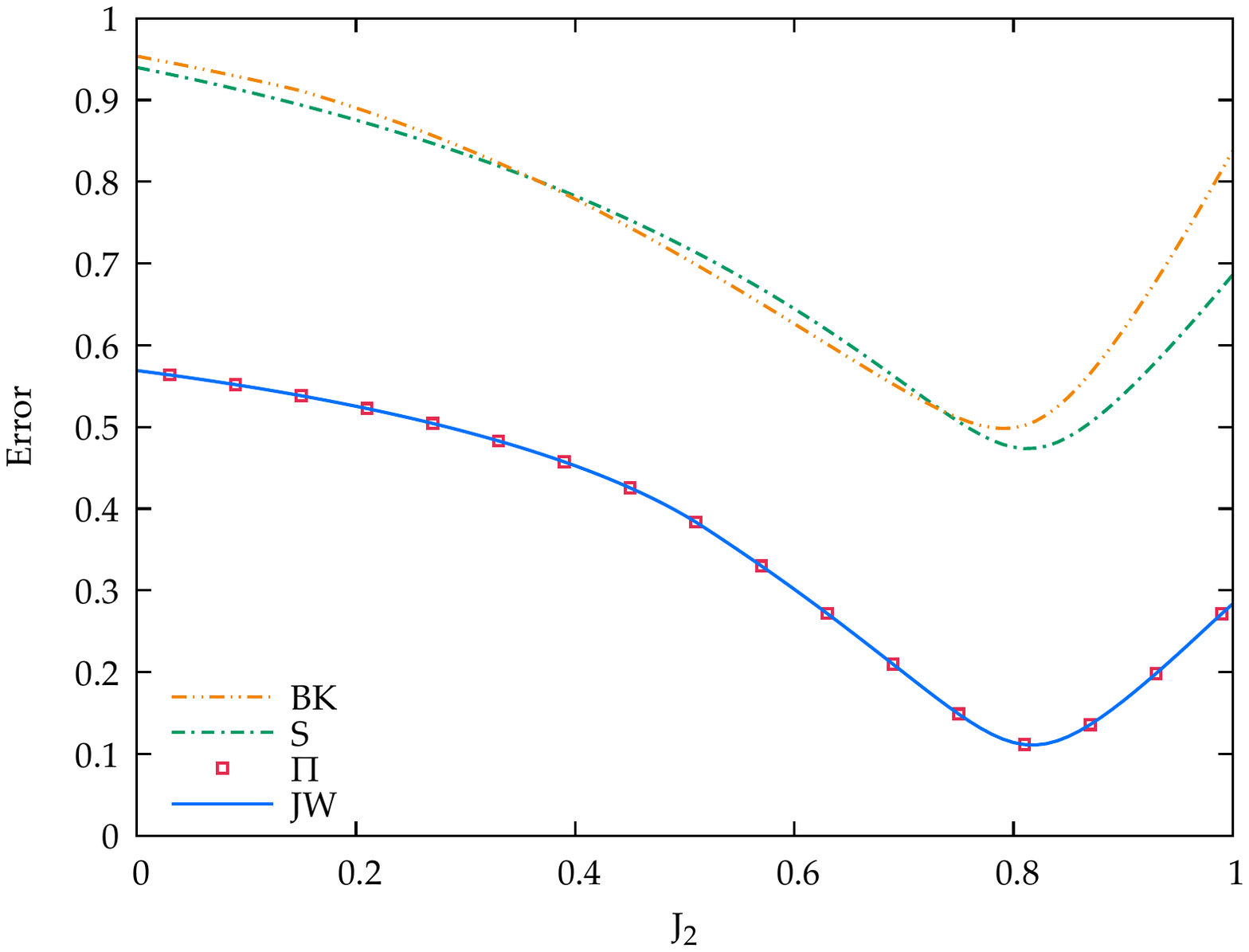}%
\hfill
\includegraphics[width=0.45\textwidth]{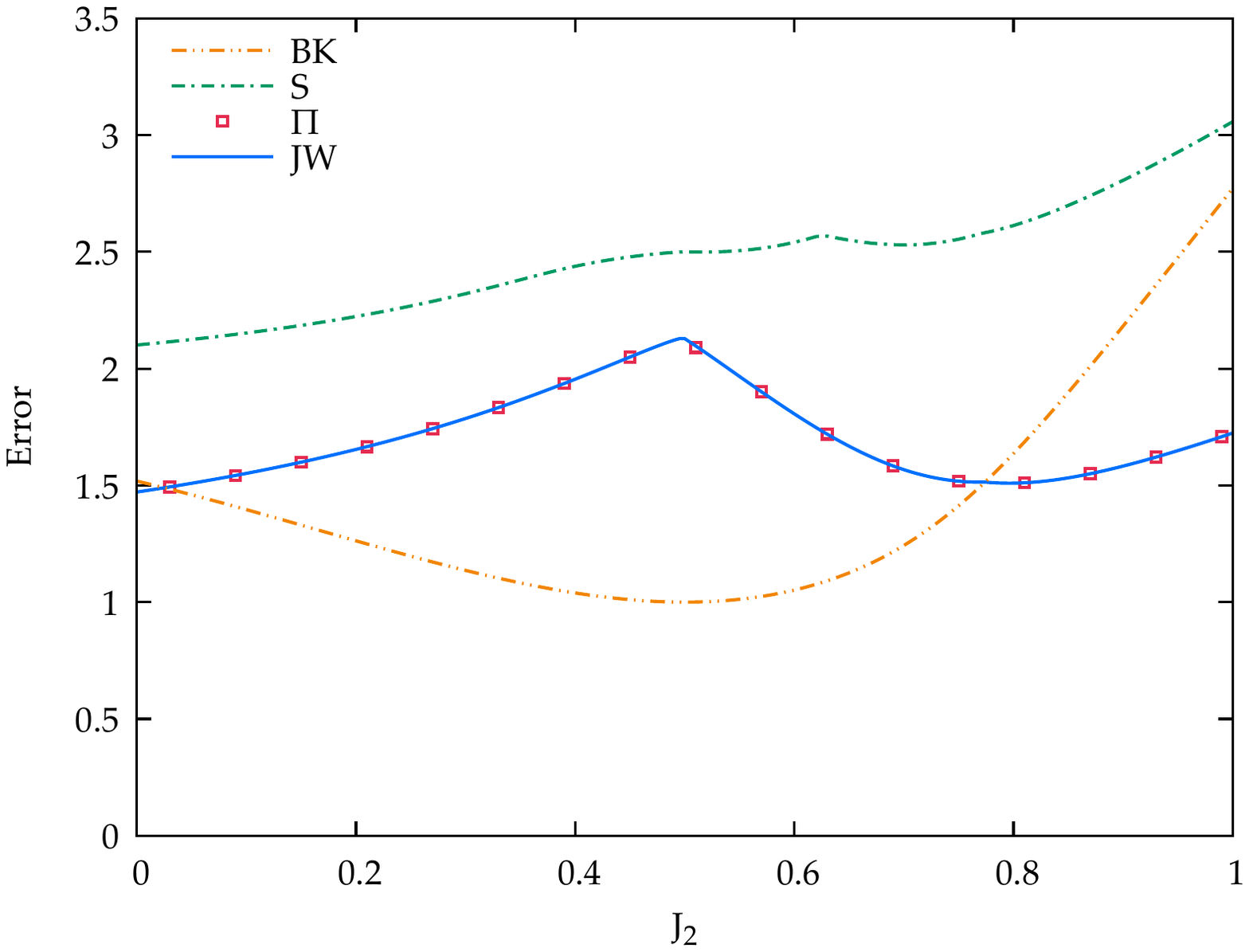}
\caption{HFBF mean-field energy errors relative to the exact diagonalizations in the fermionized $2 \times 4$ $J_1\text{--}J_2$ model with $J_1 = 1$ and either open boundary conditions (OBC) or periodic boundary conditions (PBC).
Top left:		$2 \times 4$ labeling with OBC.
Top right:		$2 \times 4$ labeling with PBC.
Bottom left:		$4 \times 2$ labeling with OBC.
Bottom right:		$4 \times 2$ labeling with PBC.
\label{Fig:J1J2Results}}
\end{figure*}

\section{Results}
Now that we have outlined the various fermionizations we will consider and have touched on the mean-field methods we will employ, we can proceed to analyze the various techniques.  With the exception of the {\JW} transformation, the fermionizations map spin Hamiltonians which possess $S_z$ symmetry onto fermionic Hamiltonians which have neither number symmetry nor number parity symmetry, so we must use the general HFBF mean-field.  In contrast, for {\JW} the transformed Hamiltonian has number symmetry; this means that in the {\JW} case we may try HF and HFB in addition to HFBF.  To eliminate one variable from consideration, we will use HFBF for the {\JW}-transformed Hamiltonian as well, except when indicated otherwise.  Generally, however, we find that for the {\JW}-transformed Hamiltonian, number parity symmetry does not spontaneously break at the mean-field level so that HFBF and HFB are entirely equivalent.

In this work, we will consider three $\mathfrak{su}(2)$ Hamiltonians: the $J_1\text{--}J_2$ Heisenberg model, the spin XXZ model, and the pairing Hamiltonian which we write in terms of spin operators.  We will discuss results for each in turn.  The two Heisenberg Hamiltonians are commonly studied textbook models (see, for example, Ref. \onlinecite{Bishop2004}); although they can be used to model certain physical systems \cite{Carretta2002,Bombardi2004,Rams2020}, their practical utility is as simple models which exhibit spin frustration.  The pairing Hamiltonian essentially models Bardeen-Cooper-Schrieffer (BCS) \cite{Bardeen1957} superconductivity (and is also known as the reduced BCS Hamiltonian), but has also been used to describe, for example, ultrasmall superconducting grains \cite{Sierra2000}.
All of these Hamiltonians have $S_z$ symmetry, and we will always work in the $S_z = 0$ sector for simplicity.  In the parameter regimes we will consider, this is always the global ground state.  In the thermodynamic limit, these Hamiltonians exhibit phase transitions which are of course absent in the exact solution for the finite systems we will consider here.  Nevertheless, the presence of these multiple spin arrangements is reflected in mean-field solutions which frequently just cross one another.  This leads to error curves which are not entirely smooth, because we have plotted, at each Hamiltonian parameter value, the lowest energy mean-field solution we could find.

Because our results depend on the labeling scheme chosen, we must say a bit about this first.  In one dimension, we use the natural labeling scheme where sites connected in the lattice have sequential numbers, e.g., site 1 connects to site 2, site 2 to site 3, and so on.  For the quasi-1D spin ladders, however, we have a ``$2 \times n$'' labeling scheme and a ``$n \times2$'' labeling scheme, depicted in Fig. \ref{Fig:Labeling}.  Put simply, the first number (``2'' or ``$n$'') denotes the faster-moving index in the 2D lattice.

\subsection{The $\boldsymbol{J_1}\text{--}\boldsymbol{J_2}$ Hamiltonian}
The $J_1\text{--}J_2$ model has both nearest-neighbor and next-nearest neighbor interactions:
\begin{equation}
H_{J_1\text{--}J_2} = J_1 \sum_{\langle pq \rangle} \vec{S}_p \cdot \vec{S}_q + J_2 \, \sum_{\langle\langle pq \rangle \rangle} \vec{S}_p \cdot \vec{S}_q,
\end{equation}
where $\langle pq \rangle$ denotes nearest neighbors (sites connected in the lattice) and $\langle \langle pq \rangle\rangle$ denotes next-nearest neighbors.  The physics is driven by the ratio $J_2 / J_1$, where we will take both to be positive.  For the 2D square lattice in the thermodynamic limit, the ground state is a N\'eel antiferromagnet for $J_2 \lesssim 0.4\,  J_1$ and a striped antiferromagnet for $J_2 \gtrsim 0.6 \, J_1$.  In between, the magnetic structure is more complicated \cite{Darradi2008,Gong2014,Richter2015}.  In this work we will focus on the quasi-1D spin ladders, i.e., $2 \times n$ systems.

As we have shown elsewhere \cite{Henderson2022}, results using mean-field theory in combination with the {\JW} transformation are not quantitative but are greatly superior to what one obtains with an $\mathfrak{su}(2)$ mean-field technique, being roughly comparable to configuration interaction with double and quadruple excitations atop the symmetry-adapted spin mean-field.  To get some sense of the difference in quality between fermionic mean-field and $\mathfrak{su}(2)$ mean-field, Fig. \ref{Fig:J1J2TotalEnergies} compares total energies from the spin and fermionic mean-fields to the exact results for the $2 \times 4$ model.

Figure \ref{Fig:J1J2Results} shows a comparison of the total energy errors with respect to the exact diagonalizations, resulting from the mean-field solutions with the four different fermionization schemes, both for open boundary conditions (OBC) and periodic boundary conditions (PBC), with $J_1 = 1$.  We can immediately see that the {\JW} and {\parity} transformations give completely identical mean-field energies, which is sensible since, as we have noted, these two transformations are loosely speaking dual companions of one another.  More formally, as we will discuss in future publications, the {\JW}-transformed fermionic operators of Eqn. \ref{Eqn:JW} and the {\rotatedparity}-transformed operators of Eqn. \ref{Eqn:PiRot} are, themselves, related by a mean-field canonical transformation of a form equivalent to HFBF; accordingly, the two seemingly different Hamiltonians have identical mean-field solutions.

\begin{figure}
\includegraphics[width=\columnwidth]{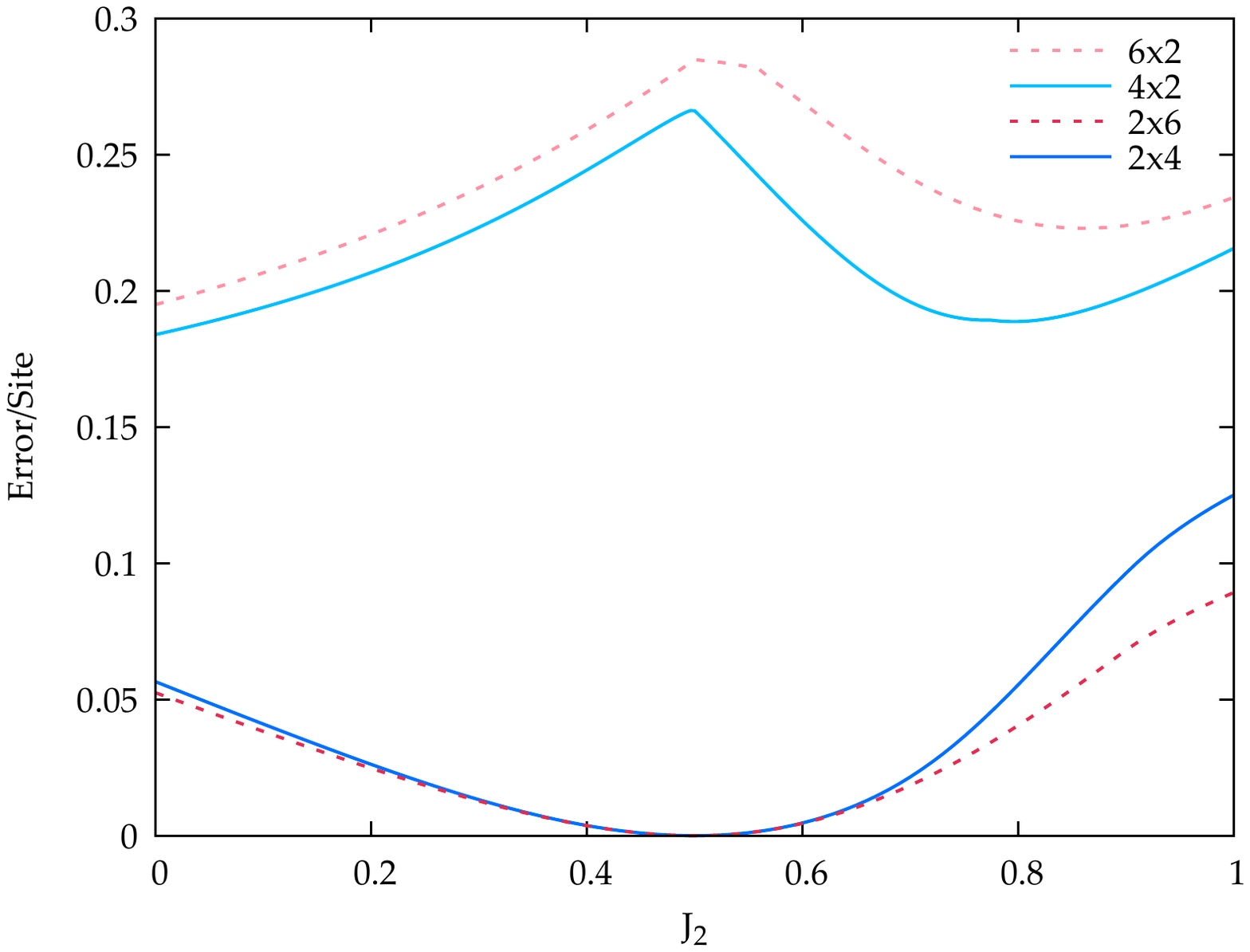}
\caption{HFBF mean-field energy errors per site relative to the exact diagonalization in the $2 \times 4$ and $2 \times 6$ {\JW}-transformed $J_1\text{--}J_2$ Hamiltonian.
\label{Fig:12SiteJ1J2}}
\end{figure}

The {\BK} and {\sierpinski} mappings generally yield significantly worse mean-field results than do the other transformations we have considered.  Mappings designed to minimize gate depth when mapping fermions to spins are not, evidently, the same as mappings designed to yield optimal mean-field solutions when mapping spins to fermions.  The one apparent exception is in the $4 \times 2$ case with PBC, where the {\BK} mapping is, overall, superior in the physically interesting regime around $J_2 = 1/2$.

We also see the expected dependence on the labeling scheme, which we recall can be eliminated in the {\JW} case (and, presumably, in the {\parity} case) by generalizing the strings $\tilde{\phi}_p$.  It is interesting to note that we generally obtain superior results for OBC with the $4 \times 2$ labeling and for PBC with the $2 \times 4$ labeling where {\JW} and {\parity} are exact at $J_2 = 1/2$, presumably related to the Majumdar--Ghosh point in the 1D $J_1\text{--}J_2$ model where the ground state at $J_2 = 1/2$ is exactly dimerized \cite{Majumdar1969}.

Finally, we compare results for the $2 \times 4$ and $2 \times 6$ models in Fig. \ref{Fig:12SiteJ1J2}.  These results (consistent with those of Ref. \onlinecite{Henderson2024a}) suggest that increasing system size does not necessarily degrade the quality of the mean-field results.  In other words, as is the usual case for mean-field theory, our results are extensive; see also the supplementary material.
%\textcolor{red}{Formally, the mean-field theories are extensive, but they are applied to quite unusual Hamiltonians which may have very high operator rank which increases with system size.  For example, in a $k \times k$ Hamiltonian transformed with Jordan-Wigner, we may have a $k^2$-body operator coupling sites $(1,1)$ and $(k,k)$: the $S_+$ and $S_-$ operators each convert into a single creation or annihilation operator, together with the JW strings, which are a product of $(kl-1)$ objects of the form $(1 - 2 \, n_p)$ for sites $p$ running from 1 to $k^2-1$.  The odd structure of these Hamiltonians complicates analyzing their thermodynamic limits at the mean-field level.  Note, however, that when expressed in the language of spins, these Hamiltonians have well-behaved therodynamic limits.  Moreover, because the operator transformations are exact, full configuration interaction calculations on the fermionic Hamiltonian simply reproduce the exact diagonalization of the spin Hamiltonian.}

\begin{figure*}[t]
\includegraphics[width=0.45\textwidth]{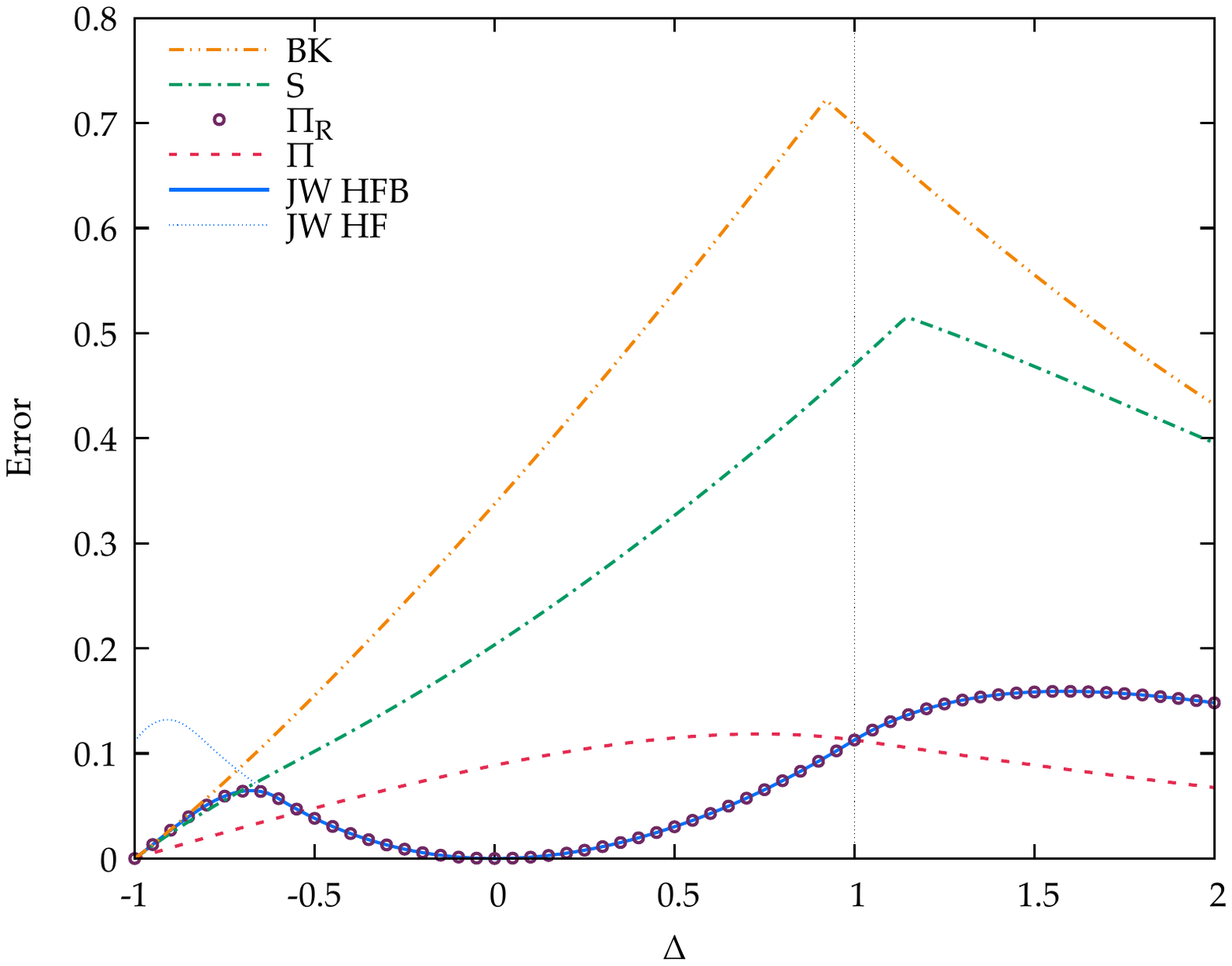}%
\hfill
\includegraphics[width=0.45\textwidth]{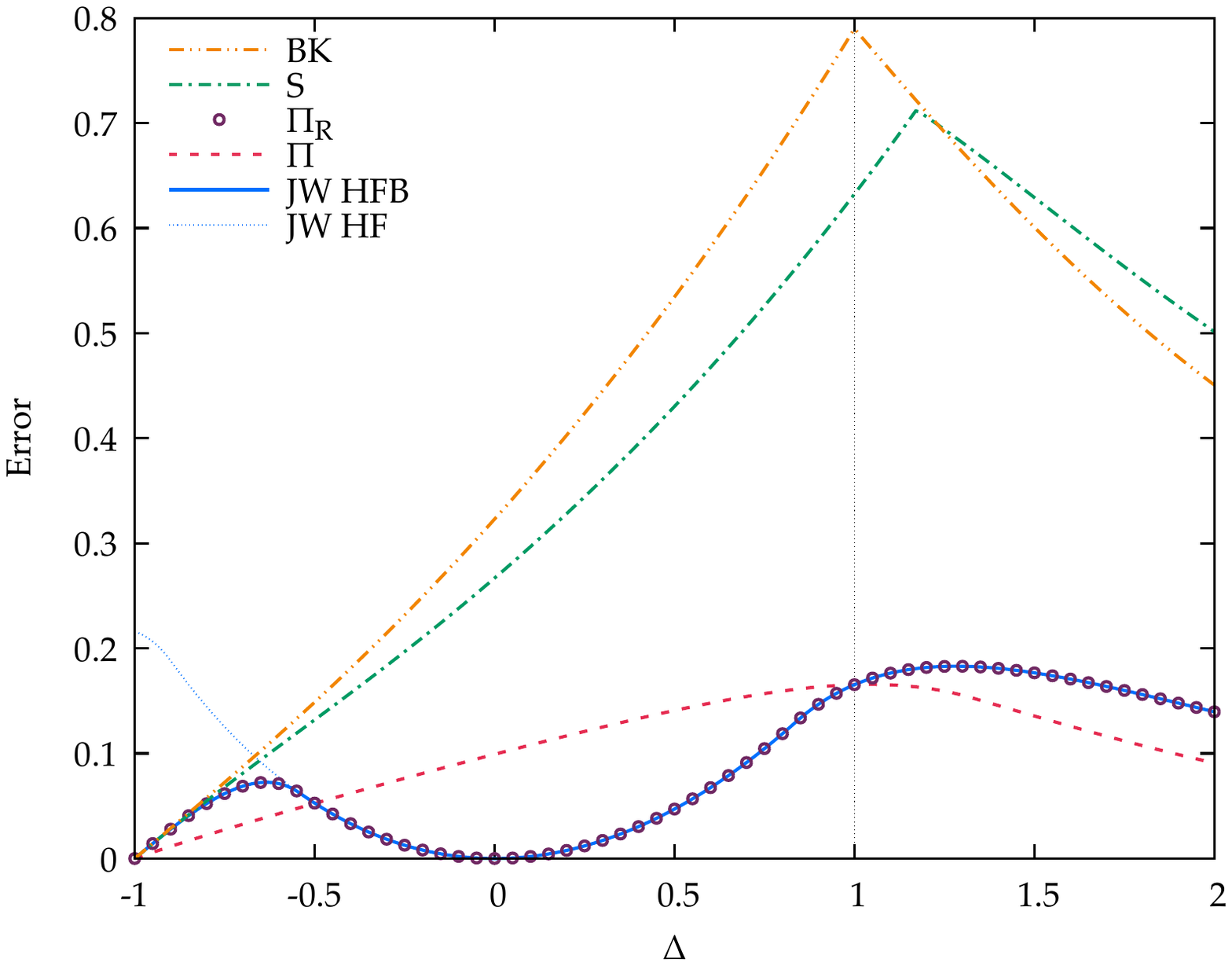}
\caption{HFBF ({\BK}, {\sierpinski}, {\rotatedparity}, {\parity}) and HF and HFB ({\JW}) mean-field energy errors relative to the exact diagonalizations in the fermionized 8-site 1D XXZ model.  Left panel: OBC.  Right panel: PBC.
\label{Fig:XXZ1D}}
\end{figure*}

\subsection{The XXZ Hamiltonian}
The XXZ Hamiltonian is also a spin lattice model, this time given by
\begin{equation}
H_{\mathrm{XXZ}} = \sum_{\langle pq \rangle} \left(S_p^x \, S_q^x + S_p^y \, S_q^z + \Delta \, S_p^z \, S_q^z\right).
\end{equation}
Where the physics of the $J_1\text{--}J_2$ model depends on $J_2/J_1$, here it depends on $\Delta$.  In one dimension, at $\Delta = 0$ the {\JW}-transformed XXZ model is exactly solved by HF with sites labeled in a natural sequential way.  This is not the case in two or more dimensions, so we will consider both the 1D system and the 2D rectangular lattice.  The point at $\Delta = 1$ is the Heisenberg point, at which the Hamiltonian has $S^2$ symmetry, while $\Delta = -1$ is also a special point in which the exact ground state is given by the extreme antisymmetrized geminal power \cite{Massaccesi2021,Liu2023}.

The ground state over all $S_z$ sectors occurs at $S_z = 0$ for $\Delta > -1$, while the different $S_z$ sectors are all degenerate with one another at $\Delta = -1$ and the global minimum of the energy occurs at maximal $S_z$ for $\Delta < -1$.  Since we are focusing on $S_z = 0$ we will only consider $\Delta \ge -1$ in this work.  In the thermodynamic limit, for rectangular lattices, there is a phase transition at $\Delta = 1$, where for $\Delta > 1$ the ground state is antiferromagnetic with magnetization along the $z$ axis and for $|\Delta| < 1$ the ground state is instead magnetized in the $xy$ plane \cite{Farnell2004}.

We begin with the 1D case, depicted in Fig. \ref{Fig:XXZ1D}, in which the {\JW} strings do not contribute.  Again, the {\parity} and {\JW} transformations provide uniformly more accurate results than we obtain with the other two mappings considered, although all methods are exact at $\Delta = -1$ where, as we have indicated, the ground state energy for each $S_z$ sector is the same and spin mean field is already exact.  As expected, we get the exact result at $\Delta = 0$ with the {\JW} transformation.

\begin{figure}
\includegraphics[width=\columnwidth]{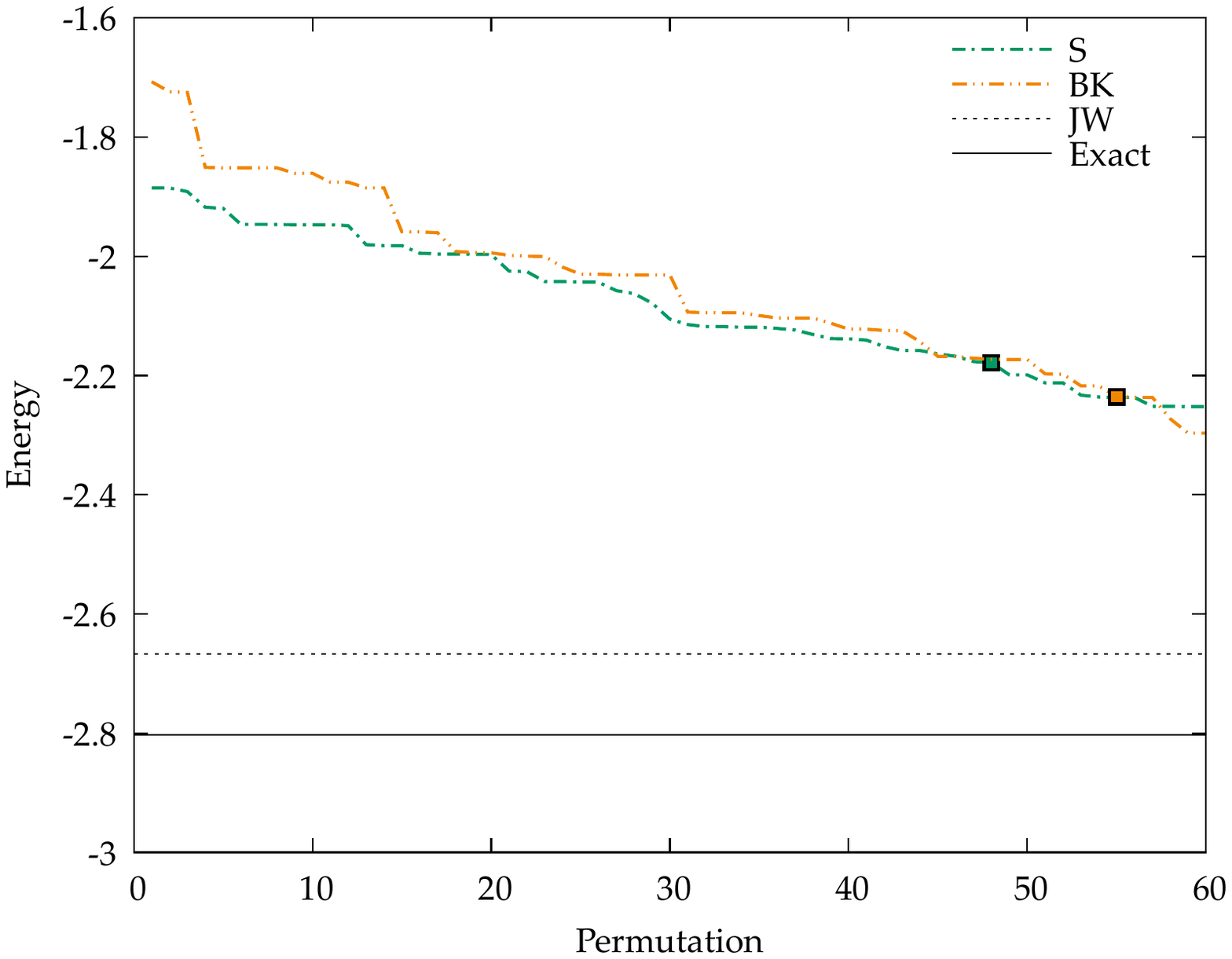}
\caption{HFBF mean-field energy for the unique labelings of the 6-site 1D XXZ model at $\Delta = 1$ with PBC, based on the {\BK} and {\sierpinski} encodings.  The lines marked ``{\JW}'' and ``Exact'' denote, respectively, the energy of the mean-field solution based on the {\JW} transformation with the natural labeling scheme 1-2-3-4-5-6, and the exact ground state energy.  The squares indicate the result based on the natural labeling scheme.
\label{Fig:XXZ1DLabeling}}
\end{figure}

Unlike with the $J_1\text{--}J_2$ model, the {\JW} and {\parity} transformations do not generally give the same result, except at $\Delta = \pm 1$.  At $\Delta = 1$ the XXZ model is isotropic, but elsewhere it treats the $z$ component of spin differently from the other two.  For this reason, we also show the {\rotatedparity} transformation which, like {\JW} and like the Hamiltonian, treats the $z$ component of spin differently (because it does not include the string operator $\tilde{\phi}_p$).  Indeed, the {\rotatedparity} transformation yields results completely equivalent to those we obtain with {\JW}.  We have also verified that a rotated {\JW} transformation which has strings for the $y$ and $z$ spin components but not the $x$ component yields results identical to the original {\parity} transformation (data not shown).

We must mention one other consideration.  For most $\Delta$ values, the fermionic mean-field for the {\JW}-transformed Hamiltonian is HF.  Near $\Delta = -1$, however, number symmetry breaks spontaneously and we can distinguish HF and HFB solutions.  Where these two methods differ, it is the lower energy of the two (the HFB) with which the HFBF solution for the {\rotatedparity} transformation agrees.

One may wonder whether the poor performance of the {\BK} or {\sierpinski} encodings in this context is simply due to having chosen the wrong labeling scheme.  We have seen that labeling schemes matter for all of these encodings, and previous work has found that the natural labeling scheme we have chosen is, on the whole, the best available for the 1D XXZ model with the {\JW} transformation.  But is that the case for these encodings?

\begin{figure*}[t]
\includegraphics[width=0.45\textwidth]{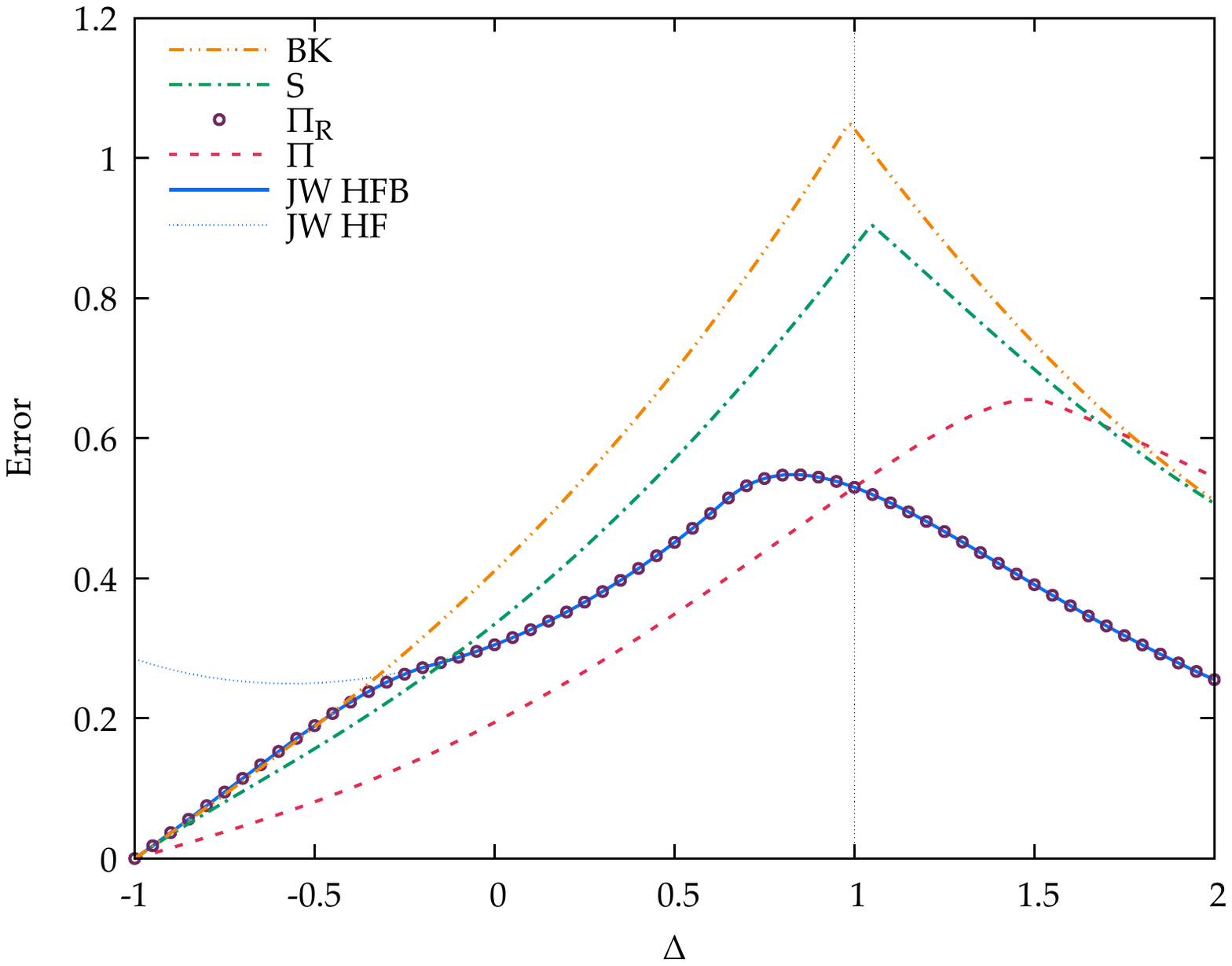}%
\hfill
\includegraphics[width=0.45\textwidth]{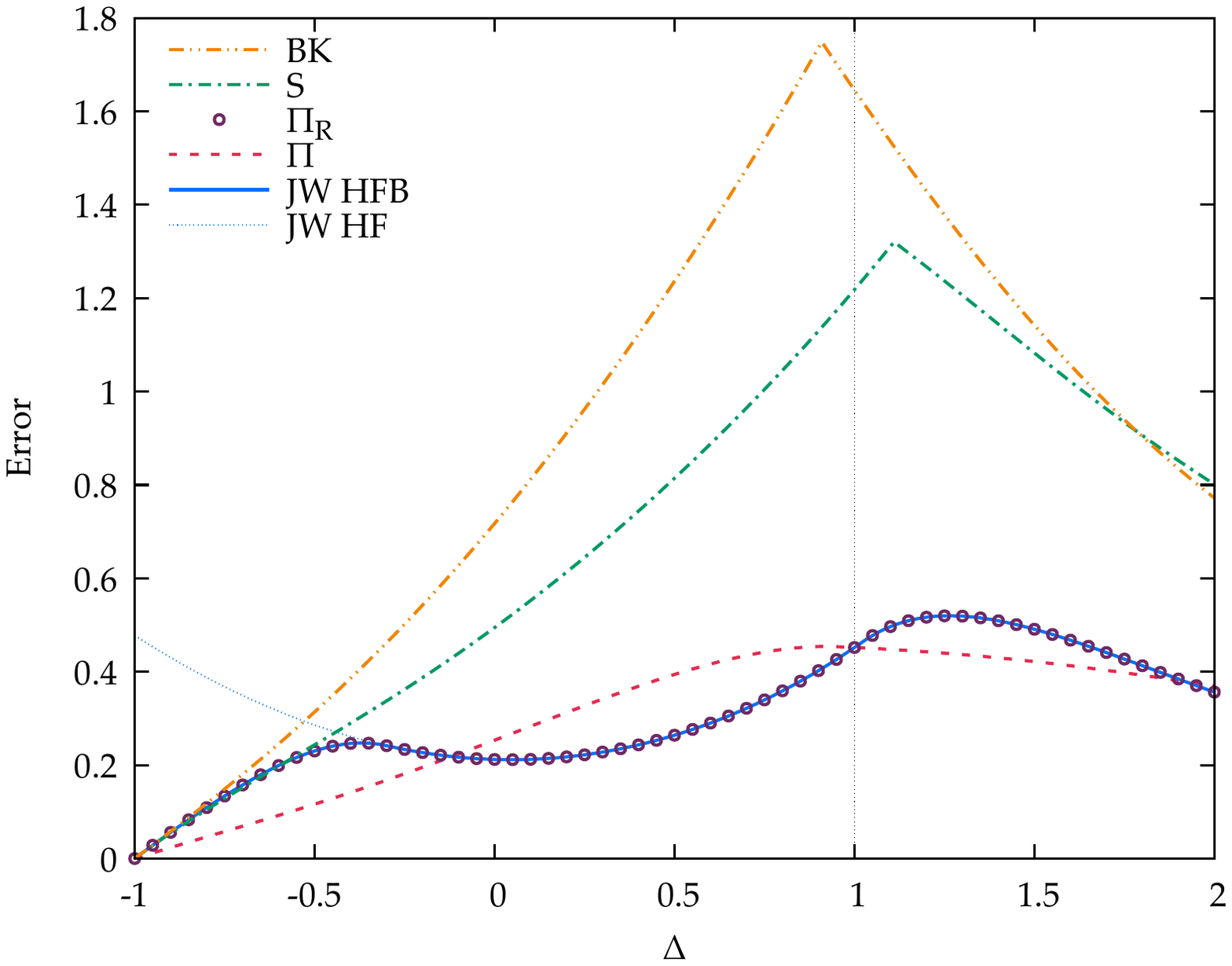}
\\
\includegraphics[width=0.45\textwidth]{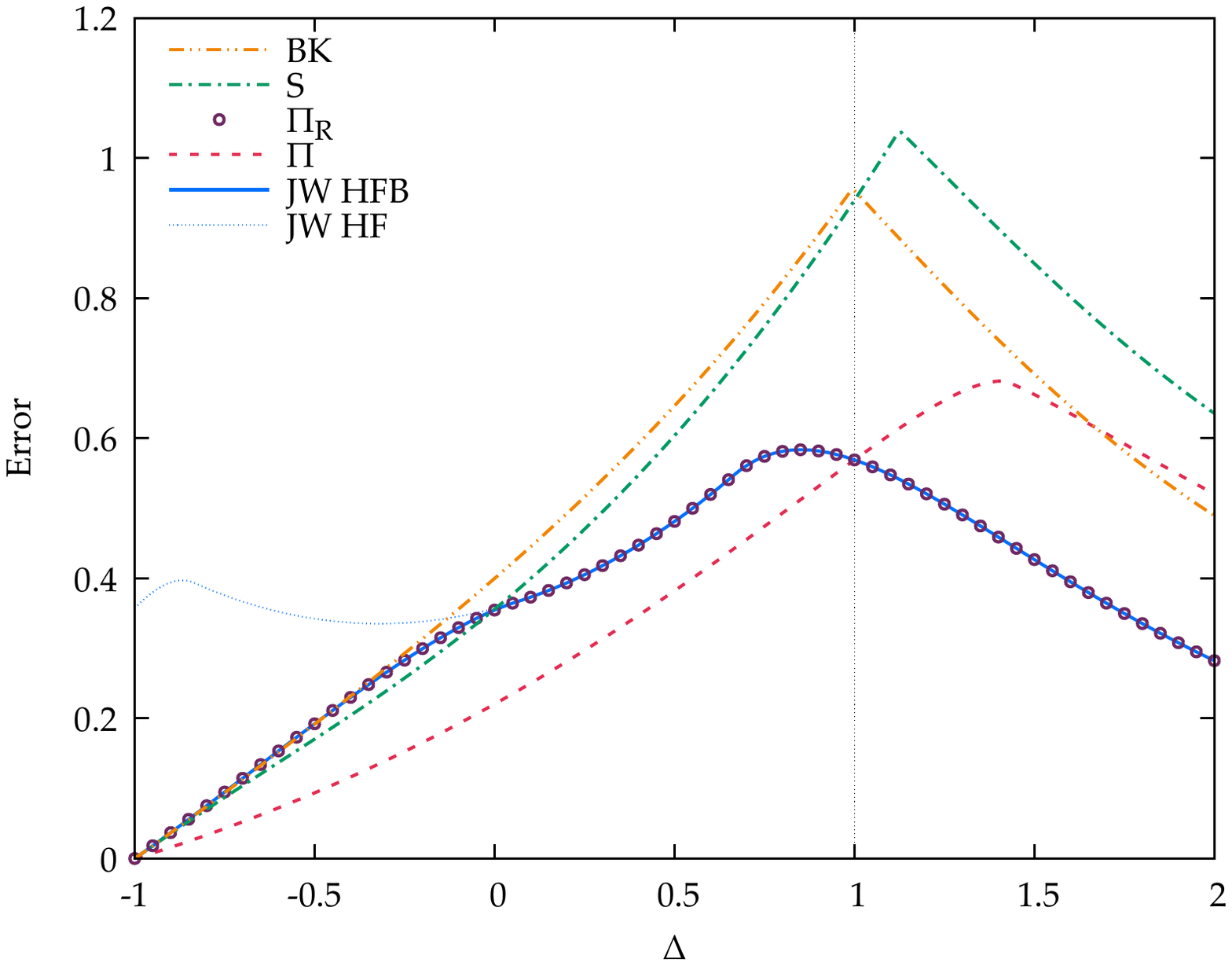}%
\hfill
\includegraphics[width=0.45\textwidth]{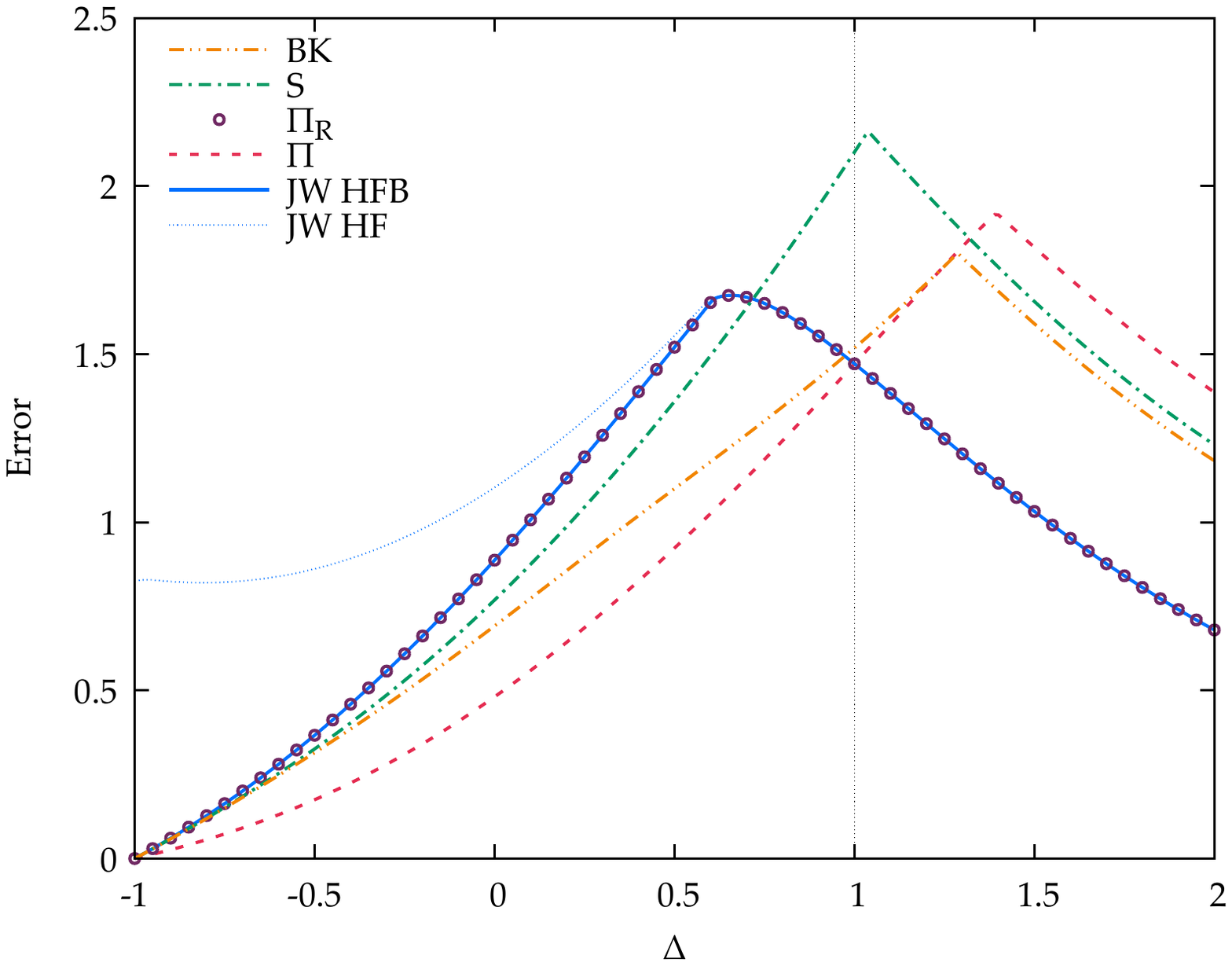}
\caption{HFBF ({\BK}, {\sierpinski}, {\rotatedparity}, {\parity}) and HF and HFB ({\JW}) mean-field energy errors relative to the exact diagonalizations in the fermionized $2 \times 4$ XXZ model with either open boundary conditions (OBC) or periodic boundary conditions (PBC).
Top left:		$2 \times 4$ labeling with OBC.
Top right:		$2 \times 4$ labeling with PBC.
Bottom left:	$4 \times 2$ labeling with OBC.
Bottom right:	$4 \times 2$ labeling with PBC.
\label{Fig:XXZ2D}}
\end{figure*}

To answer that question, we have looked at the 6-site 1D XXZ model with PBC.  Here, there are 60 distinct labeling schemes ($6!$ ways of labeling the sites, reduced by a factor of 6 due to translational symmetry and by another factor of 2 due to reflection symmetry) and we have tried the {\BK} and {\sierpinski} encodings at $\Delta = 1$ for all 60 distinct labelings.  The results in Fig. \ref{Fig:XXZ1DLabeling} make clear that no relabeling of the sites will suffice to remedy the poor performance of mean-field methods based on these encodings, at least at $\Delta = 1$ but presumably also for other values of $\Delta$.

Results for the 2D XXZ model are roughly similar to those for the 1D case (see Fig. \ref{Fig:XXZ2D}) although the mean-field methods are generally less accurate.  For the {\JW} and {\parity} transformations, this is presumably because the strings $\tilde{\phi}_p$ which vanished in 1D now contribute in 2D.  The {\BK} and {\sierpinski} encodings are more competitive with the {\JW} and {\parity} encodings in this quasi-1D case than they are in the genuinely 1D case, but the {\JW} and {\parity} encodings still generally appear to yield superior mean-field results.

\subsection{The Pairing Hamiltonian}
The pairing Hamiltonian, also known as the reduced Bardeen--Cooper--Schrieffer Hamiltonian, is not really a Hamiltonian of spins at all.  Instead, it is a Hamiltonian consisting of electron pair creation, pair annihilation, and number operators.  As these operators satisfy the same $\mathfrak{su}(2)$ algebra as do the spin operators, however, we can write the pairing model as a spin Hamiltonian, in which case it takes the form
\begin{equation}
H_P = \sum_p \epsilon_p \, \left(2 \, S_p^z - 1\right) - G \, \sum_{pq} S_p^+ \, S_q^+.
\end{equation}
This Hamiltonian is exactly solvable with a form of Bethe ansatz \cite{Richardson1963,Richardson1964,Richardson1965}, for any choice of $\epsilon_p$ and $G$.

We can eliminate the $p=q$ term in the interaction, using $S_p^+ \, S_p^- = \frac{1}{2} + S_p^z$, thereby absorbing this diagonal contribution into the first, Zeeman-like, term.  As a practical matter, we choose to work at half filling, in which case we can rewrite the Hamiltonian as
\begin{equation}
H_P^\prime = 2 \, \sum_p \epsilon_p \, S_p^z - G \, \sum_{p \ne q} S_p^+ \, S_q^+ = H_P - \lambda \, S_z
\end{equation}
and pick the $\epsilon$ to be equally spaced and centered around 0.  The chemical potential $\lambda$ enforces that the ground state is $S_z = 0$ everywhere, and the physics is driven by the parameter $G$.

If we disregard the chemical potential for a moment and consider only the first form of the Hamiltonian, we can see that as $|G|$ tends to $\infty$, so that the Zeeman-like term is irrelevant, then
\begin{equation}
\frac{1}{|G|} \, H_P \to \mp \, S_+ \, S_- = \mp \left(S^2 - S_z^2 + S_z\right),
\label{Eqn:Pairing}
\end{equation}
where $S_+ = \sum_p S_p^+$ and similarly for $S_-$ and $S_z$; the sign is $\mp$ as $G \to \pm \infty$.  Accordingly, the $S_z = 0$ ground state occurs for maximal $S^2$ in the $G \to \infty$ limit and for $S^2 = 0$ in the $G \to -\infty$ limit.  Respectively, these amount to an extreme form of the antisymmetrized geminal power, and a kind of dimerized state obtained by taking one of the (many) degenerate $S^2 = 0$ states.  See Refs. \onlinecite{Yuzbashyan2003,Faribault2010,johnson2023richardsongaudinstates} for details.

\begin{figure}
\includegraphics[width=\columnwidth]{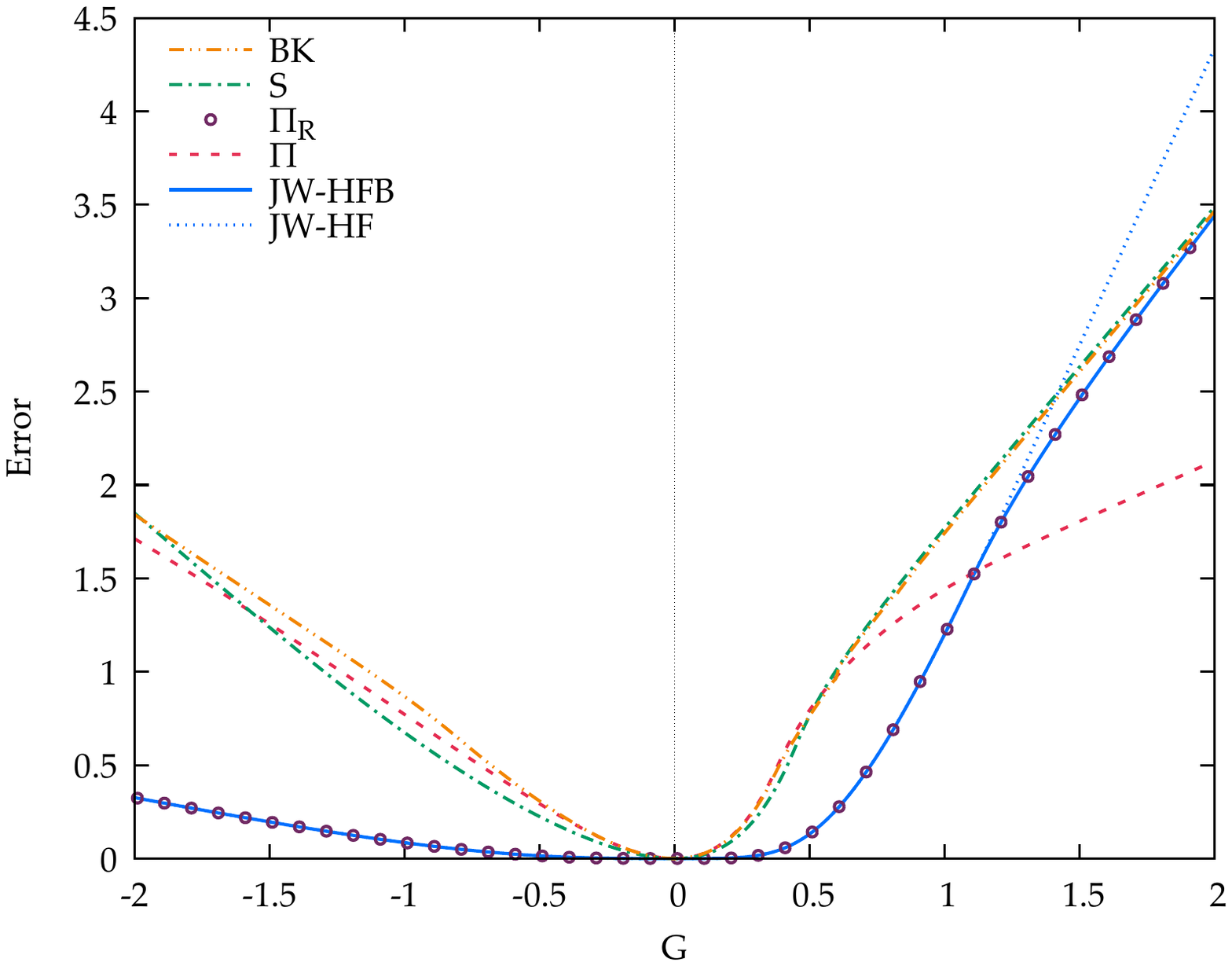}
\caption{HFBF ({\BK}, {\sierpinski}, {\rotatedparity}, {\parity}) and HF and HFB ({\JW}) mean-field energy errors relative to the exact diagonalizations in the fermionized 8-site pairing model.
\label{Fig:Pairing}}
\end{figure}

The pairing model presents a particular challenge.  Where in the 2D XXZ models, each spin interacted with at most three other spins, in the pairing model, each spin interacts equivalently with \textit{every} other spin.  Indeed, Fig. \ref{Fig:Pairing} shows that for the pairing model, our results are generally rather poor away from $G=0$, though again the {\JW} and (rotated) {\parity} transformations are superior to the other fermionizations on the repulsive side ($G < 0$).

\begin{figure*}
\includegraphics[width=0.45\textwidth]{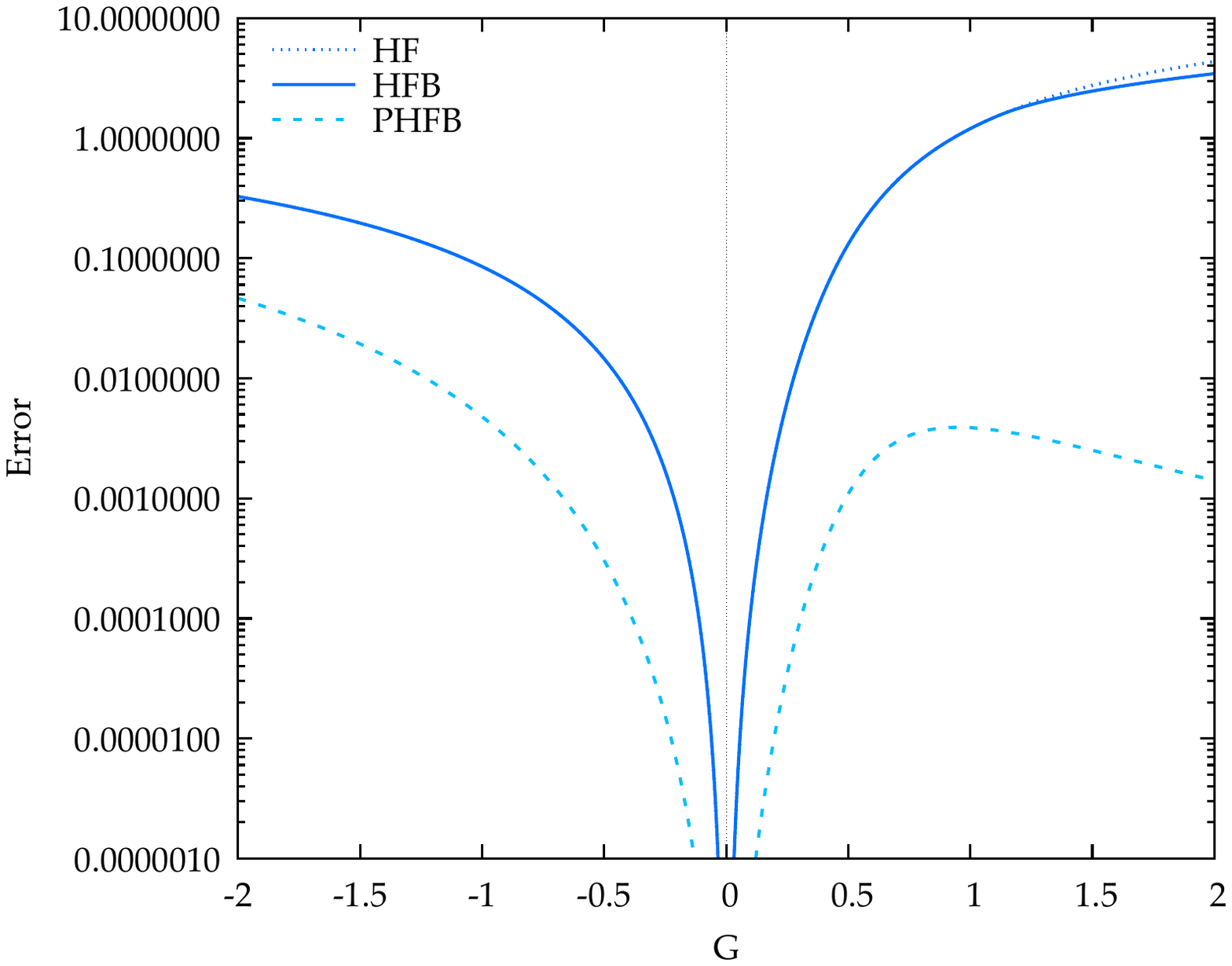}
\hfill
\includegraphics[width=0.45\textwidth]{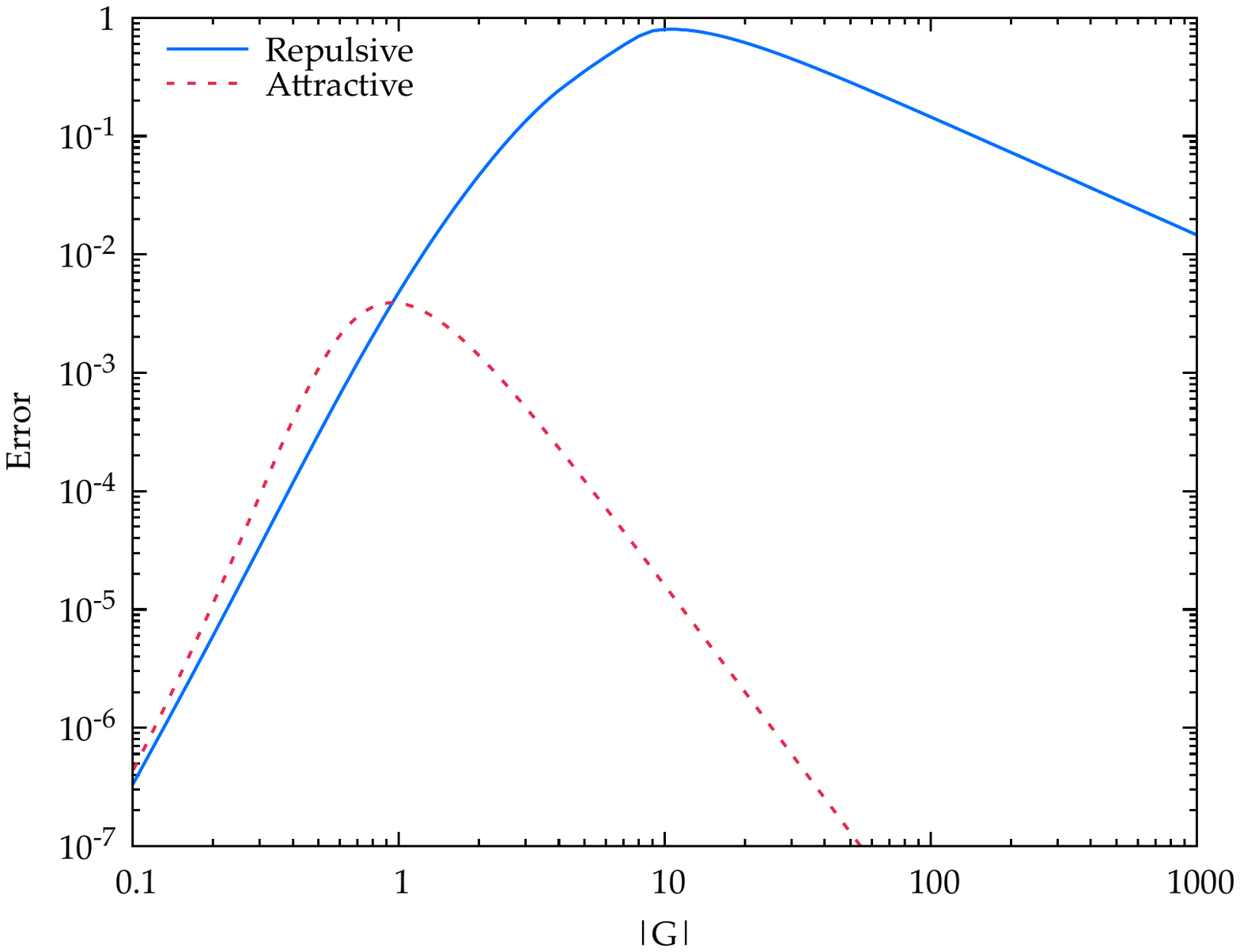}
\caption{Mean-field energy errors relative to the exact diagonalizations in the 8-site pairing model transformed with {\JW}.  Left panel: Comparison between HF, HFB, and PHFB.  Right panel: PHFB errors on a log-log scale to show the large $|G|$ behavior.
\label{Fig:PairingPHFB}}
\end{figure*}

\subsection{Fermionic Number Projection}
For most of these fermionizations, we have gone about as far as we can go with mean-field theory.  For the {\JW} transformation, however, we can do a bit better.  This is because the {\JW} transformation maps the global $S_z$ symmetry of the $\mathfrak{su}(2)$ Hamiltonian to global number symmetry of the fermionic Hamiltonian, and we can deliberately break and then projectively restore this number symmetry to do a {\JW}-transformed number-projected HFB.  For details about number projection, see Refs. \onlinecite{Sheikh2000,Schmid2004,Scuseria2011}, but the gist is that we write the projected HFB (PHFB) wave function as an HFB state and then project out those components with the incorrect particle number.  We could do something analagous for any of the other encodings -- the global $S_z$ symmetry maps under all of them to \textit{some} symmetry of the fermionic Hamiltonian -- but {\JW} has the advantage that this symmetry is both obvious and, more importantly, given by a one-body operator.  As such, the projection can be efficiently implemented in a self-consistent field code.

\begin{figure*}
\includegraphics[width=0.45\textwidth]{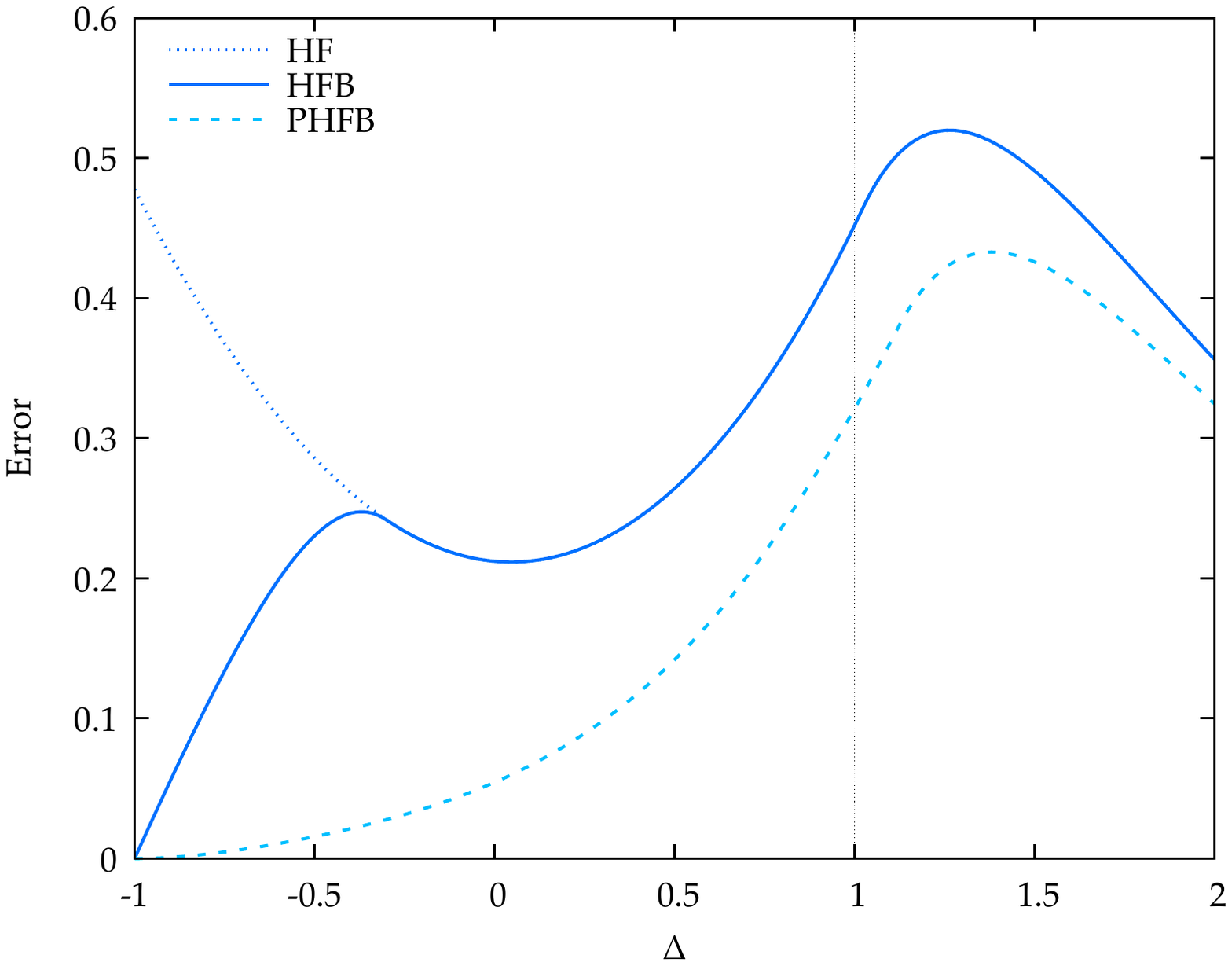}
\hfill
\includegraphics[width=0.45\textwidth]{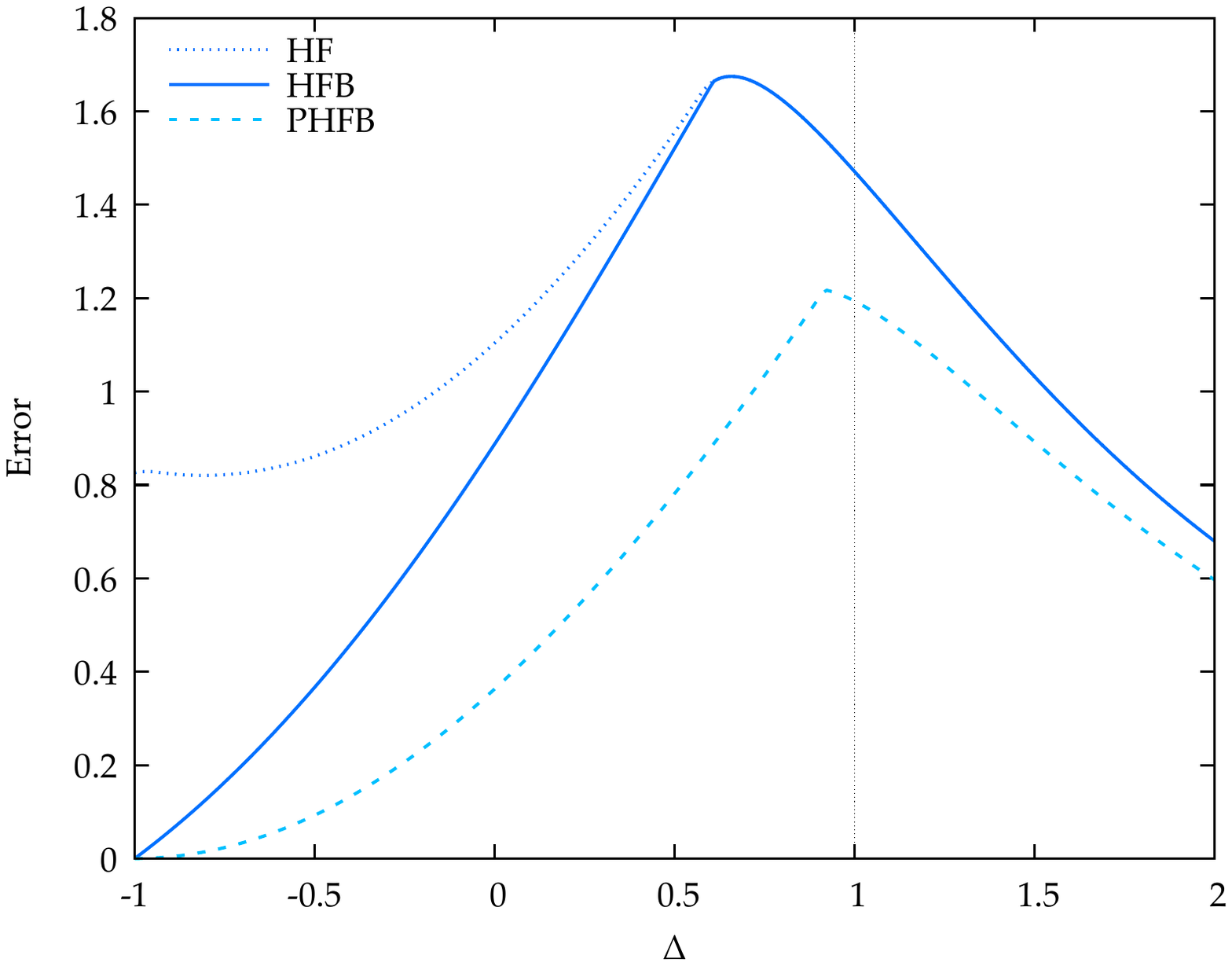}
\caption{Energy errors relative to the exact diagonalizations in the $2 \times 4$ {\JW}-transformed XXZ model with PBC for HF, HFB, and PHFB.  Left panel: $2 \times 4$ labeling.  Right panel: $4 \times 2$ labeling.
\label{Fig:XXZPHFB}}
\end{figure*}

Fortunately, PHFB provides quite reasonable accuracy for the {\JW}-transformed pairing Hamiltonian, as seen in Fig. \ref{Fig:PairingPHFB}.  In fact, it is energetically exact in both the strongly attractive ($G \to \infty$) and strongly repulsive ($G \to -\infty$) limits.  We prove this in the appendix.

Figure \ref{Fig:XXZPHFB} shows that PHFB is also a useful improvement upon HF or HFB in the {\JW}-transformed XXZ model.  Indeed, PHFB is variationally bounded from above by HFB (at least when we are looking at the ground state number sector), so it must be at least as good as the results we obtained for the $J_1\text{--}J_2$ and XXZ models with HFB on the {\JW}-transformed Hamiltonian.

\section{Discussion}
Each of the fermionizations we have considered preserves the exact spectrum of the spin Hamiltonian, but they generally give different results at the mean-field level.  In other words, all fermionizations are equivalent in the full Fock space, but some fermionizations are more compatible with mean-field approaches than others.  %For the Hamiltonians we have considered, those fermion to $\mathfrak{su}(2)$ mappings designed to produce more efficient encodings (i.e., encodings involving fewer Pauli operators overall) seem generally to yield fermionic Hamiltonians whose mean-field solutions are simply not as energetically accurate as are those from the {\JW} or {\parity} transformations, even though these latter transformations involve the very nonlocal {\JW} strings $\tilde{\phi}_p$.  
For the Hamiltonians we have considered, those fermion to $\mathfrak{su}(2)$ mappings designed to produce more efficient encodings seem to yield fermionic Hamiltonians whose mean-field solutions are energetically less accurate.  This is so even though the {\JW} or {\parity} transformations involve the very nonlocal {\JW} strings $\tilde{\phi}_p$, which the BK and Sierpinski encodings eliminate.  Fortunately, 
%Moreover, from a practical implementation perspective, 
the action of the {\JW} string on a fermionic mean-field state is to produce another mean-field state by virtue of the Thouless theorem \cite{Thouless1960}.  This means that it is actually straightforward to account for the {\JW} strings when evaluating matrix elements, by simply using the non-orthogonal Wick's theorem \cite{Balian1969}, though some care must be taken to implement this idea in a numerically robust and efficient way \cite{Chen2023}.

Another advantage of the {\JW} transformation is that by generalizing the string from $\tilde{\phi_p} = \exp(\mathrm{i} \, \pi \, \sum_{q<p} n_q)$ to
\begin{equation}
\hat{\phi}_p = \mathrm{e}^{\mathrm{i} \, \sum_{q} \theta_{qp} \, n_q} \, \tilde{\phi}_p
\end{equation}
where the matrix of parameters $\boldsymbol{\theta}$ is symmetric with vanishing diagonals, we can obtain results invariant to the labeling of the spin lattice \cite{Henderson2024a} when $\boldsymbol{\theta}$ is optimized.  This generalization, too, can be readily implemented using nonorthogonal Wick's theorem.  The same should be true for the {\parity} transformation.

There is one additional advantage which suggests we should prefer to use the {\JW} transformation to the {\parity} mapping: the {\JW} mapping converts an $\mathfrak{su}(2)$ Hamiltonian with $S_z$ symmetry to a fermionic Hamiltonian with number symmetry.  This permits us to select several different levels of mean-field theory: HF, HFB, HFBF, as well as their number-projected counterparts.  The other mappings all require us to use the rather more complicated HFBF mean-field theory or to employ Colpa's trick to replace HFBF for $M$ fermions with HFB for $M+1$ fermions (and the latter is pragmatically simpler) and generally do not permit symmetry projection at all.  This is not to say that these fermionized Hamiltonians do not inherit the symmetries of the underlying spin Hamiltonian, because naturally they do.  Rather, these symmetries are not generally one-body symmetries and therefore are difficult to project efficiently.

For all of these reasons, it seems clear that the {\JW} transformation, in addition to being the oldest mapping relating spins to fermions, is (in the context of mapping spin Hamiltonians to fermions for mean-field solution) also the most powerful (at least among those we have tried).

\section*{Supplementary Material}
See supplementary material for the proof that spin mean-field maps, under Jordan-Wigner transformation, to a special case of the fermionic HFBF mean-field.  We also provide additional numerical evidence of the extensivity of our results.

\section*{Acknowledgments}
This work was supported by the U.S. Department of Energy, Office of Basic Energy Sciences, under Award DE-SC0019374.  G.E.S. is a Welch Foundation Chair (C-0036). JDW holds concurrent appointments at Dartmouth College and as an Amazon Visiting Academic. This paper describes work performed at Dartmouth College and is not associated with Amazon.
%JN, AMP, BH and JDW thank ... This work was performed at Dartmouth...Amazon.
%\TMH{Note to self: Make sure acknowledgments are finalized before submission}

\appendix
\section{Exactness of JW-PHFB for the Pairing Model}
Here we demonstrate exactness of number-projected HFB in the $G \to \pm \infty$ limit of the {\JW}--transformed pairing model.  As we have noted, in the $S_z = 0$ sector we can work with the simpler Hamiltonian
\begin{equation}
H_\mathrm{eff} = \mp S^2;
\end{equation}
see Eqn. \ref{Eqn:Pairing}.

Let us start with the attractive limit.  In this case, we know that the ground state we seek has $S_z = 0$ and maximal $S^2$.  We can build this by starting with the state with all spins $\downarrow$ and applying the global raising operator the appropriate number of times to reach
\begin{equation}
\ket{\Psi_A} = \frac{1}{N!} \, \left(\sum_p S_p^+\right)^N \ket{\Downarrow} = P_{S_z} \, \mathrm{e}^{\sum_p S_p^+} \ket{\Downarrow}
\end{equation}
where $N$ is the number of $\uparrow$ spins and $\ket{\Downarrow}$ is the vacuum in which each spin is $\downarrow$; $P_{S_z}$ is the projector onto the appropriate $S_z$ sector.  The wave function $\exp(\sum_p S_p^+) \ket{\Downarrow}$ is a spin mean-field state.

Now transform this wave function with the {\JW} transformation.  Because $S_z$ becomes particle number, $P_{S_z}$ transforms to the fermionic number projector.  We have argued elsewhere \cite{Henderson2024b} (and prove in the supplementary material) that a spin mean-field state maps into a fermionic HFBF state, which means that $\ket{\Psi_A}$ maps, under {\JW} transformation, into a number-projected HFBF state.  But a number-projected HFBF state is also a number-projected HFB state since HFBF is just a linear combination of an even-particle number HFB and an odd-particle number HFB.  Ergo, the exact ground state in the $G \to \infty$ limit of the pairing model maps, under {\JW}, into PHFB.

The repulsive limit is more complicated.  As we have noted, it is an $S^2 = 0$ state, which we can get by the standard machinery of coupling angular momenta.  One way to write (one of) the ground states is in a dimerized form:
\begin{subequations}
\begin{align}
\ket{\Psi_R} &= \ket{s_{12}} \otimes \ket{s_{34}} \ldots \ket{s_{M-1,M}},
\\
\ket{s_{pq}} &= \ket{\uparrow_p \downarrow_q} - \ket{\downarrow_p \uparrow_q},
\end{align}
\end{subequations}
where we have a total of $M$ spins.  That is, we divide the system into disjoint pairs and place each pair in a separate singlet state.  We can express this wave function as
\begin{equation}
\ket{\Psi_R} = \left[\prod_{k=1}^{M/2} \left(1 - S_{2k}^+ \, S_{2k-1}^-\right)\right] \ket{\uparrow \downarrow \uparrow \downarrow \ldots}.
\end{equation}

Now we transform the wave function with the {\JW} transformation.  The state $\ket{\uparrow \downarrow \uparrow \downarrow \ldots}$ becomes a fermionic state $\ket{1 0 1 0 \ldots}$ with odd levels occupied and even levels empty.  The operator $S_{2k}^+ \, S_{2k-1}^-$ transforms as
\begin{subequations}
\begin{align}
S_{2k}^+ \, S_{2k-1}^-
    &\underset{\text{JW}}{\mapsto} c_{2k}^\dagger \, \tilde{\phi}_{2k} \, \tilde{\phi}_{2k-1} \, c_{2k-1}
\\
    &= c_{2k}^\dagger \, c_{2k-1}
\end{align}
\end{subequations}
where we have used that $\tilde{\phi}_{2k} \, \tilde{\phi}_{2k+1} = 1 - 2 \, c^\dagger_{2k-1} \, c_{2k-1}$ so that {\JW} strings between sequential fermionic operators cancel out.

All of this means that
\begin{subequations}
\begin{align}
\ket{\Psi_R} 
 &\underset{\text{JW}}{\mapsto} \left[\prod_{k=1}^{M/2} \left(1 - c_{2k}^\dagger \, c_{2k-1}\right)\right]] \ket{1010\ldots}
\\
 &= \left[\prod_{k=1}^{M/2} \mathrm{e}^{-c_{2k}^\dagger \, c_{2k-1}}\right] \ket{1010\ldots}
\\
 &= \mathrm{e}^{-\sum_{k=1}^{M/2} c_{2k}^\dagger \, c_{2k-1}} \ket{1010\ldots},
\end{align}
\end{subequations}
which is the Thouless representation of an HF state.  Thus, HF is already energetically exact in the $G\to -\infty$ limit of the {\JW}--transformed pairing Hamiltonian, and as HF is a special case of PHFB, this means PHFB is also energetically exact.

\section{The 8-Site Bravyi-Kitaev and Sierpinski Transformations}
For the {\JW} and {\parity} transformations, it is straightforward to write down both the mapping from fermions to spins and the inverse mapping from spins back to fermions.  For the {\BK} and {\sierpinski} mappings, this is less easy.  Accordingly, we note the 8-site mappings we have used here.  More specifically, we record the mappings from the Pauli operators $\sigma^x$ and $\sigma^z$ to the Majoranas; we extract $\sigma^y$ from
\begin{equation}
\sigma_p^z \, \sigma_p^x = \mathrm{i} \, \sigma_p^y.
\end{equation}

For brevity we define the Majorana products
\begin{subequations}
\begin{align}
A_p &= \mathrm{i} \, \gamma_{1,p} \, \gamma_{2,p} = 1 - 2 \, c_p^\dagger \, c_p,
\\
B_p &= \mathrm{i} \, \gamma_{2,p} \, \gamma_{1,p+1}.
\end{align}
\end{subequations}
These objects are Hermitian, mutually commuting, and satisfy $A_p^2 = B_p^2 = 1$.  They also commute with individual Majorana operators, except that they anticommute with their constitutents, i.e., $A_p$ anticommutes with $\gamma_{1,p}$ and $\gamma_{2,p}$ while $B_p$ anticommutes with $\gamma_{1,p+1}$ and $\gamma_{2,p}$.

The 8-site {\BK} spin-to-fermion mapping is
\begin{subequations}
\begin{alignat}{2}
\sigma_1^x &\underset{\text{BK}}{\mapsto} B_1                                  \qquad&     \sigma_1^z &\underset{\text{BK}}{\mapsto} A_1,
\\
\sigma_2^x &\underset{\text{BK}}{\mapsto} B_2 \, B_3                           \qquad&     \sigma_2^z &\underset{\text{BK}}{\mapsto} A_1 \, A_2,
\\
\sigma_3^x &\underset{\text{BK}}{\mapsto} B_3                                  \qquad&     \sigma_3^z &\underset{\text{BK}}{\mapsto} A_3,
\\
\sigma_4^x &\underset{\text{BK}}{\mapsto} B_4 \, B_5 \, B_6 \, B_7             \qquad&     \sigma_4^z &\underset{\text{BK}}{\mapsto} A_1 \, A_2 \, A_3 \, A_4,
\\
\sigma_5^x &\underset{\text{BK}}{\mapsto} B_5                                  \qquad&     \sigma_5^z &\underset{\text{BK}}{\mapsto} A_5,
\\
\sigma_6^x &\underset{\text{BK}}{\mapsto} B_6 \, B_7                           \qquad&     \sigma_6^z &\underset{\text{BK}}{\mapsto} A_5 \, A_6,
\\
\sigma_7^x &\underset{\text{BK}}{\mapsto} B_7                                  \qquad&     \sigma_7^z &\underset{\text{BK}}{\mapsto} A_7,
\\
\sigma_8^x &\underset{\text{BK}}{\mapsto} -\mathrm{i} \, \Pi \, \gamma_{2,8}   \qquad&     \sigma_8^z &\underset{\text{BK}}{\mapsto} \Pi,
\end{alignat}
\end{subequations}
while the 8-site {\sierpinski} spin-to-fermion mapping is instead
\begin{subequations}
\begin{alignat}{2}
\sigma^x_1 &\underset{\text{S}}{\mapsto}	B_1		                                  \qquad&    \sigma^z_1 &\underset{\text{S}}{\mapsto} A_1,
\\
\sigma^x_2 &\underset{\text{S}}{\mapsto}	B_2 \, B_3 \, B_4		                  \qquad&    \sigma^z_2 &\underset{\text{S}}{\mapsto} A_1\,  A_2 \, A_3,
\\
\sigma^x_3 &\underset{\text{S}}{\mapsto}	B_2		                                  \qquad&    \sigma^z_3 &\underset{\text{S}}{\mapsto} A_3,
\\
\sigma^x_4 &\underset{\text{S}}{\mapsto}	B_4		                                  \qquad&    \sigma^z_4 &\underset{\text{S}}{\mapsto} A_4,
\\
\sigma^x_5 &\underset{\text{S}}{\mapsto}	B_1 \, B_2 \, B_3 \, B_4 \, \gamma_{1,1}  \qquad&    \sigma^z_5 &\underset{\text{S}}{\mapsto} \Pi \, A_7 \, A_8,
\\
\sigma^x_6 &\underset{\text{S}}{\mapsto}	B_5                                       \qquad&    \sigma^z_6 &\underset{\text{S}}{\mapsto} A_6,
\\
\sigma^x_7 &\underset{\text{S}}{\mapsto}	B_7		                                  \qquad&    \sigma^z_7 &\underset{\text{S}}{\mapsto} A_7,
\\
\sigma^x_8 &\underset{\text{S}}{\mapsto}	\mathrm{i} \, \Pi \, \gamma_{2,8}	      \qquad&    \sigma^z_8 &\underset{\text{S}}{\mapsto} A_7 \,  A_8.
\end{alignat}
\end{subequations}
Overall signs are arbitrary, but factors of $\mathrm{i}$ are not.  Ultimately, this is because we can multiply any two Paulis by $-1$ without changing the $\mathfrak{su}(2)$ commutation rules.

Recall that $\Pi$ is the number parity operator (c.f. Eqn. \eqref{Eqn:DefNumberParity}), given in this notation as
\begin{equation}
\Pi = \prod_{p} A_p.
\end{equation}

\pagebreak
\onecolumngrid
\begin{center}
\Large
Supplementary Material for Fermionic Mean-Field Theory as a Tool for Studying Spin Hamiltonians
\end{center}
\renewcommand{\theequation}{S\arabic{equation}}
\renewcommand{\thefigure}{S\arabic{figure}}
\setcounter{equation}{0}
\setcounter{figure}{0}

\section*{Proof That a Spin Mean-Field Wave Function Maps to a Fermionic Mean-Field Wave Function}
In this note we prove that a spin mean-field wave function of the form $\ket{\mathrm{BCS}} = \exp(\sum_p t_p \, S_p^+) \ket{\Downarrow}$ maps, under Jordan--Wigner transformation, to a special case of the Hartree--Fock--Bogoliubov--Fukutome transformation.  Specifically, we show that
\begin{equation}
\mathrm{e}^{\sum_p t_p \, S_p^+} \ket{\Downarrow} \underset{\text{JW}}{\mapsto} \left(1 + \sum t_p \, c_p^\dagger\right) \, \mathrm{e}^{\sum_{p<q} t_p \, t_q \, c_p^\dagger \, c_q^\dagger} \ket{-}.
\end{equation}
\addtolength{\parskip}{\baselineskip}

We begin by expanding the spin mean-field state in different $S_z$ sectors.  This gives us
\begin{equation}
\mathrm{e}^{\sum_p t_p \, S_p^+} \ket{\Downarrow} = \left(1 + \sum_p t_p \, S_p^+  + \frac{1}{2} \, \sum_{pq} t_p \, t_q \, S_p^+ \, S_q^+ + \frac{1}{3!} \sum_{pqr} t_p \, t_q \, t_r \, S_p^+ \, S_q^+ \, S_r^+ + \ldots\right) \ket{\Downarrow}.
\end{equation}
Because the individual $S_p^+$ commute and are nilpotent ($S_p^+ \, S_p^+ = 0$) we can write equivalently
\begin{subequations}
\begin{align}
\mathrm{e}^{\sum_p t_p \, S_p^+} \ket{\Downarrow} 
	&= \left(1 + \sum_p t_p \, S_p^+  + \frac{1}{2} \, \sum_{p \ne q} t_p \, t_q \, S_p^+ \, S_q^+ + \frac{1}{3!} \sum_{p \ne q \ne r} t_p \, t_q \, t_r \, S_p^+ \, S_q^+ \, S_r^+ + \ldots\right) \ket{\Downarrow}
\\
	&= \left(1 + \sum_p t_p \, S_p^+  +  \sum_{p<q} t_p \, t_q \, S_p^+ \, S_q^+ +\sum_{p <q < r} t_p \, t_q \, t_r \, S_p^+ \, S_q^+ \, S_r^+ + \ldots\right) \ket{\Downarrow}.
\end{align}
\end{subequations}

Now we Jordan--Wigner-transform this wave function.  The spin vacuum $\ket{\Downarrow}$ transforms into the fermionic physical vacuum $\ket{-}$, while
\begin{subequations}
\begin{align}
S_p^+ &\underset{\text{JW}}{\mapsto} c_p^\dagger \, \tilde{\phi}_p,
\\
\tilde{\phi}_p &= \prod_{k<p} \left(1 - 2 \, n_k\right),
\\
n_k &= c_k^\dagger \, c_k.
\end{align}
\end{subequations}
This means that
\begin{equation}
\ket{\mathrm{BCS}} \underset{\text{JW}}{\mapsto} \left(1 + \sum_p t_p \, c_p^\dagger \, \tilde{\phi}_p + \sum_{p<q} t_p \, t_q \, c_p^\dagger \, \tilde{\phi}_p \, c_q^\dagger \, \tilde{\phi}_q + \sum_{p<q<r} t_p \, t_q \, t_r \, c_p^\dagger \, \tilde{\phi}_p \, c_q^\dagger \, \tilde{\phi}_q \, c_r^\dagger \, \tilde{\phi}_r + \ldots\right) \ket{-}.
\end{equation}

Because $\tilde{\phi}_p$ commutes with $c_q^\dagger$ for $q \ge p$ and we have written the operators in ascending order, the JW strings $\tilde{\phi}_p$ commute to the right where they meet the physical vacuum $\ket{-}$, for which we can use
\begin{equation}
\tilde{\phi}_p \, \ket{-} = \ket{-}.
\end{equation}
Thus, we have
\begin{equation}
\ket{\mathrm{BCS}} \underset{\text{JW}}{\mapsto} \left(1 + \sum_p t_p \, c_p^\dagger + \sum_{p<q} t_p \, t_q \, c_p^\dagger \, c_q^\dagger + \sum_{p<q<r} t_p \, t_q \, t_r \, c_p^\dagger \, c_q^\dagger \, c_r^\dagger + \ldots\right) \ket{-}.
\end{equation}
We must prove that
\begin{equation}
\left(1 + \sum_p t_p \, c_p^\dagger + \sum_{p<q} t_p \, t_q \, c_p^\dagger \, c_q^\dagger + \sum_{p<q<r} t_p \, t_q \, t_r \, c_p^\dagger \, c_q^\dagger \, c_r^\dagger + \ldots\right) \ket{-} = \left(1 + \sum t_p \, c_p^\dagger\right) \, \mathrm{e}^{\sum_{p<q} t_p \, t_q \, c_p^\dagger \, c_q^\dagger} \ket{-}.
\label{ToProve}
\end{equation}

\subsection*{A Useful Identity}
\vspace{-2\baselineskip}
Let us begin with a useful identity:
\begin{equation}
\sum_{p_1 < \ldots < p_m} \sum_{p_{m+1} > p_k} t_{p_1} \ldots t_{p_{m+1}} \, c_{p_1}^\dagger \ldots c_{p_{m+1}}^\dagger
 = 
\begin{cases}
\sum_{p_1 < \ldots < p_{m+1}} t_{p_1} \ldots t_{p_{m+1}} \, c_{p_1}^\dagger \ldots \, c_{p_{m+1}}^\dagger  & \qquad m-k \text{ even},
\\
\\
0  & \qquad m-k \text{ odd},
\end{cases}
\label{IdentityFinal}
\end{equation}
with $p_k \in \{0, p_1, \ldots, p_m\}$.

To see this, we begin by examining the simpler case where the second summation is restricted by an upper limit of $p_{k+1}$, which enforces an ordering between the indices in the two summations:
\begin{subequations}
\label{MasterIdentity}
\begin{align}
\sum_{p_1 < \ldots < p_m} \sum_{p_{m+1} > p_k}^{p_{k+1}} t_{p_1} \ldots t_{p_{m+1}} \, c_{p_1}^\dagger \ldots c_{p_{m+1}}^\dagger
 &= (-1)^{m-k} \sum_{p_1 < \ldots < p_m} \sum_{p_{m+1} > p_k}^{p_{k+1}} t_{p_1} \ldots t_{p_{m+1}} \, c_{p_1}^\dagger \ldots c_{p_k}^\dagger \, c_{p_{m+1}}^\dagger \, c_{p_{k+1}}^\dagger \ldots \, c_{p_m}^\dagger
\\
 &= (-1)^{m-k} \sum_{p_1 < \ldots < p_{m+1}}  t_{p_1} \ldots t_{p_{m+1}} \, c_{p_1}^\dagger \ldots \, c_{p_{m+1}}^\dagger,
\end{align}
\end{subequations}
where we used the fact that fermionic creation operators are nilpotent to eliminate the $p_{m+1} = p_{k+1}$ term and strictly order the indices.  In the special case that $k = m$, the upper limit of the summation over $p_{m+1}$ is $M$, where $M$ is the number of levels; if $p_k=0$ then the lower limit of the summation over $p_{m+1}$ is 1.  In the first line, we have anticommuted $c_{p_{m+1}}^\dagger$ to the left $m-k$ times to sort the creation operators in ascending order of their indices.  This gives us a plus or minus sign, depending on whether $m-k$ is an even (plus sign) or an odd (minus sign) number.  In the second line, we have relabeled the indices, renaming $p_{m+1} \to p_{k+1}$, $p_{k+1} \to p_{k+2}$, and so on.  The relabeling naturally has no effect on the coefficient since it is completely factorized.

Using Eqn. \eqref{MasterIdentity}, we can prove Eqn. \eqref{IdentityFinal}.

% When $m-k$ is even ($k=m-2j$), we have
% \begin{equation}
% \sum_{p_1 < \ldots < p_m} \sum_{p_{m+1} > p_{m-2j}}^{p_{m-2j+1}} t_{p_1} \ldots t_{p_{m+1}} \, c_{p_1}^\dagger \ldots c_{p_{m+1}}^\dagger
%  = \sum_{p_1 < \ldots < p_{m+1}} t_{p_1} \ldots t_{p_{m+1}} \, c_{p_1}^\dagger \ldots \, c_{p_{m+1}}^\dagger.
% \end{equation}
% That is, there are $2j$ operators between $c_{p_{m-2j}}^\dagger$ and $c_{p_{m+1}}^\dagger$ in the left-hand-side of the foregoing equation, namely $c_{p_{m-2j+1}}^\dagger$ through $c_{p_m}^\dagger$.  With an even number of operators, we get an overall $+$ sign from anticommutation. 
Let us start with the expression on the left-hand side of Eqn. \eqref{IdentityFinal}, in which the second summation runs over $p_{m+1} \in (p_k,M]$.  We divide this range into pieces: $(p_k,p_{k+1}]$, $(p_{k+1},p_{k+2}]$, and so on, ending with $(p_m,M]$. There are two cases to consider. If $m-k$ is even, i.e., $k=m-2j$,  there are $2j+1$ such pieces.  Applying the identity of Eqn. \eqref{MasterIdentity} to each, we see that all these pieces are identical, save that $j+1$ of them appear with a plus sign and $j$ of them appear with a minus sign.  All but one of these pieces thus cancel each other, and we are left with
\begin{equation}
\sum_{p_1 < \ldots < p_m} \sum_{p_{m+1} > p_{m-2j}} t_{p_1} \ldots t_{p_{m+1}} \, c_{p_1}^\dagger \ldots c_{p_{m+1}}^\dagger
 = \sum_{p_1 < \ldots < p_{m+1}}  t_{p_1} \ldots t_{p_{m+1}} \, c_{p_1}^\dagger \ldots \, c_{p_{m+1}}^\dagger.
\label{Identity:EvenDelta}
\end{equation}

When $m-k$ is odd ($k = m-2j-1)$, we instead divide the range $(p_k,M]$ into an even number of pieces.  These pieces are identical, save that half of them appear with a plus sign and half with a minus sign.  The entire summation thus adds to zero:
\begin{equation}
\sum_{p_1 < \ldots < p_{m}} \sum_{p_{m+1} > p_{m-2j-1}} t_{p_1} \ldots t_{p_{m+1}} \, c_{p_1}^\dagger \ldots c_{p_{m+1}}^\dagger
 = 0.
\label{Identity:OddDelta}
\end{equation}

Equations \eqref{Identity:EvenDelta} and \eqref{Identity:OddDelta} together prove our target, Eqn. \eqref{IdentityFinal}.

\subsection*{Terms in the Left-Hand-Side of Eqn. (\ref{ToProve}) With an Even Number of Fermionic Operators}
\vspace{-2\baselineskip}
Our task here is to prove that
\begin{equation}
\frac{1}{n!} \, \left(\sum_{p<q} t_p \, t_q \, c_p^\dagger \, c_q^\dagger\right)^n = \sum_{p_1 < \ldots < p_{2n}} t_{p_1} \ldots t_{p_{2n}} \, c_{p_1}^\dagger \ldots c_{p_{2n}}^\dagger.
\label{ToProve1}
\end{equation}
We will do so inductively.

For $n=1$ this is trivially true.  We must show that Eqn. \eqref{ToProve1} is true for $n+1$ if it is true for $n$.  That is, we must show that
\begin{equation}
\frac{1}{(n+1)!} \, \left(\sum_{p<q} t_p \, t_q \, c_p^\dagger \, c_q^\dagger\right)^{n+1} = \sum_{p_1 < \ldots < p_{2n+2}} t_{p_1} \ldots t_{p_{2n+2}} \, c_{p_1}^\dagger \ldots c_{p_{2n+2}}^\dagger.
\label{InductionTarget1}
\end{equation}

Now, assuming that Eqn. \eqref{ToProve1} is true for $n$, we have
\begin{equation}
\frac{1}{(n+1)!} \, \left(\sum_{p<q} t_p \, t_q \, c_p^\dagger \, c_q^\dagger\right)^{n+1} = \frac{1}{n+1} \, \sum_{p_1 < \ldots < p_{2n}} t_{p_1} \ldots t_{p_{2n}} \, c_{p_1}^\dagger \ldots c_{p_{2n}}^\dagger \, \sum_{p_{2n+1} < p_{2n+2}} t_{p_{2n+1}} \, t_{p_{2n+2}} \, c_{p_{2n+1}}^\dagger \, c_{p_{2n+2}}^\dagger.
\label{Eqn:Induction}
\end{equation}

Divide the summation over $p_{2n+1}$ into the $2n+1$ distinct ranges: $p_{2n+1} < p_1$, and $p_1 < p_{2n+1} < p_2$, and so on.  After anticommuting $c_{p_{2n+1}}^\dagger$ into the appropriate position so that the operators $c_{p_1}^\dagger \ldots c_{p_{2n+1}}^\dagger$ appear in ascending order, we have
\begin{align}
\frac{1}{(n+1)!} \, \left(\sum_{p<q} t_p \, t_q \, c_p^\dagger \, c_q^\dagger\right)^{n+1}
	&= \frac{1}{n+1} \, \sum_{p_1 < \ldots < p_{2n+1}} t_{p_1} \ldots t_{p_{2n+2}} \, c_{p_1}^\dagger \ldots c_{p_{2n+1}}^\dagger \, \sum_{p_{2n+2} > p_{2n+1}} \, c_{p_{2n+2}}^\dagger
\\
	&- \frac{1}{n+1} \, \sum_{p_1 < \ldots < p_{2n+1} < p_{2n}} t_{p_1} \ldots t_{p_{2n+2}} \, c_{p_1}^\dagger \ldots c_{p_{2n+1}}^\dagger \,  c_{p_{2n}}^\dagger \, \sum_{p_{2n+2} > p_{2n+1}} \, c_{p_{2n+2}}^\dagger
\nonumber
\\
	&+\ldots
\nonumber
\\
	&+ \frac{1}{n+1} \, \sum_{p_{2n+1} < p_1 < \ldots < p_{2n}} t_{p_1} \ldots t_{p_{2n+2}} \, c_{p_{2n+1}}^\dagger \, c_{p_1}^\dagger \ldots c_{p_{2n}}^\dagger \, \sum_{p_{2n+2} > p_{2n+1}} \, c_{p_{2n+2}}^\dagger.
\nonumber
\end{align}
It will prove helpful to relabel indices $p_1$ through $p_{2n+1}$ at this point:
\begin{align}
\frac{1}{(n+1)!} \, \left(\sum_{p<q} t_p \, t_q \, c_p^\dagger \, c_q^\dagger\right)^{n+1}
	&= \frac{1}{n+1} \, \sum_{p_1 < \ldots < p_{2n+1}} t_{p_1} \ldots t_{p_{2n+2}} \, c_{p_1}^\dagger \ldots c_{p_{2n+1}}^\dagger \, \sum_{p_{2n+2} > p_{2n+1}} \, c_{p_{2n+2}}^\dagger
\label{Relabeled}
\\
	&- \frac{1}{n+1} \, \sum_{p_1 < \ldots < p_{2n+1}} t_{p_1} \ldots t_{p_{2n+2}} \, c_{p_1}^\dagger \ldots c_{p_{2n+1}}^\dagger \, \sum_{p_{2n+2} > p_{2n}} \, c_{p_{2n+2}}^\dagger
\nonumber
\\
	&+\ldots
\nonumber
\\
	&+ \frac{1}{n+1} \, \sum_{p_1 < \ldots < p_{2n+1}} t_{p_1} \ldots t_{p_{2n+2}} \, c_{p_1}^\dagger \ldots c_{p_{2n+1}}^\dagger \, \sum_{p_{2n+2} > p_{1}} \, c_{p_{2n+2}}^\dagger.
\nonumber
\end{align}

Now we apply the identity of Eqn. \eqref{IdentityFinal} to each line sequentially.  Doing so reveals that the odd lines (those with plus signs) are all identical, and the even lines (those with minus signs) all vanish.  For example, the first line on the right-hand-side of Eqn. \eqref{Relabeled} has $m=2n+1$ and $k=m$ so that $m-k=0$; the second line has $m-k=1$, and so on.  We have $n+1$ odd lines, and this factor of $n+1$ exactly cancels the $\frac{1}{n+1}$ on each line, completing the proof.

Let us give an example for concreteness.  We will consider the case $n=2$, so we have
\begin{equation}
\mathcal{O} = \frac{1}{2} \sum_{p_1 < p_2} \, \sum_{p_3 < p_4} t_{p_1} \, \ldots t_{p_4} \, c_{p_1}^\dagger \, c_{p_2}^\dagger \, c_{p_3}^\dagger \, c_{p_4}^\dagger.
\end{equation}
We break the summation over $p_3$ into 3 distinct ranges, resulting in
\begin{align}
\mathcal{O}
	&= \frac{1}{2} \sum_{p_1 < p_2 < p_3} \sum_{p_4 > p_3} t_{p_1} \, \ldots t_{p_4} \, c_{p_1}^\dagger \, c_{p_2}^\dagger \, c_{p_3}^\dagger \, c_{p_4}^\dagger
\\
	&+ \frac{1}{2} \sum_{p_1 < p_3 < p_2} \sum_{p_4 > p_3} t_{p_1} \, \ldots t_{p_4} \, c_{p_1}^\dagger \, c_{p_2}^\dagger \, c_{p_3}^\dagger \, c_{p_4}^\dagger
\nonumber
\\
	&+ \frac{1}{2} \sum_{p_3 < p_1 < p_2} \sum_{p_4 > p_3} t_{p_1} \, \ldots t_{p_4} \, c_{p_1}^\dagger \, c_{p_2}^\dagger \, c_{p_3}^\dagger \, c_{p_4}^\dagger.
\nonumber
\end{align}

Anticommute $c_{p_3}^\dagger$ into the correct position:
\begin{align}
\mathcal{O}
	&= \frac{1}{2} \sum_{p_1 < p_2 < p_3} \sum_{p_4 > p_3} t_{p_1} \, \ldots t_{p_4} \, c_{p_1}^\dagger \, c_{p_2}^\dagger \, c_{p_3}^\dagger \, c_{p_4}^\dagger
\\
	&- \frac{1}{2} \sum_{p_1 < p_3 < p_2} \sum_{p_4 > p_3} t_{p_1} \, \ldots t_{p_4} \, c_{p_1}^\dagger \, c_{p_3}^\dagger \, c_{p_2}^\dagger \, c_{p_4}^\dagger
\nonumber
\\
	&+ \frac{1}{2} \sum_{p_3 < p_1 < p_2} \sum_{p_4 > p_3} t_{p_1} \, \ldots t_{p_4} \, c_{p_3}^\dagger \, c_{p_1}^\dagger \, c_{p_2}^\dagger \, c_{p_4}^\dagger.
\nonumber
\end{align}

Relabel for convenience:
\begin{align}
\mathcal{O}
	&= \frac{1}{2} \sum_{p_1 < p_2 < p_3} \sum_{p_4 > p_3} t_{p_1} \, \ldots t_{p_4} \, c_{p_1}^\dagger \, c_{p_2}^\dagger \, c_{p_3}^\dagger \, c_{p_4}^\dagger
\label{WorkedExampleRelabel}
\\
	&- \frac{1}{2} \sum_{p_1 < p_2 < p_3} \sum_{p_4 > p_2} t_{p_1} \, \ldots t_{p_4} \, c_{p_1}^\dagger \, c_{p_2}^\dagger \, c_{p_3}^\dagger \, c_{p_4}^\dagger
\nonumber
\\
	&+ \frac{1}{2} \sum_{p_1 < p_2 < p_3} \sum_{p_4 > p_1} t_{p_1} \, \ldots t_{p_4} \, c_{p_1}^\dagger \, c_{p_2}^\dagger \, c_{p_3}^\dagger \, c_{p_4}^\dagger.
\nonumber
\end{align}

Now break the summations over $p_4$ into separate ranges:
\begin{align}
\mathcal{O}
	&= \Bigg(\frac{1}{2} \sum_{p_1 < p_2 < p_3} \sum_{p_4 > p_3} t_{p_1} \, \ldots t_{p_4} \, c_{p_1}^\dagger \, c_{p_2}^\dagger \, c_{p_3}^\dagger \, c_{p_4}^\dagger\Bigg)
\\
	&- \Bigg(\frac{1}{2} \sum_{p_1 < p_2 < p_3} \sum_{p_4 > p_2}^{p_3} t_{p_1} \, \ldots t_{p_4} \, c_{p_1}^\dagger \, c_{p_2}^\dagger \, c_{p_3}^\dagger \, c_{p_4}^\dagger
     + \frac{1}{2} \, \sum_{p_1 < p_2 < p_3} \sum_{p_4 > p_3} t_{p_1} \, \ldots t_{p_4} \, c_{p_1}^\dagger \, c_{p_2}^\dagger \, c_{p_3}^\dagger \, c_{p_4}^\dagger\Bigg)
\nonumber
\\
	&+ \Bigg(\frac{1}{2} \sum_{p_1 < p_2 < p_3} \sum_{p_4 > p_1}^{p_2} t_{p_1} \, \ldots t_{p_4} \, c_{p_1}^\dagger \, c_{p_2}^\dagger \, c_{p_3}^\dagger \, c_{p_4}^\dagger
    + \frac{1}{2} \sum_{p_1 < p_2 < p_3} \sum_{p_4 > p_2}^{p_3} t_{p_1} \, \ldots t_{p_4} \, c_{p_1}^\dagger \, c_{p_2}^\dagger \, c_{p_3}^\dagger \, c_{p_4}^\dagger
\nonumber
\\
   &+ \sum_{p_1 < p_2 < p_3} \sum_{p_4 > p_3} t_{p_1} \, \ldots t_{p_4} \, c_{p_1}^\dagger \, c_{p_2}^\dagger \, c_{p_3}^\dagger \, c_{p_4}^\dagger\Bigg).
\nonumber
\end{align}
Note that each set of terms enclosed in parentheses corresponds to a single line of Eqn. \eqref{WorkedExampleRelabel}.

Anticommute $c_{p_4}^\dagger$ into the appropriate positions:
\begin{align}
\mathcal{O}
	&= \Bigg(\frac{1}{2} \sum_{p_1 < p_2 < p_3} \sum_{p_4 > p_3} t_{p_1} \, \ldots t_{p_4} \, c_{p_1}^\dagger \, c_{p_2}^\dagger \, c_{p_3}^\dagger \, c_{p_4}^\dagger\Bigg)
\\
	&+ \Bigg(\frac{1}{2} \sum_{p_1 < p_2 < p_3} \sum_{p_4 > p_2}^{p_3} t_{p_1} \, \ldots t_{p_4} \, c_{p_1}^\dagger \, c_{p_2}^\dagger \, c_{p_4}^\dagger \, c_{p_3}^\dagger
	- \frac{1}{2} \sum_{p_1 < p_2 < p_3} \sum_{p_4 > p_3} t_{p_1} \, \ldots t_{p_4} \, c_{p_1}^\dagger \, c_{p_2}^\dagger \, c_{p_3}^\dagger \, c_{p_4}^\dagger\Bigg)
\nonumber
\\
	&+ \Bigg(\frac{1}{2} \sum_{p_1 < p_2 < p_3} \sum_{p_4 > p_1}^{p_2} t_{p_1} \, \ldots t_{p_4} \, c_{p_1}^\dagger \, c_{p_4}^\dagger \, c_{p_2}^\dagger \, c_{p_3}^\dagger
    - \frac{1}{2} \sum_{p_1 < p_2 < p_3} \sum_{p_4 > p_2}^{p_3} t_{p_1} \, \ldots t_{p_4} \, c_{p_1}^\dagger \, c_{p_2}^\dagger \, c_{p_4}^\dagger \, c_{p_3}^\dagger \, c_{p_4}^\dagger
\nonumber
\\
	&+ \frac{1}{2} \sum_{p_1 < p_2 < p_3} \sum_{p_4 > p_3} t_{p_1} \, \ldots t_{p_4} \, c_{p_1}^\dagger \, c_{p_2}^\dagger \, c_{p_3}^\dagger \, c_{p_4}^\dagger\Bigg).
\nonumber
\end{align}
After relabeling, we see that the first and third set of terms are identical and the second set of terms vanishes, so we have
\begin{equation}
\mathcal{O}	= \sum_{p_1 < p_2 < p_3 < p_4} t_{p_1} \, \ldots t_{p_4} \, c_{p_1}^\dagger \, c_{p_2}^\dagger \, c_{p_3}^\dagger \, c_{p_4}^\dagger.
\end{equation}

\subsection*{Terms in the Left-Hand-Side of Eqn. (\ref{ToProve}) With an Odd Number of Fermionic Operators}
\vspace{-2\baselineskip}
It remains to prove that
\begin{equation}
\left(\sum_{p_{2n+1}} t_{p_{2n+1}} \, c_{p_{2n+1}}^\dagger\right) \, \sum_{p_1 < \ldots < p_{2n}} \, t_{p_1} \ldots t_{p_{2n}} \, c_{p_1}^\dagger \ldots c_{p_{2n}}^\dagger = \sum_{p_1 < p_2 \ldots < p_{2n+1}} t_{p_1} \ldots t_{p_{2n+1}} \, c_{p_1}^\dagger \ldots c_{p_{2n+1}}^\dagger.
\end{equation}
After anticommuting $c_{p_{2n+1}}^\dagger$ all the way to the right (using the fact that if any two indices are equal the term vanishes, and otherwise we have anticommuted through an even number of operators), this is a straightforward application of Eqn. \ref{IdentityFinal} where $m=2n$ and $k=0$.

\subsection*{The Final Result}
\vspace{-2\baselineskip}
We have seen that
\begin{subequations}
\begin{align}
\sum_{p_1 < \ldots < p_{2n}} t_{p_1} \ldots t_{p_{2n}} \, c_{p_1}^\dagger \ldots c_{p_{2n}}^\dagger &= \frac{1}{n!} \, \left(\sum_{p<q} t_p \, t_q \, c_p^\dagger \, c_q^\dagger\right)^n,
\\
\sum_{p_1 < \ldots < p_{2n+1}} t_{p_1} \ldots t_{p_{2n+1}} \, c_{p_1}^\dagger \ldots c_{p_{2n+1}}^\dagger &= \left(\sum_{p_{2n+1}} t_{p_{2n+1}} \, c_{p_{2n+1}}^\dagger\right) \, \sum_{p_1 < p_2 \ldots < p_{2n}} \, t_{p_1} \ldots t_{p_{2n}} \, c_{p_1}^\dagger \ldots c_{p_{2n}}^\dagger.
\end{align}
\end{subequations}
From the first result, we have that
\begin{equation}
\mathrm{e}^{\sum_{p<q} t_p \, t_q \, c_p^\dagger \, c_q^\dagger} = 1 + \sum_{p<q} t_p \, t_q \, c_p^\dagger \, c_q^\dagger + \sum_{p<q<r<s} t_p \, t_q \, t_r \, t_s \, c_p^\dagger \, c_q^\dagger \, c_r^\dagger \, c_s^\dagger + \ldots.
\end{equation}
The two lines together mean that
\begin{equation}
\sum_p t_p \, c_p^\dagger \, \mathrm{e}^{\sum_{p<q} t_p \, t_q \, c_p^\dagger \, c_q^\dagger} = \sum_p t_p \, c_p^\dagger + \sum_{p<q<r} t_p \, t_q \, t_r \, c_p^\dagger \, c_q^\dagger \, c_r^\dagger + \sum_{p<q<r<s<t} t_p \, t_q \, t_r \, t_s \, t_t \, c_p^\dagger \, c_q^\dagger \, c_r^\dagger \, c_s^\dagger \, c_t^\dagger + \ldots
\end{equation}
Overall, we therefore have
\begin{equation}
\left(1 + \sum_p t_p \, c_p^\dagger\right) \, \mathrm{e}^{\sum_{p<q} t_p \, t_q \, c_p^\dagger \, c_q^\dagger} \ket{-}
 = \left(1 + \sum_p t_p \, c_p^\dagger + \sum_{p<q} t_p \, t_q \, c_p^\dagger \, c_q^\dagger + \sum_{p<q<r} t_p \, t_q \, t_r \, c_p^\dagger \, c_q^\dagger \, c_r^\dagger + \ldots\right) \ket{-}.
\end{equation}

This is what we wished, ultimately, to prove: the right-hand-side is the JW-transformation of the spin mean-field state, and the left-hand-side is a special case of an HFBF fermionic mean-field.  This means that any $\mathfrak{su}(2)$ problem exactly solved by a spin mean-field of the form $\ket{\mathrm{BCS}} = \mathrm{e}^{\sum_p t_p \, S_p^+} \ket{\Downarrow}$ is also exactly solved, upon JW-transformation, by a fermionic HFBF.  Further, because a general HFBF is actually of the form
\begin{equation}
\ket{\mathrm{HFBF}} = \left(1 + \sum_p t_p \, c_p^\dagger\right) \, \mathrm{e}^{\sum_{p<q} \mathcal{T}_{pq} \, c_p^\dagger \, c_q^\dagger} \ket{-},
\end{equation}
it also proves that the mean-field solution of a JW-transformed $\mathfrak{su}(2)$ Hamiltonian is variationally bound from above by the spin mean-field solution.

\section*{Additional Plots Demonstrating Extensivity}
\begin{figure*}
\includegraphics[width=0.45\textwidth]{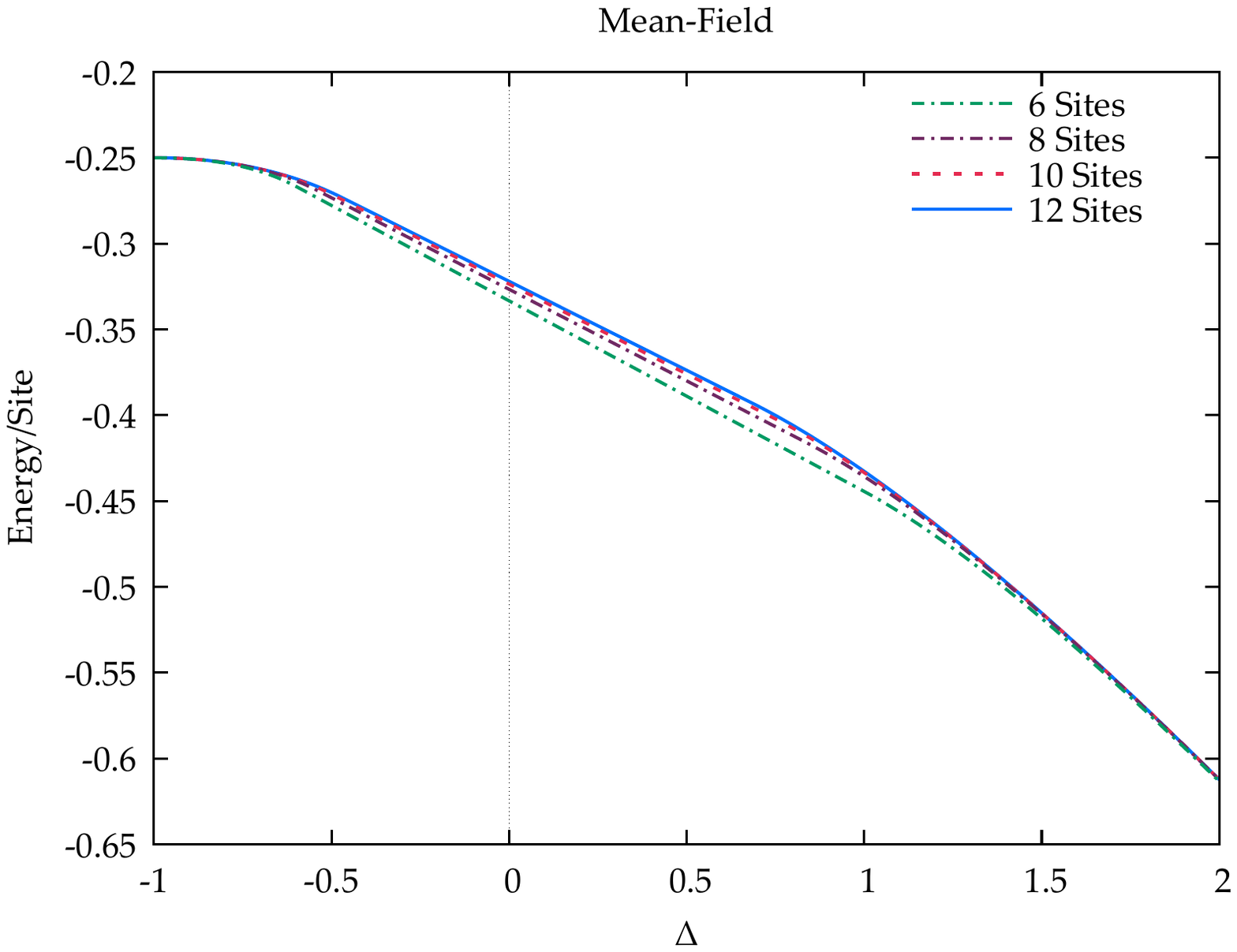}
\hfill
\includegraphics[width=0.45\textwidth]{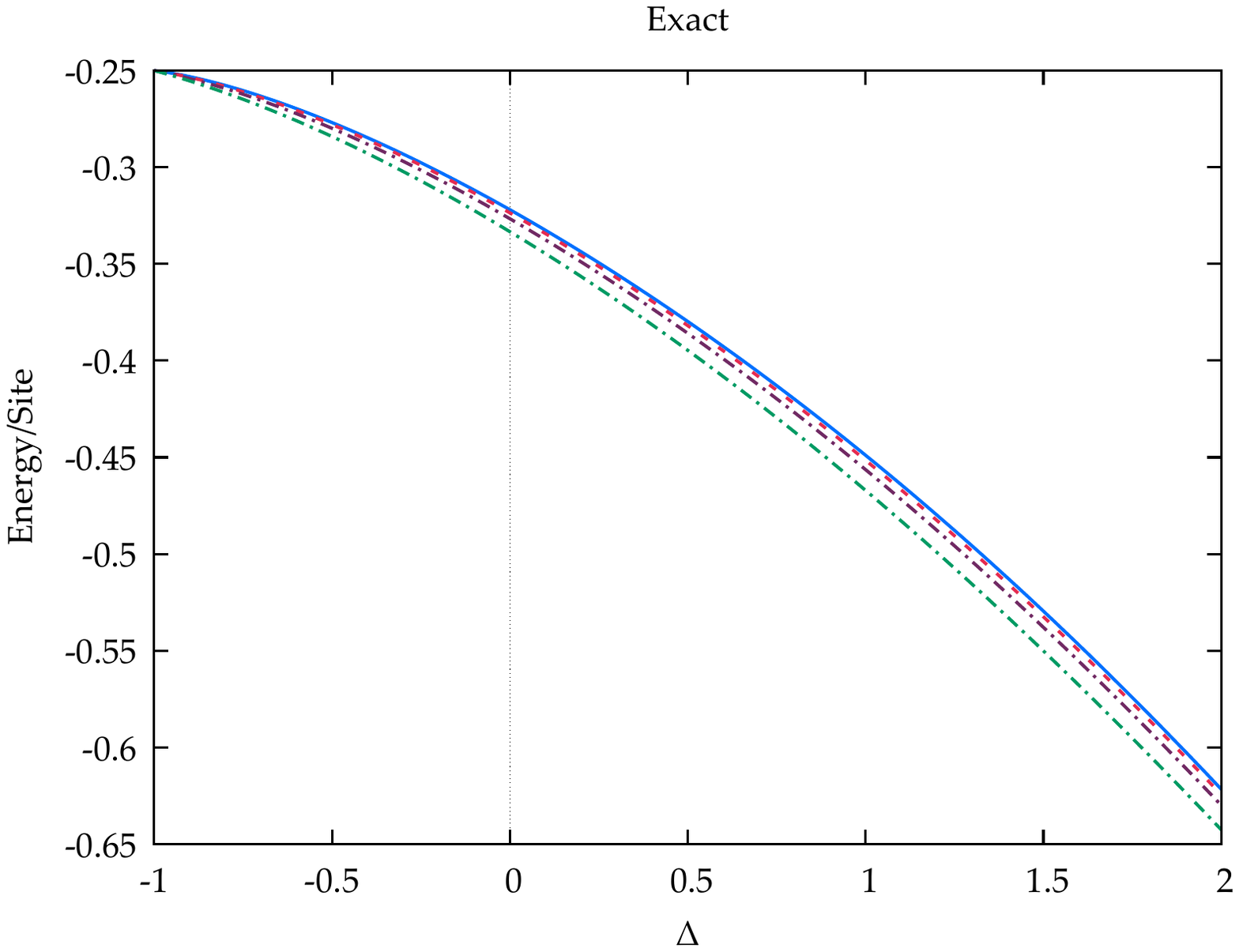}
\caption{Total energy per site in the 1D XXZ Hamiltonian with periodic boundary conditions, comparing the fermionic mean-field after Jordan-Wigner transformation (left panel) to the exact results (right panel).
\label{Fig:Extensivity}}
\end{figure*}

\vspace{-2\baselineskip}
We have noted in the main manuscript that our results are extensive and have shown data for the 2D $J_1\text{--}J_2$ Hamiltonian to support this notion.  Here we show results for the 1D XXZ model.  As Fig. \ref{Fig:Extensivity} shows, total energies/site in the fermionic mean-field behave, as a function of system size, essentially as do the results of the exact diagonalization of the spin Hamiltonian, and converge for increasing lattice sizes.

\begin{thebibliography}{63}%
\makeatletter
\providecommand \@ifxundefined [1]{%
 \@ifx{#1\undefined}
}%
\providecommand \@ifnum [1]{%
 \ifnum #1\expandafter \@firstoftwo
 \else \expandafter \@secondoftwo
 \fi
}%
\providecommand \@ifx [1]{%
 \ifx #1\expandafter \@firstoftwo
 \else \expandafter \@secondoftwo
 \fi
}%
\providecommand \natexlab [1]{#1}%
\providecommand \enquote  [1]{``#1''}%
\providecommand \bibnamefont  [1]{#1}%
\providecommand \bibfnamefont [1]{#1}%
\providecommand \citenamefont [1]{#1}%
\providecommand \href@noop [0]{\@secondoftwo}%
\providecommand \href [0]{\begingroup \@sanitize@url \@href}%
\providecommand \@href[1]{\@@startlink{#1}\@@href}%
\providecommand \@@href[1]{\endgroup#1\@@endlink}%
\providecommand \@sanitize@url [0]{\catcode `\\12\catcode `\$12\catcode
  `\&12\catcode `\#12\catcode `\^12\catcode `\_12\catcode `\%12\relax}%
\providecommand \@@startlink[1]{}%
\providecommand \@@endlink[0]{}%
\providecommand \url  [0]{\begingroup\@sanitize@url \@url }%
\providecommand \@url [1]{\endgroup\@href {#1}{\urlprefix }}%
\providecommand \urlprefix  [0]{URL }%
\providecommand \Eprint [0]{\href }%
\providecommand \doibase [0]{http://dx.doi.org/}%
\providecommand \selectlanguage [0]{\@gobble}%
\providecommand \bibinfo  [0]{\@secondoftwo}%
\providecommand \bibfield  [0]{\@secondoftwo}%
\providecommand \translation [1]{[#1]}%
\providecommand \BibitemOpen [0]{}%
\providecommand \bibitemStop [0]{}%
\providecommand \bibitemNoStop [0]{.\EOS\space}%
\providecommand \EOS [0]{\spacefactor3000\relax}%
\providecommand \BibitemShut  [1]{\csname bibitem#1\endcsname}%
\let\auto@bib@innerbib\@empty
%</preamble>
\bibitem [{\citenamefont {Batista}\ and\ \citenamefont
  {Ortiz}(2001)}]{Batista2001}%
  \BibitemOpen
  \bibfield  {author} {\bibinfo {author} {\bibfnamefont {C.~D.}\ \bibnamefont
  {Batista}}\ and\ \bibinfo {author} {\bibfnamefont {G.}~\bibnamefont
  {Ortiz}},\ }in\ \href@noop {} {\emph {\bibinfo {booktitle} {Condensed Matter
  Theories}}},\ Vol.~\bibinfo {volume} {16},\ \bibinfo {editor} {edited by\
  \bibinfo {editor} {\bibfnamefont {S.}~\bibnamefont {Hernandez}}\ and\
  \bibinfo {editor} {\bibfnamefont {W.~J.}\ \bibnamefont {Clark}}}\ (\bibinfo
  {publisher} {Nova Science Publishers, Inc},\ \bibinfo {address} {Huntington,
  New York},\ \bibinfo {year} {2001})\ pp.\ \bibinfo {pages}
  {1--15}\BibitemShut {NoStop}%
\bibitem [{\citenamefont {Nishimori}\ and\ \citenamefont
  {Ortiz}(2011)}]{Nishimori2011}%
  \BibitemOpen
  \bibfield  {author} {\bibinfo {author} {\bibfnamefont {H.}~\bibnamefont
  {Nishimori}}\ and\ \bibinfo {author} {\bibfnamefont {G.}~\bibnamefont
  {Ortiz}},\ }\href@noop {} {\emph {\bibinfo {title} {Elements of Phase
  Transitions and Critical Phenomena}}}\ (\bibinfo  {publisher} {Oxford
  University Press},\ \bibinfo {address} {Oxford},\ \bibinfo {year} {2011})\
  p.\ \bibinfo {pages} {220}\BibitemShut {NoStop}%
\bibitem [{\citenamefont {Jordan}\ and\ \citenamefont
  {Wigner}(1928)}]{Jordan1928}%
  \BibitemOpen
  \bibfield  {author} {\bibinfo {author} {\bibfnamefont {P.}~\bibnamefont
  {Jordan}}\ and\ \bibinfo {author} {\bibfnamefont {E.}~\bibnamefont
  {Wigner}},\ }\href {\doibase 10.1007/BF01331938} {\bibfield  {journal}
  {\bibinfo  {journal} {Zeitschrift f\"ur Physik}\ }\textbf {\bibinfo {volume}
  {47}},\ \bibinfo {pages} {631} (\bibinfo {year} {1928})}\BibitemShut
  {NoStop}%
\bibitem [{\citenamefont {Gon{\c{c}}alves}\ \emph {et~al.}(2005)\citenamefont
  {Gon{\c{c}}alves}, \citenamefont {Coutinho},\ and\ \citenamefont {{de
  Lima}}}]{Goncalves2005}%
  \BibitemOpen
  \bibfield  {author} {\bibinfo {author} {\bibfnamefont {L.~L.}\ \bibnamefont
  {Gon{\c{c}}alves}}, \bibinfo {author} {\bibfnamefont {L.~P.~S.}\ \bibnamefont
  {Coutinho}}, \ and\ \bibinfo {author} {\bibfnamefont {J.~P.}\ \bibnamefont
  {{de Lima}}},\ }\href {\doibase 10.1016/j.physa.2004.06.066} {\bibfield
  {journal} {\bibinfo  {journal} {Physica A}\ }\textbf {\bibinfo {volume}
  {345}},\ \bibinfo {pages} {71} (\bibinfo {year} {2005})}\BibitemShut
  {NoStop}%
\bibitem [{\citenamefont {Verkholyak}\ \emph {et~al.}(2006)\citenamefont
  {Verkholyak}, \citenamefont {Honecker},\ and\ \citenamefont
  {Brenig}}]{Verkholyak2006}%
  \BibitemOpen
  \bibfield  {author} {\bibinfo {author} {\bibfnamefont {T.}~\bibnamefont
  {Verkholyak}}, \bibinfo {author} {\bibfnamefont {A.}~\bibnamefont
  {Honecker}}, \ and\ \bibinfo {author} {\bibfnamefont {W.}~\bibnamefont
  {Brenig}},\ }\href {\doibase 10.1140/epjb/e2006-00077-1} {\bibfield
  {journal} {\bibinfo  {journal} {Eur. Phys. J. B}\ }\textbf {\bibinfo {volume}
  {49}},\ \bibinfo {pages} {283} (\bibinfo {year} {2006})}\BibitemShut
  {NoStop}%
\bibitem [{\citenamefont {Kitaev}\ and\ \citenamefont
  {Laumann}(2008)}]{Kitaev2009}%
  \BibitemOpen
  \bibfield  {author} {\bibinfo {author} {\bibfnamefont {A.}~\bibnamefont
  {Kitaev}}\ and\ \bibinfo {author} {\bibfnamefont {C.}~\bibnamefont
  {Laumann}},\ }\href {\doibase 10.48550/arXiv.0904.2771} {\enquote {\bibinfo
  {title} {Topological phases and quantum computation},}\ }\bibinfo
  {howpublished} {Lectures given by Alexei Kitaev at the 2008 Les Houches
  summer school ``Exact methods in low-dimensional physics and quantum
  computing.''} (\bibinfo {year} {2008})\BibitemShut {NoStop}%
\bibitem [{\citenamefont {Verkholyak}\ \emph {et~al.}(2010)\citenamefont
  {Verkholyak}, \citenamefont {Stre\v{c}ka}, \citenamefont {Ja\v{s}\v{c}ur},\
  and\ \citenamefont {Richter}}]{Verkholyak2010}%
  \BibitemOpen
  \bibfield  {author} {\bibinfo {author} {\bibfnamefont {T.}~\bibnamefont
  {Verkholyak}}, \bibinfo {author} {\bibfnamefont {J.}~\bibnamefont
  {Stre\v{c}ka}}, \bibinfo {author} {\bibfnamefont {M.}~\bibnamefont
  {Ja\v{s}\v{c}ur}}, \ and\ \bibinfo {author} {\bibfnamefont {J.}~\bibnamefont
  {Richter}},\ }\href {\doibase 10.12693/APhysPolA.118.978} {\bibfield
  {journal} {\bibinfo  {journal} {Acta Phys. Pol. A}\ }\textbf {\bibinfo
  {volume} {118}},\ \bibinfo {pages} {978} (\bibinfo {year}
  {2010})}\BibitemShut {NoStop}%
\bibitem [{\citenamefont {Bardyn}\ and\ \citenamefont
  {\.{I}mamo\v{g}lu}(2012)}]{Bardyn2012}%
  \BibitemOpen
  \bibfield  {author} {\bibinfo {author} {\bibfnamefont {C.-E.}\ \bibnamefont
  {Bardyn}}\ and\ \bibinfo {author} {\bibfnamefont {A.}~\bibnamefont
  {\.{I}mamo\v{g}lu}},\ }\href {\doibase 10.1103/PhysRevLett.109.253606}
  {\bibfield  {journal} {\bibinfo  {journal} {Phys. Rev. Lett.}\ }\textbf
  {\bibinfo {volume} {109}},\ \bibinfo {pages} {253606} (\bibinfo {year}
  {2012})}\BibitemShut {NoStop}%
\bibitem [{\citenamefont {Zvyagin}(2013)}]{Zvyagin2013}%
  \BibitemOpen
  \bibfield  {author} {\bibinfo {author} {\bibfnamefont {A.~A.}\ \bibnamefont
  {Zvyagin}},\ }\href {\doibase 10.1103/PhysRevLett.110.217207} {\bibfield
  {journal} {\bibinfo  {journal} {Phs. Rev. Lett.}\ }\textbf {\bibinfo {volume}
  {110}},\ \bibinfo {pages} {217207} (\bibinfo {year} {2013})}\BibitemShut
  {NoStop}%
\bibitem [{\citenamefont {Greiter}\ \emph {et~al.}(2014)\citenamefont
  {Greiter}, \citenamefont {Schnells},\ and\ \citenamefont
  {Thomale}}]{Greiter2014}%
  \BibitemOpen
  \bibfield  {author} {\bibinfo {author} {\bibfnamefont {M.}~\bibnamefont
  {Greiter}}, \bibinfo {author} {\bibfnamefont {V.}~\bibnamefont {Schnells}}, \
  and\ \bibinfo {author} {\bibfnamefont {R.}~\bibnamefont {Thomale}},\ }\href
  {\doibase 10.1016/j.aop.2014.08.013} {\bibfield  {journal} {\bibinfo
  {journal} {Ann. Phys.}\ }\textbf {\bibinfo {volume} {351}},\ \bibinfo {pages}
  {1026} (\bibinfo {year} {2014})}\BibitemShut {NoStop}%
\bibitem [{\citenamefont {Gebhard}\ \emph {et~al.}(2022)\citenamefont
  {Gebhard}, \citenamefont {Bauerbach},\ and\ \citenamefont
  {Legeza}}]{Gebhard2022}%
  \BibitemOpen
  \bibfield  {author} {\bibinfo {author} {\bibfnamefont {F.}~\bibnamefont
  {Gebhard}}, \bibinfo {author} {\bibfnamefont {K.}~\bibnamefont {Bauerbach}},
  \ and\ \bibinfo {author} {\bibfnamefont {{\"O}.}~\bibnamefont {Legeza}},\
  }\href {\doibase 10.1103/PhysRevB.106.205133} {\bibfield  {journal} {\bibinfo
   {journal} {Phys. Rev. B}\ }\textbf {\bibinfo {volume} {106}},\ \bibinfo
  {pages} {205133} (\bibinfo {year} {2022})}\BibitemShut {NoStop}%
\bibitem [{\citenamefont {Henderson}\ \emph {et~al.}(2022)\citenamefont
  {Henderson}, \citenamefont {Chen},\ and\ \citenamefont
  {Scuseria}}]{Henderson2022}%
  \BibitemOpen
  \bibfield  {author} {\bibinfo {author} {\bibfnamefont {T.~M.}\ \bibnamefont
  {Henderson}}, \bibinfo {author} {\bibfnamefont {G.~P.}\ \bibnamefont {Chen}},
  \ and\ \bibinfo {author} {\bibfnamefont {G.~E.}\ \bibnamefont {Scuseria}},\
  }\href {\doibase 10.1063/5.0125124} {\bibfield  {journal} {\bibinfo
  {journal} {J. Chem. Phys.}\ }\textbf {\bibinfo {volume} {157}},\ \bibinfo
  {pages} {194114} (\bibinfo {year} {2022})}\BibitemShut {NoStop}%
\bibitem [{\citenamefont {Henderson}\ \emph
  {et~al.}(2024{\natexlab{a}})\citenamefont {Henderson}, \citenamefont {Gao},\
  and\ \citenamefont {Scuseria}}]{Henderson2024a}%
  \BibitemOpen
  \bibfield  {author} {\bibinfo {author} {\bibfnamefont {T.~M.}\ \bibnamefont
  {Henderson}}, \bibinfo {author} {\bibfnamefont {F.}~\bibnamefont {Gao}}, \
  and\ \bibinfo {author} {\bibfnamefont {G.~E.}\ \bibnamefont {Scuseria}},\
  }\href {\doibase 10.1080/00268976.2023.2254857} {\bibfield  {journal}
  {\bibinfo  {journal} {Mol. Phys.}\ }\textbf {\bibinfo {volume} {122}},\
  \bibinfo {pages} {e2254857} (\bibinfo {year}
  {2024}{\natexlab{a}})}\BibitemShut {NoStop}%
\bibitem [{\citenamefont {Bravyi}\ and\ \citenamefont
  {Kitaev}(2002)}]{Bravyi2002}%
  \BibitemOpen
  \bibfield  {author} {\bibinfo {author} {\bibfnamefont {S.~B.}\ \bibnamefont
  {Bravyi}}\ and\ \bibinfo {author} {\bibfnamefont {A.~Y.}\ \bibnamefont
  {Kitaev}},\ }\href {\doibase 10.1006/aphy.2002.6254} {\bibfield  {journal}
  {\bibinfo  {journal} {Ann. Phys.}\ }\textbf {\bibinfo {volume} {298}},\
  \bibinfo {pages} {210} (\bibinfo {year} {2002})}\BibitemShut {NoStop}%
\bibitem [{\citenamefont {Cao}\ \emph {et~al.}(2019)\citenamefont {Cao},
  \citenamefont {Romero}, \citenamefont {Olson}, \citenamefont {Degroote},
  \citenamefont {Johnson}, \citenamefont {Kieferov{\'a}}, \citenamefont
  {Kivlichan}, \citenamefont {Menke}, \citenamefont {Peropadre}, \citenamefont
  {Sawaya}, \citenamefont {Sim}, \citenamefont {Veis},\ and\ \citenamefont
  {{Aspuru-Guzik}}}]{Cao2019}%
  \BibitemOpen
  \bibfield  {author} {\bibinfo {author} {\bibfnamefont {Y.}~\bibnamefont
  {Cao}}, \bibinfo {author} {\bibfnamefont {J.}~\bibnamefont {Romero}},
  \bibinfo {author} {\bibfnamefont {J.~P.}\ \bibnamefont {Olson}}, \bibinfo
  {author} {\bibfnamefont {M.}~\bibnamefont {Degroote}}, \bibinfo {author}
  {\bibfnamefont {P.~D.}\ \bibnamefont {Johnson}}, \bibinfo {author}
  {\bibfnamefont {M.}~\bibnamefont {Kieferov{\'a}}}, \bibinfo {author}
  {\bibfnamefont {I.~D.}\ \bibnamefont {Kivlichan}}, \bibinfo {author}
  {\bibfnamefont {T.}~\bibnamefont {Menke}}, \bibinfo {author} {\bibfnamefont
  {B.}~\bibnamefont {Peropadre}}, \bibinfo {author} {\bibfnamefont {N.~P.~D.}\
  \bibnamefont {Sawaya}}, \bibinfo {author} {\bibfnamefont {S.}~\bibnamefont
  {Sim}}, \bibinfo {author} {\bibfnamefont {L.}~\bibnamefont {Veis}}, \ and\
  \bibinfo {author} {\bibfnamefont {A.}~\bibnamefont {{Aspuru-Guzik}}},\ }\href
  {\doibase 10.1021/acs.chemrev.8b00803} {\bibfield  {journal} {\bibinfo
  {journal} {Chem. Rev.}\ }\textbf {\bibinfo {volume} {119}},\ \bibinfo {pages}
  {10856} (\bibinfo {year} {2019})}\BibitemShut {NoStop}%
\bibitem [{\citenamefont {Bauer}\ \emph {et~al.}(2020)\citenamefont {Bauer},
  \citenamefont {Bravyi}, \citenamefont {Mottao},\ and\ \citenamefont
  {Chan}}]{Bauer2020}%
  \BibitemOpen
  \bibfield  {author} {\bibinfo {author} {\bibfnamefont {B.}~\bibnamefont
  {Bauer}}, \bibinfo {author} {\bibfnamefont {S.}~\bibnamefont {Bravyi}},
  \bibinfo {author} {\bibfnamefont {M.}~\bibnamefont {Mottao}}, \ and\ \bibinfo
  {author} {\bibfnamefont {G.~K.}\ \bibnamefont {Chan}},\ }\href {\doibase
  10.1021/acs.chemrev.9b00829} {\bibfield  {journal} {\bibinfo  {journal}
  {Chem. Rev.}\ }\textbf {\bibinfo {volume} {120}},\ \bibinfo {pages} {12685}
  (\bibinfo {year} {2020})}\BibitemShut {NoStop}%
\bibitem [{\citenamefont {McArdle}\ \emph {et~al.}(2020)\citenamefont
  {McArdle}, \citenamefont {Endo}, \citenamefont {{Aspuru-Guzik}},
  \citenamefont {Benjamin},\ and\ \citenamefont {Yuan}}]{McArdle2020}%
  \BibitemOpen
  \bibfield  {author} {\bibinfo {author} {\bibfnamefont {S.}~\bibnamefont
  {McArdle}}, \bibinfo {author} {\bibfnamefont {S.}~\bibnamefont {Endo}},
  \bibinfo {author} {\bibfnamefont {A.}~\bibnamefont {{Aspuru-Guzik}}},
  \bibinfo {author} {\bibfnamefont {S.~C.}\ \bibnamefont {Benjamin}}, \ and\
  \bibinfo {author} {\bibfnamefont {X.}~\bibnamefont {Yuan}},\ }\href {\doibase
  10.1103/RevModPhys.92.015003} {\bibfield  {journal} {\bibinfo  {journal}
  {Rev. Mod. Phys.}\ }\textbf {\bibinfo {volume} {92}},\ \bibinfo {pages}
  {015003} (\bibinfo {year} {2020})}\BibitemShut {NoStop}%
\bibitem [{\citenamefont {Verstraete}\ and\ \citenamefont
  {Cirac}(2005)}]{Verstraete2005}%
  \BibitemOpen
  \bibfield  {author} {\bibinfo {author} {\bibfnamefont {F.}~\bibnamefont
  {Verstraete}}\ and\ \bibinfo {author} {\bibfnamefont {J.~I.}\ \bibnamefont
  {Cirac}},\ }\href {\doibase 10.1088/1742-5468/2005/09/P09012} {\bibfield
  {journal} {\bibinfo  {journal} {J. Stat. Mech.}\ }\textbf {\bibinfo {volume}
  {2005}},\ \bibinfo {pages} {P09012} (\bibinfo {year} {2005})}\BibitemShut
  {NoStop}%
\bibitem [{\citenamefont {Seeley}\ \emph {et~al.}(2012)\citenamefont {Seeley},
  \citenamefont {Richard},\ and\ \citenamefont {Love}}]{Seeley2012}%
  \BibitemOpen
  \bibfield  {author} {\bibinfo {author} {\bibfnamefont {J.~T.}\ \bibnamefont
  {Seeley}}, \bibinfo {author} {\bibfnamefont {M.~J.}\ \bibnamefont {Richard}},
  \ and\ \bibinfo {author} {\bibfnamefont {P.~J.}\ \bibnamefont {Love}},\
  }\href {\doibase 10.1063/1.4768229} {\bibfield  {journal} {\bibinfo
  {journal} {J. Chem. Phys.}\ }\textbf {\bibinfo {volume} {137}},\ \bibinfo
  {pages} {224109} (\bibinfo {year} {2012})}\BibitemShut {NoStop}%
\bibitem [{\citenamefont {Havl\'{\i}\v{c}ek}\ \emph {et~al.}(2017)\citenamefont
  {Havl\'{\i}\v{c}ek}, \citenamefont {Troyer},\ and\ \citenamefont
  {Whitfield}}]{Havlicek2017}%
  \BibitemOpen
  \bibfield  {author} {\bibinfo {author} {\bibfnamefont {V.}~\bibnamefont
  {Havl\'{\i}\v{c}ek}}, \bibinfo {author} {\bibfnamefont {M.}~\bibnamefont
  {Troyer}}, \ and\ \bibinfo {author} {\bibfnamefont {J.~D.}\ \bibnamefont
  {Whitfield}},\ }\href {\doibase 10.1103/physreva.95.032332} {\bibfield
  {journal} {\bibinfo  {journal} {Phys. Rev. A}\ }\textbf {\bibinfo {volume}
  {95}},\ \bibinfo {pages} {032332} (\bibinfo {year} {2017})}\BibitemShut
  {NoStop}%
\bibitem [{\citenamefont {Bravyi}\ \emph {et~al.}(2017)\citenamefont {Bravyi},
  \citenamefont {Gambetta}, \citenamefont {Mezzacapo},\ and\ \citenamefont
  {Temme}}]{Bravyi2017}%
  \BibitemOpen
  \bibfield  {author} {\bibinfo {author} {\bibfnamefont {S.}~\bibnamefont
  {Bravyi}}, \bibinfo {author} {\bibfnamefont {J.~M.}\ \bibnamefont
  {Gambetta}}, \bibinfo {author} {\bibfnamefont {A.}~\bibnamefont {Mezzacapo}},
  \ and\ \bibinfo {author} {\bibfnamefont {K.}~\bibnamefont {Temme}},\ }\href
  {https://arxiv.org/abs/1701.08213} {\enquote {\bibinfo {title} {Tapering off
  qubits to simulate fermionic hamiltonians},}\ } (\bibinfo {year} {2017}),\
  \Eprint {http://arxiv.org/abs/1701.08213} {arXiv:1701.08213 [quant-ph]}
  \BibitemShut {NoStop}%
\bibitem [{\citenamefont {Setia}\ \emph {et~al.}(2019)\citenamefont {Setia},
  \citenamefont {Bravyi}, \citenamefont {Mezzacapo},\ and\ \citenamefont
  {Whitfield}}]{Setia2019}%
  \BibitemOpen
  \bibfield  {author} {\bibinfo {author} {\bibfnamefont {K.}~\bibnamefont
  {Setia}}, \bibinfo {author} {\bibfnamefont {S.}~\bibnamefont {Bravyi}},
  \bibinfo {author} {\bibfnamefont {A.}~\bibnamefont {Mezzacapo}}, \ and\
  \bibinfo {author} {\bibfnamefont {J.~D.}\ \bibnamefont {Whitfield}},\ }\href
  {\doibase 10.1103/physrevresearch.1.033033} {\bibfield  {journal} {\bibinfo
  {journal} {Phys. Rev. Res.}\ }\textbf {\bibinfo {volume} {1}},\ \bibinfo
  {pages} {033033} (\bibinfo {year} {2019})}\BibitemShut {NoStop}%
\bibitem [{\citenamefont {Jiang}\ \emph {et~al.}(2020)\citenamefont {Jiang},
  \citenamefont {Kalev}, \citenamefont {Mruczkiewicz},\ and\ \citenamefont
  {Neven}}]{Jiang2020}%
  \BibitemOpen
  \bibfield  {author} {\bibinfo {author} {\bibfnamefont {Z.}~\bibnamefont
  {Jiang}}, \bibinfo {author} {\bibfnamefont {A.}~\bibnamefont {Kalev}},
  \bibinfo {author} {\bibfnamefont {W.}~\bibnamefont {Mruczkiewicz}}, \ and\
  \bibinfo {author} {\bibfnamefont {H.}~\bibnamefont {Neven}},\ }\href
  {\doibase 10.22331/q-2020-06-04-276} {\bibfield  {journal} {\bibinfo
  {journal} {Quantum}\ }\textbf {\bibinfo {volume} {4}},\ \bibinfo {pages}
  {276} (\bibinfo {year} {2020})}\BibitemShut {NoStop}%
\bibitem [{\citenamefont {Picozzi}\ and\ \citenamefont
  {Tennyson}(2023)}]{Picozzi2023}%
  \BibitemOpen
  \bibfield  {author} {\bibinfo {author} {\bibfnamefont {D.}~\bibnamefont
  {Picozzi}}\ and\ \bibinfo {author} {\bibfnamefont {J.}~\bibnamefont
  {Tennyson}},\ }\href {\doibase 10.1088/2058-9565/acd86c} {\bibfield
  {journal} {\bibinfo  {journal} {Quantum Sci. Technol.}\ }\textbf {\bibinfo
  {volume} {8}},\ \bibinfo {pages} {035026} (\bibinfo {year}
  {2023})}\BibitemShut {NoStop}%
\bibitem [{\citenamefont {Liu}\ \emph {et~al.}(2024)\citenamefont {Liu},
  \citenamefont {Che}, \citenamefont {Zhou}, \citenamefont {Shi},\ and\
  \citenamefont {Li}}]{Liu2024}%
  \BibitemOpen
  \bibfield  {author} {\bibinfo {author} {\bibfnamefont {Y.}~\bibnamefont
  {Liu}}, \bibinfo {author} {\bibfnamefont {S.}~\bibnamefont {Che}}, \bibinfo
  {author} {\bibfnamefont {J.}~\bibnamefont {Zhou}}, \bibinfo {author}
  {\bibfnamefont {Y.}~\bibnamefont {Shi}}, \ and\ \bibinfo {author}
  {\bibfnamefont {G.}~\bibnamefont {Li}},\ }in\ \href@noop {} {\emph {\bibinfo
  {booktitle} {ASPLOS 2024}}}\ (\bibinfo {year} {2024})\BibitemShut {NoStop}%
\bibitem [{\citenamefont {Harrison}\ \emph {et~al.}(2023)\citenamefont
  {Harrison}, \citenamefont {Nelson}, \citenamefont {Adamiak},\ and\
  \citenamefont {Whitfield}}]{Harrison2023}%
  \BibitemOpen
  \bibfield  {author} {\bibinfo {author} {\bibfnamefont {B.}~\bibnamefont
  {Harrison}}, \bibinfo {author} {\bibfnamefont {D.}~\bibnamefont {Nelson}},
  \bibinfo {author} {\bibfnamefont {D.}~\bibnamefont {Adamiak}}, \ and\
  \bibinfo {author} {\bibfnamefont {J.}~\bibnamefont {Whitfield}},\ }\href
  {https://arxiv.org/abs/2211.04501} {\enquote {\bibinfo {title} {Reducing the
  qubit requirement of {Jordan-Wigner} encodings of $n$-mode, $k$-fermion
  systems from $n$ to $\lceil \log_2 {N \choose K} \rceil$},}\ } (\bibinfo
  {year} {2023}),\ \Eprint {http://arxiv.org/abs/2211.04501} {arXiv:2211.04501
  [quant-ph]} \BibitemShut {NoStop}%
\bibitem [{\citenamefont {O'Brien}\ and\ \citenamefont
  {Strelchuk}(2024)}]{OBrien2024}%
  \BibitemOpen
  \bibfield  {author} {\bibinfo {author} {\bibfnamefont {O.}~\bibnamefont
  {O'Brien}}\ and\ \bibinfo {author} {\bibfnamefont {S.}~\bibnamefont
  {Strelchuk}},\ }\href {\doibase 10.1103/PhysRevB.109.115149} {\bibfield
  {journal} {\bibinfo  {journal} {Phys. Rev. B}\ }\textbf {\bibinfo {volume}
  {109}},\ \bibinfo {pages} {115149} (\bibinfo {year} {2024})}\BibitemShut
  {NoStop}%
\bibitem [{\citenamefont {Harrison}\ \emph {et~al.}(2024)\citenamefont
  {Harrison}, \citenamefont {Chiew}, \citenamefont {Necaise}, \citenamefont
  {Projansky}, \citenamefont {Strelchuk},\ and\ \citenamefont
  {Whitfield}}]{Harrison2024b}%
  \BibitemOpen
  \bibfield  {author} {\bibinfo {author} {\bibfnamefont {B.}~\bibnamefont
  {Harrison}}, \bibinfo {author} {\bibfnamefont {M.}~\bibnamefont {Chiew}},
  \bibinfo {author} {\bibfnamefont {J.}~\bibnamefont {Necaise}}, \bibinfo
  {author} {\bibfnamefont {A.}~\bibnamefont {Projansky}}, \bibinfo {author}
  {\bibfnamefont {S.}~\bibnamefont {Strelchuk}}, \ and\ \bibinfo {author}
  {\bibfnamefont {J.~D.}\ \bibnamefont {Whitfield}},\ }\href
  {https://arxiv.org/abs/2409.04348} {\enquote {\bibinfo {title} {A
  {S}ierpinski triangle fermion-to-qubit transform},}\ } (\bibinfo {year}
  {2024}),\ \Eprint {http://arxiv.org/abs/2409.04348} {arXiv:2409.04348
  [quant-ph]} \BibitemShut {NoStop}%
\bibitem [{\citenamefont {Fukutome}\ \emph {et~al.}(1977)\citenamefont
  {Fukutome}, \citenamefont {Yamamura},\ and\ \citenamefont
  {Nishiyama}}]{Fukutome1977}%
  \BibitemOpen
  \bibfield  {author} {\bibinfo {author} {\bibfnamefont {H.}~\bibnamefont
  {Fukutome}}, \bibinfo {author} {\bibfnamefont {M.}~\bibnamefont {Yamamura}},
  \ and\ \bibinfo {author} {\bibfnamefont {S.}~\bibnamefont {Nishiyama}},\
  }\href {\doibase 10.1143/PTP.57.1554} {\bibfield  {journal} {\bibinfo
  {journal} {Prog. Theor. Phys.}\ }\textbf {\bibinfo {volume} {57}},\ \bibinfo
  {pages} {1554} (\bibinfo {year} {1977})}\BibitemShut {NoStop}%
\bibitem [{\citenamefont {Fukutome}(1977)}]{Fukutome1977b}%
  \BibitemOpen
  \bibfield  {author} {\bibinfo {author} {\bibfnamefont {H.}~\bibnamefont
  {Fukutome}},\ }\href {\doibase 10.1143/PTP.58.1692} {\bibfield  {journal}
  {\bibinfo  {journal} {Prog. Theor. Phys.}\ }\textbf {\bibinfo {volume}
  {58}},\ \bibinfo {pages} {1692} (\bibinfo {year} {1977})}\BibitemShut
  {NoStop}%
\bibitem [{\citenamefont {Moussa}(2012)}]{Moussa2018}%
  \BibitemOpen
  \bibfield  {author} {\bibinfo {author} {\bibfnamefont {J.~E.}\ \bibnamefont
  {Moussa}},\ }\href {https://arxiv.org/abs/1208.1086} {\enquote {\bibinfo
  {title} {{Generalized unitary Bogoliubov transformation that breaks fermion
  number parity}},}\ } (\bibinfo {year} {2012}),\ \Eprint
  {http://arxiv.org/abs/1208.1086} {arXiv:1208.1086 [cond-mat.str-el]}
  \BibitemShut {NoStop}%
\bibitem [{\citenamefont {Henderson}\ \emph
  {et~al.}(2024{\natexlab{b}})\citenamefont {Henderson}, \citenamefont
  {{Ghassemi Tabrizi}}, \citenamefont {Chen},\ and\ \citenamefont
  {Scuseria}}]{Henderson2024b}%
  \BibitemOpen
  \bibfield  {author} {\bibinfo {author} {\bibfnamefont {T.~M.}\ \bibnamefont
  {Henderson}}, \bibinfo {author} {\bibfnamefont {S.}~\bibnamefont {{Ghassemi
  Tabrizi}}}, \bibinfo {author} {\bibfnamefont {G.~P.}\ \bibnamefont {Chen}}, \
  and\ \bibinfo {author} {\bibfnamefont {G.~E.}\ \bibnamefont {Scuseria}},\
  }\href {\doibase 10.1063/5.0188155} {\bibfield  {journal} {\bibinfo
  {journal} {J. Chem. Phys.}\ }\textbf {\bibinfo {volume} {160}},\ \bibinfo
  {pages} {064103} (\bibinfo {year} {2024}{\natexlab{b}})}\BibitemShut
  {NoStop}%
\bibitem [{\citenamefont {Chiew}\ and\ \citenamefont
  {Strelchuk}(2023)}]{Chiew2023}%
  \BibitemOpen
  \bibfield  {author} {\bibinfo {author} {\bibfnamefont {M.}~\bibnamefont
  {Chiew}}\ and\ \bibinfo {author} {\bibfnamefont {S.}~\bibnamefont
  {Strelchuk}},\ }\href {\doibase 10.22331/q-2023-10-18-1145} {\bibfield
  {journal} {\bibinfo  {journal} {{Quantum}}\ }\textbf {\bibinfo {volume}
  {7}},\ \bibinfo {pages} {1145} (\bibinfo {year} {2023})}\BibitemShut
  {NoStop}%
\bibitem [{\citenamefont {Wang}(1991)}]{Wang1990}%
  \BibitemOpen
  \bibfield  {author} {\bibinfo {author} {\bibfnamefont {Y.~R.}\ \bibnamefont
  {Wang}},\ }\href {\doibase 10.1103/PhysRevB.43.3786} {\bibfield  {journal}
  {\bibinfo  {journal} {Phys. Rev. B}\ }\textbf {\bibinfo {volume} {43}},\
  \bibinfo {pages} {3786} (\bibinfo {year} {1991})}\BibitemShut {NoStop}%
\bibitem [{\citenamefont {Tranter}\ \emph {et~al.}(2015)\citenamefont
  {Tranter}, \citenamefont {Sofia}, \citenamefont {Seeley}, \citenamefont
  {Kaicher}, \citenamefont {McClean}, \citenamefont {Babbush}, \citenamefont
  {Coveney}, \citenamefont {Mintert}, \citenamefont {Wilhelm},\ and\
  \citenamefont {Love}}]{bk_prop}%
  \BibitemOpen
  \bibfield  {author} {\bibinfo {author} {\bibfnamefont {A.}~\bibnamefont
  {Tranter}}, \bibinfo {author} {\bibfnamefont {S.}~\bibnamefont {Sofia}},
  \bibinfo {author} {\bibfnamefont {J.}~\bibnamefont {Seeley}}, \bibinfo
  {author} {\bibfnamefont {M.}~\bibnamefont {Kaicher}}, \bibinfo {author}
  {\bibfnamefont {J.}~\bibnamefont {McClean}}, \bibinfo {author} {\bibfnamefont
  {R.}~\bibnamefont {Babbush}}, \bibinfo {author} {\bibfnamefont {P.~V.}\
  \bibnamefont {Coveney}}, \bibinfo {author} {\bibfnamefont {F.}~\bibnamefont
  {Mintert}}, \bibinfo {author} {\bibfnamefont {F.}~\bibnamefont {Wilhelm}}, \
  and\ \bibinfo {author} {\bibfnamefont {P.~J.}\ \bibnamefont {Love}},\ }\href
  {\doibase 10.1002/qua.24969} {\bibfield  {journal} {\bibinfo  {journal}
  {International Journal of Quantum Chemistry}\ }\textbf {\bibinfo {volume}
  {115}},\ \bibinfo {pages} {1431} (\bibinfo {year} {2015})},\ \Eprint
  {http://arxiv.org/abs/https://onlinelibrary.wiley.com/doi/pdf/10.1002/qua.24969}
  {https://onlinelibrary.wiley.com/doi/pdf/10.1002/qua.24969} \BibitemShut
  {NoStop}%
\bibitem [{\citenamefont {Thouless}(1960)}]{Thouless1960}%
  \BibitemOpen
  \bibfield  {author} {\bibinfo {author} {\bibfnamefont {D.~J.}\ \bibnamefont
  {Thouless}},\ }\href {\doibase 10.1016/0029-5582(60)90048-1} {\bibfield
  {journal} {\bibinfo  {journal} {Nucl. Phys.}\ }\textbf {\bibinfo {volume}
  {21}},\ \bibinfo {pages} {225} (\bibinfo {year} {1960})}\BibitemShut
  {NoStop}%
\bibitem [{\citenamefont {Bach}\ \emph {et~al.}(1994)\citenamefont {Bach},
  \citenamefont {Lieb},\ and\ \citenamefont {Solovej}}]{Bach1994}%
  \BibitemOpen
  \bibfield  {author} {\bibinfo {author} {\bibfnamefont {V.}~\bibnamefont
  {Bach}}, \bibinfo {author} {\bibfnamefont {E.~H.}\ \bibnamefont {Lieb}}, \
  and\ \bibinfo {author} {\bibfnamefont {J.~P.}\ \bibnamefont {Solovej}},\
  }\href {\doibase 10.1007/BF02188656} {\bibfield  {journal} {\bibinfo
  {journal} {J. Stat. Phys.}\ }\textbf {\bibinfo {volume} {76}},\ \bibinfo
  {pages} {3} (\bibinfo {year} {1994})}\BibitemShut {NoStop}%
\bibitem [{\citenamefont {Colpa}(1979)}]{Colpa1979}%
  \BibitemOpen
  \bibfield  {author} {\bibinfo {author} {\bibfnamefont {J.~H.~P.}\
  \bibnamefont {Colpa}},\ }\href {\doibase 10.1088/0305-4470/12/4/008}
  {\bibfield  {journal} {\bibinfo  {journal} {J. Phys. A: Math. Gen.}\ }\textbf
  {\bibinfo {volume} {12}},\ \bibinfo {pages} {49} (\bibinfo {year}
  {1979})}\BibitemShut {NoStop}%
\bibitem [{\citenamefont {Nishiyama}\ and\ \citenamefont {{da
  Provid{\^e}ncia}}(2019)}]{Nishiyama2019}%
  \BibitemOpen
  \bibfield  {author} {\bibinfo {author} {\bibfnamefont {S.}~\bibnamefont
  {Nishiyama}}\ and\ \bibinfo {author} {\bibfnamefont {J.}~\bibnamefont {{da
  Provid{\^e}ncia}}},\ }\href {\doibase 10.1142/S0219887819501846} {\bibfield
  {journal} {\bibinfo  {journal} {Internat. J. Geom. Meth. in Mod. Phys.}\
  }\textbf {\bibinfo {volume} {16}},\ \bibinfo {pages} {1950184} (\bibinfo
  {year} {2019})}\BibitemShut {NoStop}%
\bibitem [{\citenamefont {Balian}\ and\ \citenamefont
  {Brezin}(1969)}]{Balian1969}%
  \BibitemOpen
  \bibfield  {author} {\bibinfo {author} {\bibfnamefont {R.}~\bibnamefont
  {Balian}}\ and\ \bibinfo {author} {\bibfnamefont {E.}~\bibnamefont
  {Brezin}},\ }\href {\doibase 10.1007/BF02710281} {\bibfield  {journal}
  {\bibinfo  {journal} {Il Nuovo Cimento B}\ }\textbf {\bibinfo {volume}
  {64}},\ \bibinfo {pages} {37} (\bibinfo {year} {1969})}\BibitemShut {NoStop}%
\bibitem [{\citenamefont {Chen}\ and\ \citenamefont
  {Scuseria}(2023)}]{Chen2023}%
  \BibitemOpen
  \bibfield  {author} {\bibinfo {author} {\bibfnamefont {G.~P.}\ \bibnamefont
  {Chen}}\ and\ \bibinfo {author} {\bibfnamefont {G.~E.}\ \bibnamefont
  {Scuseria}},\ }\href {\doibase doi.org/10.1063/5.0156124} {\bibfield
  {journal} {\bibinfo  {journal} {J. Chem. Phys.}\ }\textbf {\bibinfo {volume}
  {158}},\ \bibinfo {pages} {231102} (\bibinfo {year} {2023})}\BibitemShut
  {NoStop}%
\bibitem [{\citenamefont {Schollw{\"o}ck}\ \emph {et~al.}(2004)\citenamefont
  {Schollw{\"o}ck}, \citenamefont {Richter}, \citenamefont {Farnell},\ and\
  \citenamefont {Bishop}}]{Bishop2004}%
  \BibitemOpen
  \bibinfo {editor} {\bibfnamefont {U.}~\bibnamefont {Schollw{\"o}ck}},
  \bibinfo {editor} {\bibfnamefont {J.}~\bibnamefont {Richter}}, \bibinfo
  {editor} {\bibfnamefont {D.~J.~J.}\ \bibnamefont {Farnell}}, \ and\ \bibinfo
  {editor} {\bibfnamefont {R.~F.}\ \bibnamefont {Bishop}},\ eds.,\ \href
  {\doibase 10.1007/BFb0119597} {\emph {\bibinfo {title} {Quantum Magnetism}}}\
  (\bibinfo  {publisher} {Springer},\ \bibinfo {address} {Berlin, Heidelberg},\
  \bibinfo {year} {2004})\BibitemShut {NoStop}%
\bibitem [{\citenamefont {Carretta}\ \emph {et~al.}(2002)\citenamefont
  {Carretta}, \citenamefont {Melzi}, \citenamefont {Papinutto},\ and\
  \citenamefont {Millet}}]{Carretta2002}%
  \BibitemOpen
  \bibfield  {author} {\bibinfo {author} {\bibfnamefont {P.}~\bibnamefont
  {Carretta}}, \bibinfo {author} {\bibfnamefont {R.}~\bibnamefont {Melzi}},
  \bibinfo {author} {\bibfnamefont {N.}~\bibnamefont {Papinutto}}, \ and\
  \bibinfo {author} {\bibfnamefont {P.}~\bibnamefont {Millet}},\ }\href
  {\doibase 10.1103/PhysRevLett.88.047601} {\bibfield  {journal} {\bibinfo
  {journal} {Phys. Rev. Lett.}\ }\textbf {\bibinfo {volume} {88}},\ \bibinfo
  {pages} {047601} (\bibinfo {year} {2002})}\BibitemShut {NoStop}%
\bibitem [{\citenamefont {Bombardi}\ \emph {et~al.}(2004)\citenamefont
  {Bombardi}, \citenamefont {Rodriguez-Carvajal}, \citenamefont {Di~Matteo},
  \citenamefont {de~Bergevin}, \citenamefont {Paolasini}, \citenamefont
  {Carretta}, \citenamefont {Millet},\ and\ \citenamefont
  {Caciuffo}}]{Bombardi2004}%
  \BibitemOpen
  \bibfield  {author} {\bibinfo {author} {\bibfnamefont {A.}~\bibnamefont
  {Bombardi}}, \bibinfo {author} {\bibfnamefont {J.}~\bibnamefont
  {Rodriguez-Carvajal}}, \bibinfo {author} {\bibfnamefont {S.}~\bibnamefont
  {Di~Matteo}}, \bibinfo {author} {\bibfnamefont {F.}~\bibnamefont
  {de~Bergevin}}, \bibinfo {author} {\bibfnamefont {L.}~\bibnamefont
  {Paolasini}}, \bibinfo {author} {\bibfnamefont {P.}~\bibnamefont {Carretta}},
  \bibinfo {author} {\bibfnamefont {P.}~\bibnamefont {Millet}}, \ and\ \bibinfo
  {author} {\bibfnamefont {R.}~\bibnamefont {Caciuffo}},\ }\href {\doibase
  10.1103/PhysRevLett.93.027202} {\bibfield  {journal} {\bibinfo  {journal}
  {Phys. Rev. Lett.}\ }\textbf {\bibinfo {volume} {93}},\ \bibinfo {pages}
  {027202} (\bibinfo {year} {2004})}\BibitemShut {NoStop}%
\bibitem [{\citenamefont {Rams}\ \emph {et~al.}(2020)\citenamefont {Rams},
  \citenamefont {Jochim}, \citenamefont {B{\"o}hme}, \citenamefont {Lohmiller},
  \citenamefont {Ceglarska}, \citenamefont {Rams}, \citenamefont {Schnegg},
  \citenamefont {Plass},\ and\ \citenamefont {N{\"a}ther}}]{Rams2020}%
  \BibitemOpen
  \bibfield  {author} {\bibinfo {author} {\bibfnamefont {M.}~\bibnamefont
  {Rams}}, \bibinfo {author} {\bibfnamefont {A.}~\bibnamefont {Jochim}},
  \bibinfo {author} {\bibfnamefont {M.}~\bibnamefont {B{\"o}hme}}, \bibinfo
  {author} {\bibfnamefont {T.}~\bibnamefont {Lohmiller}}, \bibinfo {author}
  {\bibfnamefont {M.}~\bibnamefont {Ceglarska}}, \bibinfo {author}
  {\bibfnamefont {M.~M.}\ \bibnamefont {Rams}}, \bibinfo {author}
  {\bibfnamefont {A.}~\bibnamefont {Schnegg}}, \bibinfo {author} {\bibfnamefont
  {W.}~\bibnamefont {Plass}}, \ and\ \bibinfo {author} {\bibfnamefont
  {C.}~\bibnamefont {N{\"a}ther}},\ }\href {\doibase 10.1002/chem.201903924}
  {\bibfield  {journal} {\bibinfo  {journal} {Chem. Eur. J}\ }\textbf {\bibinfo
  {volume} {26}},\ \bibinfo {pages} {2837} (\bibinfo {year}
  {2020})}\BibitemShut {NoStop}%
\bibitem [{\citenamefont {Bardeen}\ \emph {et~al.}(1957)\citenamefont
  {Bardeen}, \citenamefont {Cooper},\ and\ \citenamefont
  {Schrieffer}}]{Bardeen1957}%
  \BibitemOpen
  \bibfield  {author} {\bibinfo {author} {\bibfnamefont {J.}~\bibnamefont
  {Bardeen}}, \bibinfo {author} {\bibfnamefont {L.~N.}\ \bibnamefont {Cooper}},
  \ and\ \bibinfo {author} {\bibfnamefont {J.~R.}\ \bibnamefont {Schrieffer}},\
  }\href {\doibase 10.1103/PhysRev.108.1175} {\bibfield  {journal} {\bibinfo
  {journal} {Phys. Rev.}\ }\textbf {\bibinfo {volume} {108}},\ \bibinfo {pages}
  {1175} (\bibinfo {year} {1957})}\BibitemShut {NoStop}%
\bibitem [{\citenamefont {Sierra}\ \emph {et~al.}(2000)\citenamefont {Sierra},
  \citenamefont {Dukelsky}, \citenamefont {Dussel}, \citenamefont {von Delft},\
  and\ \citenamefont {Braun}}]{Sierra2000}%
  \BibitemOpen
  \bibfield  {author} {\bibinfo {author} {\bibfnamefont {G.}~\bibnamefont
  {Sierra}}, \bibinfo {author} {\bibfnamefont {J.}~\bibnamefont {Dukelsky}},
  \bibinfo {author} {\bibfnamefont {G.~G.}\ \bibnamefont {Dussel}}, \bibinfo
  {author} {\bibfnamefont {J.}~\bibnamefont {von Delft}}, \ and\ \bibinfo
  {author} {\bibfnamefont {F.}~\bibnamefont {Braun}},\ }\href {\doibase
  10.1103/PhysRevB.61.R11890} {\bibfield  {journal} {\bibinfo  {journal} {Phys.
  Rev. B}\ }\textbf {\bibinfo {volume} {61}},\ \bibinfo {pages} {R11890}
  (\bibinfo {year} {2000})}\BibitemShut {NoStop}%
\bibitem [{\citenamefont {Darradi}\ \emph {et~al.}(2008)\citenamefont
  {Darradi}, \citenamefont {Derzhko}, \citenamefont {Zinke}, \citenamefont
  {Schulenburg}, \citenamefont {Kr\"uger},\ and\ \citenamefont
  {Richter}}]{Darradi2008}%
  \BibitemOpen
  \bibfield  {author} {\bibinfo {author} {\bibfnamefont {R.}~\bibnamefont
  {Darradi}}, \bibinfo {author} {\bibfnamefont {O.}~\bibnamefont {Derzhko}},
  \bibinfo {author} {\bibfnamefont {R.}~\bibnamefont {Zinke}}, \bibinfo
  {author} {\bibfnamefont {J.}~\bibnamefont {Schulenburg}}, \bibinfo {author}
  {\bibfnamefont {S.~E.}\ \bibnamefont {Kr\"uger}}, \ and\ \bibinfo {author}
  {\bibfnamefont {J.}~\bibnamefont {Richter}},\ }\href {\doibase
  10.1103/PhysRevB.78.214415} {\bibfield  {journal} {\bibinfo  {journal} {Phys.
  Rev. B}\ }\textbf {\bibinfo {volume} {78}},\ \bibinfo {pages} {214415}
  (\bibinfo {year} {2008})}\BibitemShut {NoStop}%
\bibitem [{\citenamefont {Gong}\ \emph {et~al.}(2014)\citenamefont {Gong},
  \citenamefont {Zhu}, \citenamefont {Sheng}, \citenamefont {Motrunich},\ and\
  \citenamefont {Fisher}}]{Gong2014}%
  \BibitemOpen
  \bibfield  {author} {\bibinfo {author} {\bibfnamefont {S.-S.}\ \bibnamefont
  {Gong}}, \bibinfo {author} {\bibfnamefont {W.}~\bibnamefont {Zhu}}, \bibinfo
  {author} {\bibfnamefont {D.~N.}\ \bibnamefont {Sheng}}, \bibinfo {author}
  {\bibfnamefont {O.~I.}\ \bibnamefont {Motrunich}}, \ and\ \bibinfo {author}
  {\bibfnamefont {M.~P.~A.}\ \bibnamefont {Fisher}},\ }\href {\doibase
  10.1103/PhysRevLett.113.027201} {\bibfield  {journal} {\bibinfo  {journal}
  {Phys. Rev. Lett.}\ }\textbf {\bibinfo {volume} {113}},\ \bibinfo {pages}
  {027201} (\bibinfo {year} {2014})}\BibitemShut {NoStop}%
\bibitem [{\citenamefont {Richter}\ \emph {et~al.}(2015)\citenamefont
  {Richter}, \citenamefont {Zinke},\ and\ \citenamefont
  {Farnell}}]{Richter2015}%
  \BibitemOpen
  \bibfield  {author} {\bibinfo {author} {\bibfnamefont {J.}~\bibnamefont
  {Richter}}, \bibinfo {author} {\bibfnamefont {R.}~\bibnamefont {Zinke}}, \
  and\ \bibinfo {author} {\bibfnamefont {D.~J.~J.}\ \bibnamefont {Farnell}},\
  }\href {\doibase 10.1140/epjb/e2014-50589-x} {\bibfield  {journal} {\bibinfo
  {journal} {Eur. Phys. J. B}\ }\textbf {\bibinfo {volume} {88}},\ \bibinfo
  {pages} {2} (\bibinfo {year} {2015})}\BibitemShut {NoStop}%
\bibitem [{\citenamefont {Majumdar}\ and\ \citenamefont
  {Ghosh}(1969)}]{Majumdar1969}%
  \BibitemOpen
  \bibfield  {author} {\bibinfo {author} {\bibfnamefont {C.}~\bibnamefont
  {Majumdar}}\ and\ \bibinfo {author} {\bibfnamefont {D.~K.}\ \bibnamefont
  {Ghosh}},\ }\href {\doibase 10.1063/1.1664978} {\bibfield  {journal}
  {\bibinfo  {journal} {J. Math. Phys.}\ }\textbf {\bibinfo {volume} {10}},\
  \bibinfo {pages} {1388} (\bibinfo {year} {1969})}\BibitemShut {NoStop}%
\bibitem [{\citenamefont {Massaccesi}\ \emph {et~al.}(2021)\citenamefont
  {Massaccesi}, \citenamefont {Rubio-{G}arc{\'i}a}, \citenamefont {Capuzzi},
  \citenamefont {R{\'i}os}, \citenamefont {O{\~n}a}, \citenamefont {Dukelsky},
  \citenamefont {Lain}, \citenamefont {Torre},\ and\ \citenamefont
  {Alcoba}}]{Massaccesi2021}%
  \BibitemOpen
  \bibfield  {author} {\bibinfo {author} {\bibfnamefont {G.~E.}\ \bibnamefont
  {Massaccesi}}, \bibinfo {author} {\bibfnamefont {A.}~\bibnamefont
  {Rubio-{G}arc{\'i}a}}, \bibinfo {author} {\bibfnamefont {P.}~\bibnamefont
  {Capuzzi}}, \bibinfo {author} {\bibfnamefont {E.}~\bibnamefont {R{\'i}os}},
  \bibinfo {author} {\bibfnamefont {O.~B.}\ \bibnamefont {O{\~n}a}}, \bibinfo
  {author} {\bibfnamefont {J.}~\bibnamefont {Dukelsky}}, \bibinfo {author}
  {\bibfnamefont {L.}~\bibnamefont {Lain}}, \bibinfo {author} {\bibfnamefont
  {A.}~\bibnamefont {Torre}}, \ and\ \bibinfo {author} {\bibfnamefont {D.~R.}\
  \bibnamefont {Alcoba}},\ }\href {\doibase 10.1088/1742-5468/abd940}
  {\bibfield  {journal} {\bibinfo  {journal} {J. Stat. Mech.}\ }\textbf
  {\bibinfo {volume} {2021}},\ \bibinfo {pages} {013110} (\bibinfo {year}
  {2021})}\BibitemShut {NoStop}%
\bibitem [{\citenamefont {Liu}\ \emph {et~al.}(2023)\citenamefont {Liu},
  \citenamefont {Gao}, \citenamefont {Chen}, \citenamefont {Henderson},
  \citenamefont {Dukelsky},\ and\ \citenamefont {Scuseria}}]{Liu2023}%
  \BibitemOpen
  \bibfield  {author} {\bibinfo {author} {\bibfnamefont {Z.}~\bibnamefont
  {Liu}}, \bibinfo {author} {\bibfnamefont {F.}~\bibnamefont {Gao}}, \bibinfo
  {author} {\bibfnamefont {G.~P.}\ \bibnamefont {Chen}}, \bibinfo {author}
  {\bibfnamefont {T.~M.}\ \bibnamefont {Henderson}}, \bibinfo {author}
  {\bibfnamefont {J.}~\bibnamefont {Dukelsky}}, \ and\ \bibinfo {author}
  {\bibfnamefont {G.~E.}\ \bibnamefont {Scuseria}},\ }\href {\doibase
  10.1103/PhysRevB.108.085136} {\bibfield  {journal} {\bibinfo  {journal}
  {Phys. Rev. B}\ }\textbf {\bibinfo {volume} {108}},\ \bibinfo {pages}
  {085136} (\bibinfo {year} {2023})}\BibitemShut {NoStop}%
\bibitem [{\citenamefont {Farnell}\ and\ \citenamefont
  {Bishop}(2004)}]{Farnell2004}%
  \BibitemOpen
  \bibfield  {author} {\bibinfo {author} {\bibfnamefont {D.~J.~J.}\
  \bibnamefont {Farnell}}\ and\ \bibinfo {author} {\bibfnamefont {R.~F.}\
  \bibnamefont {Bishop}},\ }\enquote {\bibinfo {title} {The coupled cluster
  method applied to quantum magnetism},}\ in\ \href {\doibase
  10.1007/BFb0119597} {\emph {\bibinfo {booktitle} {Quantum Magnetism}}},\
  \bibinfo {editor} {edited by\ \bibinfo {editor} {\bibfnamefont
  {U.}~\bibnamefont {Schollw{\"o}ck}}, \bibinfo {editor} {\bibfnamefont
  {J.}~\bibnamefont {Richter}}, \bibinfo {editor} {\bibfnamefont {D.~J.~J.}\
  \bibnamefont {Farnell}}, \ and\ \bibinfo {editor} {\bibfnamefont {R.~F.}\
  \bibnamefont {Bishop}}}\ (\bibinfo  {publisher} {Springer},\ \bibinfo
  {address} {Berlin, Heidelberg},\ \bibinfo {year} {2004})\ pp.\ \bibinfo
  {pages} {307--348}\BibitemShut {NoStop}%
\bibitem [{\citenamefont {Richardson}(1963)}]{Richardson1963}%
  \BibitemOpen
  \bibfield  {author} {\bibinfo {author} {\bibfnamefont {R.~W.}\ \bibnamefont
  {Richardson}},\ }\href {\doibase 10.1016/0031-9163(63)90259-2} {\bibfield
  {journal} {\bibinfo  {journal} {Phys. Lett.}\ }\textbf {\bibinfo {volume}
  {3}},\ \bibinfo {pages} {277} (\bibinfo {year} {1963})}\BibitemShut {NoStop}%
\bibitem [{\citenamefont {Richardson}\ and\ \citenamefont
  {Sherman}(1964)}]{Richardson1964}%
  \BibitemOpen
  \bibfield  {author} {\bibinfo {author} {\bibfnamefont {R.~W.}\ \bibnamefont
  {Richardson}}\ and\ \bibinfo {author} {\bibfnamefont {N.}~\bibnamefont
  {Sherman}},\ }\href {\doibase 10.1016/0029-5582(64)90687-X} {\bibfield
  {journal} {\bibinfo  {journal} {Nucl. Phys.}\ }\textbf {\bibinfo {volume}
  {52}},\ \bibinfo {pages} {221} (\bibinfo {year} {1964})}\BibitemShut
  {NoStop}%
\bibitem [{\citenamefont {Richardson}(1965)}]{Richardson1965}%
  \BibitemOpen
  \bibfield  {author} {\bibinfo {author} {\bibfnamefont {R.~W.}\ \bibnamefont
  {Richardson}},\ }\href {\doibase 10.1063/1.1704367} {\bibfield  {journal}
  {\bibinfo  {journal} {J. Math. Phys.}\ }\textbf {\bibinfo {volume} {6}},\
  \bibinfo {pages} {1034} (\bibinfo {year} {1965})}\BibitemShut {NoStop}%
\bibitem [{\citenamefont {Yuzbashyan}\ \emph {et~al.}(2003)\citenamefont
  {Yuzbashyan}, \citenamefont {Baytin},\ and\ \citenamefont
  {Altshuler}}]{Yuzbashyan2003}%
  \BibitemOpen
  \bibfield  {author} {\bibinfo {author} {\bibfnamefont {E.~A.}\ \bibnamefont
  {Yuzbashyan}}, \bibinfo {author} {\bibfnamefont {A.~A.}\ \bibnamefont
  {Baytin}}, \ and\ \bibinfo {author} {\bibfnamefont {B.~L.}\ \bibnamefont
  {Altshuler}},\ }\href {\doibase 10.1103/PhysRevB.68.214509} {\bibfield
  {journal} {\bibinfo  {journal} {Phys. Rev. B}\ }\textbf {\bibinfo {volume}
  {68}},\ \bibinfo {pages} {214509} (\bibinfo {year} {2003})}\BibitemShut
  {NoStop}%
\bibitem [{\citenamefont {Faribault}\ \emph {et~al.}(2010)\citenamefont
  {Faribault}, \citenamefont {Calabrese},\ and\ \citenamefont
  {Caux}}]{Faribault2010}%
  \BibitemOpen
  \bibfield  {author} {\bibinfo {author} {\bibfnamefont {A.}~\bibnamefont
  {Faribault}}, \bibinfo {author} {\bibfnamefont {P.}~\bibnamefont
  {Calabrese}}, \ and\ \bibinfo {author} {\bibfnamefont {J.-S.}\ \bibnamefont
  {Caux}},\ }\href {\doibase 10.1103/PhysRevB.81.174507} {\bibfield  {journal}
  {\bibinfo  {journal} {Phys. Rev. B}\ }\textbf {\bibinfo {volume} {81}},\
  \bibinfo {pages} {174507} (\bibinfo {year} {2010})}\BibitemShut {NoStop}%
\bibitem [{\citenamefont {Johnson}(2023)}]{johnson2023richardsongaudinstates}%
  \BibitemOpen
  \bibfield  {author} {\bibinfo {author} {\bibfnamefont {P.~A.}\ \bibnamefont
  {Johnson}},\ }\href {https://arxiv.org/abs/2312.08804} {\enquote {\bibinfo
  {title} {{Richardson-Gaudin States}},}\ } (\bibinfo {year} {2023}),\ \Eprint
  {http://arxiv.org/abs/2312.08804} {arXiv:2312.08804 [physics.chem-ph]}
  \BibitemShut {NoStop}%
\bibitem [{\citenamefont {Sheikh}\ and\ \citenamefont
  {Ring}(2000)}]{Sheikh2000}%
  \BibitemOpen
  \bibfield  {author} {\bibinfo {author} {\bibfnamefont {J.~A.}\ \bibnamefont
  {Sheikh}}\ and\ \bibinfo {author} {\bibfnamefont {P.}~\bibnamefont {Ring}},\
  }\href {\doibase 10.1016/S0375-9474(99)00424-8} {\bibfield  {journal}
  {\bibinfo  {journal} {Nucl. Phys. A}\ }\textbf {\bibinfo {volume} {665}},\
  \bibinfo {pages} {71} (\bibinfo {year} {2000})}\BibitemShut {NoStop}%
\bibitem [{\citenamefont {Schmid}(2004)}]{Schmid2004}%
  \BibitemOpen
  \bibfield  {author} {\bibinfo {author} {\bibfnamefont {K.}~\bibnamefont
  {Schmid}},\ }\href@noop {} {\bibfield  {journal} {\bibinfo  {journal} {Prog.
  Part. Nucl. Phys.}\ }\textbf {\bibinfo {volume} {52}},\ \bibinfo {pages}
  {565} (\bibinfo {year} {2004})}\BibitemShut {NoStop}%
\bibitem [{\citenamefont {Scuseria}\ \emph {et~al.}(2011)\citenamefont
  {Scuseria}, \citenamefont {Jim{\'e}nez-Hoyos}, \citenamefont {Henderson},
  \citenamefont {Samanta},\ and\ \citenamefont {Ellis}}]{Scuseria2011}%
  \BibitemOpen
  \bibfield  {author} {\bibinfo {author} {\bibfnamefont {G.~E.}\ \bibnamefont
  {Scuseria}}, \bibinfo {author} {\bibfnamefont {C.~A.}\ \bibnamefont
  {Jim{\'e}nez-Hoyos}}, \bibinfo {author} {\bibfnamefont {T.~M.}\ \bibnamefont
  {Henderson}}, \bibinfo {author} {\bibfnamefont {K.}~\bibnamefont {Samanta}},
  \ and\ \bibinfo {author} {\bibfnamefont {J.~K.}\ \bibnamefont {Ellis}},\
  }\href {\doibase 10.1063/1.3643338} {\bibfield  {journal} {\bibinfo
  {journal} {J. Chem. Phys.}\ }\textbf {\bibinfo {volume} {135}},\ \bibinfo
  {pages} {124108} (\bibinfo {year} {2011})}\BibitemShut {NoStop}%
\end{thebibliography}
\end{document}